%% file: varenna.tex
\documentclass[varenna,seceqno]{cimento}
\usepackage{epsfig}

\newcommand{\nin}{\noindent}
\newcommand{\bc}{\begin{center}}
\newcommand{\ec}{\end{center}}
\newcommand{\be}{\begin{equation}}
\newcommand{\ee}{\end{equation}}
\newcommand{\bel}{\begin{eqnletter}}
\newcommand{\eel}{\end{eqnletter}}
\newcommand{\LD}{L_{\rm d}}
\newcommand{\dlk}{\Delta k}
\newcommand{\dlB}{\Delta B}
\newcommand{\dlT}{\delta T}

\newcommand{\gcl}{\gamma_{cl}}
\newcommand{\alc}{\alpha_{cl}}
\newcommand{\sttp}{s(\theta,\theta^{\prime} )}
\newcommand{\uttp}{u(\theta,\theta^{\prime} )}
\newcommand{\uttpt}{u({\tilde \theta},{\tilde \theta}^{\prime} )}

\newcommand{\ba}{\begin{array}}
\newcommand{\ea}{\end{array}}

\newcommand{\dosc}{{d^{\rm osc}}}

\newcommand{\Nosc}{{N^{\rm osc}}}

\newcommand{\Oosc}{{\Omega^{\rm osc}}}

\newcommand{\cl}{\chi_{\scriptscriptstyle L}}
\newcommand{\cgc}{\chi^{\rm \scriptscriptstyle GC}}
\newcommand{\chit}{\chi^{\rm \scriptscriptstyle (t)}}

\newcommand{\EF}{E_{\scriptscriptstyle F}}
\newcommand{\kf}{k_{\scriptscriptstyle F}}
\newcommand{\kb}{k_{\scriptscriptstyle B}}
\newcommand{\vf}{v_{\scriptscriptstyle F}}
\newcommand{\bsM}{{\bf \scriptscriptstyle M}}
\newcommand{\lf}{\lambda_{\scriptscriptstyle F}}
\newcommand{\lt}{l_{\scriptscriptstyle T}}
\newcommand{\taut}{\tau_{\scriptscriptstyle T}}

\newcommand{\gs}{{\sf g_s}}

\newcommand{\dif}{{\rm d}}

\newcommand{\A}{{\cal A}}
\newcommand{\C}{{\cal C}}
\newcommand{\G}{{\cal G}}
\newcommand{\SC}{{\cal S}}

\newcommand{\bM}{{\bf M}}
\newcommand{\bp}{{\bf p}}
\newcommand{\bq}{{\bf q}}
\newcommand{\bA}{{\bf A}}
\newcommand{\bB}{{\bf B}}
\newcommand{\br}{{\bf r}}
\newcommand{\brp}{{\bf r}^{\prime}}
\newcommand{\hV}{{\hat V}}

\newcommand{\smeq}{\! \! = \!}

\title{The semiclassical tool in mesoscopic physics}
\author{Rodolfo A. Jalabert}
\institute{Universit\'e Louis Pasteur, 
3-5 rue de l'Universit\'e, 67084 Strasbourg Cedex, France}
\institute{Institut de Physique et Chimie des Mat\'eriaux de Strasbourg,
23 rue du Loess, 67037 Strasbourg Cedex, France}
\begin{document}

\maketitle

\begin{abstract}
Semiclassical methods are extremely valuable in the study of transport and
thermodynamical properties of ballistic microstructures. By expressing the
conductance in terms of classical trajectories, we demonstrate that quantum
interference phenomena depend on the underlying classical dynamics of non-interacting
electrons. In particular, we are able to calculate the characteristic length of the 
ballistic conductance fluctuations and the weak localization peak in the case of
chaotic dynamics. Integrable cavities are not governed by single scales, but 
their non-generic behavior can also be obtained from semiclassical expansions 
(over isolated trajectories or families of trajectories, depending on the system).
The magnetic response of a microstructure is enhanced with respect to the bulk (Landau)
susceptibility, and the semiclassical approach shows that this enhancement is 
the largest for integrable geometries, due to the existence of families of 
periodic orbits. We show how the semiclassical tool can be adapted to describe
weak residual disorder, as well as the effects of electron-electron interactions. 
The interaction contribution to the magnetic susceptibility also depends on the 
nature of the classical dynamics of non-interacting electrons, and is parametrically 
larger in the case of integrable systems.
\end{abstract}

\newpage
%\tableofcontents

\include{var1}
\include{var2}
\include{var3}
\include{var4}
\include{var5}
\include{var6}
\include{var7}
\include{var8}
\include{varref}

\end{document}

%% file: var1.tex
%=============================================================================

%
% SECTION I: INTRODUCTION
%
% file: var1.tex
%
% last version 7/10/99
%

\section{Introduction}
\label{sec:Intro}

The field of Quantum Chaos, whose recent developments are
reviewed in this volume, deals with general and fundamental
questions, like the dependence of quantum properties on the
underlying classical dynamics of a physical system
\cite{LesHou89,Fer91,ozor:book,gutz_book}. 
Traditionally, there have been few experimental systems 
(Rydberg atoms \cite{FriWin} and microwave cavities 
\cite{Stockrev} among them) where to perform measurements 
and test the 
theoretical ideas of Quantum Chaos. In the last decade, 
low-temperature transport in mesoscopic semiconductor
structures was proposed and used as a new laboratory for
studying Quantum Chaos \cite{Jal90,Mar92} and many
interesting concepts have been developed from such a
connection \cite{Chaos,Chaost,revha,LesHouSt,csf,RUJ95,Klaus}.

The mesoscopic regime is attained in small condensed
matter systems at sufficiently low temperatures for
the electrons to propagate coherently across the 
sample \cite{ALWrev,LesHou94,Datta,Imry}. The phase coherence
of the electron wave-function is broken by an inelastic
event (coupling to an external environment, electron-phonon
or electron-electron scattering, etc) over a distance 
$L_{\Phi}$ larger than the size of the system ($a$).
In a more precise language, we should not talk of 
electrons, which are strongly interacting, but of
Landau quasiparticles, which are the weakly interacting
carriers (at low energies and small temperatures) 
moving in a self consistent field. The quasiparticle
lifetime gives the limitation on $L_{\Phi}$ arising
from electron-electron interactions. Following the
standard practice, we will refer to the carriers as 
``electrons" and we will not distinguish between the
electrostatically imposed external potential and the
self consistent field. 

The view of a mesoscopic system as a single 
phase-coherent unit allows us to deal with a one-particle
problem, where the theoretical concepts of Quantum Chaos
are more simply applied. However, this simplistic 
approach does not describe the physical reality completely 
since in real life $L_{\Phi}$ is larger than $a$ but never
strictly infinite. The fact that Mesoscopic Physics is
not such an ideal laboratory for Quantum Chaos makes
the richness of their relationship. Mesoscopic systems
are extremely useful to study the interplay between
quantum and classical mechanics, and at the same time,
we can use this interplay to test fundamental questions
of Condensed Matter Physics, like decoherence, 
dissipation and many-body effects.

Mesoscopic Physics was initially focused on disordered
metals, where the classical motion of electrons is a 
random walk between the impurities. The phase-coherence
in the multiple scattering of electrons gives rise to 
corrections to the classical (Drude) conductance. The
most studied quantum interference phenomena in disordered
metals are the {\em Aharanov-Bohm oscillations} of the 
conductance in multiply connected geometries, the
{\em weak localization} effect (a decrease in the 
average conductance around zero magnetic field), and the
{\em universal conductance fluctuations} (reproducible
fluctuations in the conductance versus magnetic field
or Fermi energy with {\em rms} of size of the order
$e^2/h$, independent of the average conductance) \cite{WashWebb}. 
A perturbative treatment of disorder, followed by an
average over impurity configurations, has provided 
the calculational tool leading to the understanding 
of those phenomena \cite{LeeRam}. The small parameter is 
$\kf l$, with $\kf=2\pi/\lf$ the Fermi wave-vector and 
$l$ the elastic mean-free-path ({\it i.e.}
the typical distance traveled by the electron between
successive collision with the impurities). Mesoscopic 
disordered conductors are then characterized by
$\lf \ll l \ll a \ll L_{\Phi}$.

It is in a ``second generation" of mesoscopic systems,
semiconductor microstructures, that the connection with 
Quantum Chaos has been more successfully developed.
Extremely pure semiconductor ($GaAs/AlGaAs$) 
heterostructures make it possible to create a 
two-dimensional electron gas (2DEG) by quantizing the 
motion perpendicular to the interphase \cite{BvHra}.
Given the crystalline perfection and the fact that the 
dopants are away from the plane of the carriers, an electron
can travel a long distance before its initial momentum is
randomized. This typical distance, the transport 
mean-free-path $\lt$, is generally larger than the 
elastic mean-free-path (due to small-angle scattering
\cite{DSS}) and it can achieve values of $5-15 \mu m$. 

Various techniques have been developed
to produce a lateral confinement in the 2DEG and
define one-dimensional (quantum wires) and 
zero-dimensional (quantum boxes or cavities) structures.
Spatial resolutions of the order of a {\em micron} allow
to define, at the level of the 2DEG, mesoscopic structures
smaller than the elastic mean-free-path, paving the way
to the {\em ballistic} regime \cite{Rouk,Ford}. When $a \ll \lt$
the classical motion of the two-dimensional electrons is
given by the collisions with the walls defining the 
cavity, with a very small drift due to the weak impurity
potential. We reserve the term of {\em clean} system for
the ideal case in which $\lt$ is strictly infinite, allowing 
us to completely ignore the effects of disorder. In usual
ballistic transport the disorder effects are very small,
and the distinction between ballistic and clean regimes is 
often skipped. On the other hand, thermodynamical 
properties are more sensitive to the residual disorder,
and we will show in this work how to include such effects
within a semiclassical approach. Changing the shape of
a clean cavity we can go from integrable to chaotic dynamics
and study the consequences at the quantum level.

It is important to realize that the constraints arising
from the measurement limit the type of problems to study.
For instance, we cannot address one of the central questions
of Quantum Chaos, the relationship between the (short range)
statistical properties of the spectrum of a quantum system 
and the nature of the underlying classical dynamics \cite{Bohigas}, 
since in the mesoscopic regime we do not have access to single-particle
energies. We do not deal with microscopic systems where the
level spacing $\Delta$ can become larger than the temperature
broadening $k_B T$. The fruitful connection between 
Quantum Chaos and Mesoscopic Physics has to be established 
from the observables that are accessible in the laboratory.

\begin{figure}
\setlength{\unitlength}{1mm}
\begin{picture}(100,110)
\put(20,55)
{\epsfxsize=9cm\epsfbox{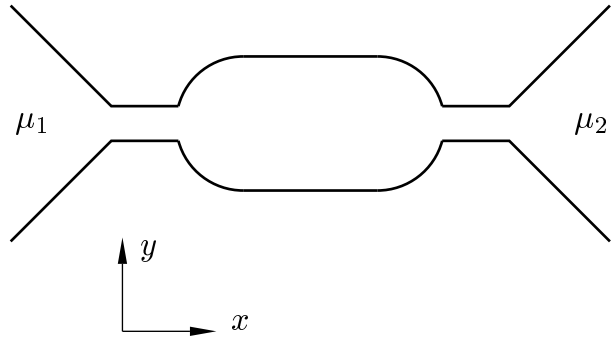}}
\end{picture}
\vspace{-5cm}
\caption{Typical ballistic cavity coupled to reservoirs 
characterized by chemical potentials $\mu_1$ and $\mu_2$.}
\label{fig:intro}
\end{figure}

The broad (long range) features of the density of states of a 
ballistic cavity can be tested by the measurement of its 
magnetic susceptibility, which is a thermodynamical property. 
This is a considerably difficult experiment due to the 
smallness of the orbital response, thus only a few experimental
results exist up to now \cite{levy93,BenMailly}. The physical
property of ballistic microstructures which is most easily
measured is their electrical resistance, and as a consequence,
an important wealth of experimental results on ballistic
transport has been obtained in the last decade 
\cite{Mar92,marcusgroup1,Kel94,berry94,chang94,marcusgroup2,
bird,lutj,lee97,zozou97,marcusgroup3,NRC}.
In order to measure the electrical resistance we have to open
the cavities, connecting them to measuring devices that are
necessarily macroscopic and can be thought as electron reservoirs
(Fig.~\ref{fig:intro}).
In the case of magnetization measurements we are dealing with
a closed system. In the transport experiments we have an
open system with a continuous spectrum. The experimental
situation concerning the existing results is to be contrasted 
with the theoretical one, since closed systems have 
traditionally captured most of the attention in Quantum Chaos.

In this work we review the theory of both, open and closed
ballistic microstructures, in connection with the two types
of cited experiments concerning transport and thermodynamical
properties. The unifying concept in our presentation
is the use of semiclassical approximations for the 
calculation of the quantum observables. The restriction to 
the semiclassical tool is made in order to provide a 
pedagogical and consistent approach to Mesoscopic Physics.
Other techniques, like supersymmetry or random matrix theory
(reviewed in this volume \cite{Mirlin,Haake}) have also 
proven to be very helpful for understanding mesoscopic systems.

Our aim is to present the semiclassical approach as a tool,
showing its principles and illustrating how it works in a few examples. 
It is expected that this introduction can provide an entry
point towards other applications of semiclassics in Mesoscopic
Physics.

Semiclassical approaches were essential at the advent of
Quantum Mechanics and have ever since remained a privileged
tool for developing our intuition on new problems and
for performing analytical calculations as well \cite{gutz_book,brack_book}.
The semiclassical approximation in one-dimension is referred
in standard textbooks as the WKB (Wentzel-Kramers-Brillouin)
method and allows to obtain closed expressions for 
eigenenergies and eigenfunctions. The extension to higher
dimensions is built from the Van Vleck approximation to the
propagator, expressed as a sum over classical
trajectories, each of them associated with a weight given by
a stability prefactor and a phase depending on the classical
action. The consistent use of the stationary-phase method
whenever an integral has to be evaluated allows us to link 
classical mechanics with other quantum protagonists, like
the Green function, the density of states, matrix elements,
scattering amplitudes, etc. The dependence of the properties
of a quantum system on the underlying classical mechanics
can then be established, and this is why the semiclassical
approach is so widely used in the studies of Quantum Chaos.

For the usual densities of the 2DEG ($3\!\times\!10^{11}$ $cm^{-2}$) 
the Fermi wave-lengths are of the order of $40 nm$. The 
semiclassical approximation is therefore justified in the 
study of mesoscopic ballistic cavities since 
$\lf \ll a \ll \lt \ll L_{\Phi}$.

The semiclassical expansions are well suited to study the
nontrivial effect of small perturbations. In this work we 
will often use the fact that a weak magnetic field can be 
accounted for by attaching an extra phase to the action 
associated with each unperturbed trajectory. Other 
perturbations, like a smooth disordered potential or a
change in the Fermi energy can also be treated by this
combination of semiclassics and classical perturbation
theory.

We remark that the use of semiclassical
expansions that we do in the theory of mesoscopic systems
differs from that of Quantum Chaos. The finite values of
$L_{\Phi}$ and $l$, that we emphasized above, and the
thermal broadening due to the Fermi distribution of 
electrons account for the fact that our semiclassical expansions are
always cut off after a certain typical trajectory length.
This observation considerably simplifies our calculations 
and keeps us away from the convergence problems that 
plague semiclassical expansions. Moreover, in certain
cases, like in the orbital magnetism of an integrable
structure, we arrive to a good physical understanding
by only considering a few orbits.

As mentioned before, the first generation of mesoscopic
systems, the disordered metals, have mainly been analyzed 
within diagrammatic perturbation theory. There has also 
been semiclassical descriptions as a way of representing
specific diagrams and providing intuitive interpretations
\cite{KaLa,ChSm}. For instance, it can be shown that the weak localization 
effect arises from the constructive interference of time-reversed 
backscattering trajectories. Since the scattering
centers of disorder metals (defects, impurities, 
interstitials, etc) are of atomic dimensions the single
scattering events have to be treated quantum mechanically.
Therefore the semiclassical description of disordered
systems is a mixed one, built from a classical propagation
between quantum scatterings. Assuming that the classical
single scattering events have a random outcome and 
invoking an ensemble average, we are lead to a diffusive
motion of electrons. The situation is then quite different
from that of ballistic systems, were the classical
trajectories are completely determined by the geometry
and the dynamics can be chaotic or integrable depending
on the shape of the cavity. Also, the notion of impurity
average, so crucial in disordered systems, is usually
replaced in the ballistic regime by averages over energy
or over samples.

At this point it is useful to establish a further clarification
concerning our use of the term ``semiclassics". In
Condensed Matter Physics the semiclassical model usually
describes the evolution of a wave-packet of Block electron
levels by classical equations of motion that take into
account the band structure effects through the dispersion
relations \cite{Ashcroft}. Such a description allows the study
of classical transport from the Boltzmann equation and 
the determination of Fermi surfaces in metals. The periodic
potential of the crystal is included quantum mechanically,
while the externally applied field is treated classically.
The band structure effects in the 2DEG are extremely simple
since we are always restricted to the bottom of the 
conduction band, which can be taken as parabolic with an
effective mass $m=0.067m_{e}$. For the usual Fermi wave-lengths
of the order of $40 nm$ the electrons are
distributed over many atomic sites, and as in standard 
condensed matter textbooks we include the periodic potential
through the dispersion relation. However, we go beyond the 
classical description of electron propagation in the applied
field in order to incorporate the quantum interference 
effects.

This review article is organized as follows. In Sec.~\ref{sec:qttccc}
we present the scattering approach to coherent transport and 
develop a semiclassical theory for conductance. The conductance
fluctuations \cite{Jal90,Blu88} and the weak localization effect \cite{Bar93} are
studied with the semiclassical tool and differences according
to the nature of the underlying classical dynamics are predicted.
Sec.~\ref{sec:saiiqt} deals with integrable cavities: the square, for which the
semiclassical expansions are organized in terms of families of 
trajectories \cite{paul} and the circle, where the effects of diffraction
and tunneling can be easily incorporated into the semiclassical
description \cite{ingold}. Sec. \ref{sec:eobtaoaott} describes a few experimental results
related with the theories previously developed and discusses
the relation of complementarity between semiclassics and random
matrix theory. In that section we also refer to the case of
cavities with mixed dynamics and the semiclassical approach to 
bulk conductivity. 

The second part of this work is devoted to orbital magnetism. 
The magnetic susceptibility of clean dots \cite{vO94,URJ95} is 
calculated in Sec.~\ref{sec:clean}. Sec.~\ref{sec:disord} studies weak-disorder effects
in ballistic cavities, and applies such a study to the magnetic
response of semiconductor microstructures \cite{RapComm}. In Sec.~\ref{sec:eeiitbr}
we include the effects of electron-electron interaction in the
orbital response of quantum dots \cite{ubroj98} summarizing the 
various contributions in different regimes and establishing a
comparison with the existing experimental results. In Sec.~\ref{sec:conc}
we conclude with a summary of the strenghts and weakness of the
semiclassical approach in Mesoscopic Physics and we point at 
some open problems.

%% file: var2.tex
%=============================================================================

%
% SECTION II: QUANTUM TRANSPORT THROUGH CLASSICALLY CHAOTIC CAVITIES
%
% file: var2.tex
%
% last version: 7/10
%

\section {Quantum transport through classically chaotic cavities}
\label{sec:qttccc}

Our understanding of quantum transport greatly owes to the
Landauer-B\"uttiker approach of viewing conductance as a
scattering problem \cite{Landauer,But86,Datta,Imry}. The
interpretation of conductance measurements in the ballistic
regime from a Quantum Chaos point of view will therefore
involve the study of the quantum and classical mechanics of
open systems. In this section we describe some features of
classical scattering relevant for quantum conductance,
we give a simple presentation of the Landauer-B\"uttiker 
formalism, and we use semiclassical methods to relate
classical and quantum properties of ballistic cavities. 

\subsection{Chaotic scattering}
\label{subsec:qs}

The study of a physical system from the Quantum Chaos point of
view usually starts with its classical dynamics. In open systems,
we have to consider
a classical scattering problem. The concept of chaos \cite{arnold:book}, 
developed for closed systems, and related to the long-time properties 
of the trajectories, has to be re-examined in open systems since the
trajectories exit the scattering region after a finite amount of
time. We will not review the field of Chaotic Scattering
\cite{jung,tel,LesHouSm}, but only present the minimal 
information needed to understand the quantum properties
of ballistic cavities. In this section we will mainly deal 
with chaotic dynamics, but not exclusively; some of the results 
are general and independent of the dynamics. In addition, we will 
stress the differences between chaotic and integrable cases.

The ``transient chaos" of a scattering problem is characterized
by the infinite set of trajectories which stay in the scattering
region forever. This set is constituted by the periodic unstable
orbits of the scattering region (the ``strange repeller") and
their stable manifold (the trajectories that converge to the
previous ones in the infinite-time limit). Chaotic scattering
is obtained when the dynamics in the neighborhood of the repeller
is chaotic in the usual sense, and this set has a fractal 
dimension in the space of classical trajectories. When an 
incoming particle enters the scattering region, it approaches
the strange repeller, bounces around close to this set for a 
while and it is eventually ejected from the scattering region 
(if it did not have the right initial conditions to be trapped). 

If we scan a set of scattering trajectories (say, we fix the 
initial position $y$ at the left entrance of the cavity of 
Fig.~\ref{fig:intro} and we vary the injection angle $\theta$)
studying the time $\tau$ that the particle spends in the interaction
region, we obtain a fractal curve for $\tau(\theta)$. The
infinitely trapped trajectories give the divergences of 
$\tau(\theta)$ and determine its self-similar structure.
The study of $\tau(\theta)$ is a quick way to determine
if our scattering is chaotic.

The rate at which particles escape from the scattering region
($\gamma$) results from a balance between the rate in which
nearby trajectories diverge away from the repeller (characterized
by its largest Lyapunov exponent $\lambda$) and the rate at
which the chaotic escaping trajectories are folded back into the
scattering region (depending on the density of the repeller,
that is measured by its fractal dimension $d$). More precisely,
if we start (or inject) particles in the scattering region,
the survival probability at time $\tau$ will be 
$P(\tau)=e^{-\gamma \tau}$, with $\gamma=\lambda(1-d)$
\cite{tel,LesHouSm,Gas89}. The {\em escape rate} may be interpreted
as the inverse of the typical time spent by the particles in
the scattering region.

\begin{figure}
\setlength{\unitlength}{1mm}
\begin{center}
\begin{picture}(110,145)
\leavevmode
\epsfxsize=0.85\textwidth
\epsfbox{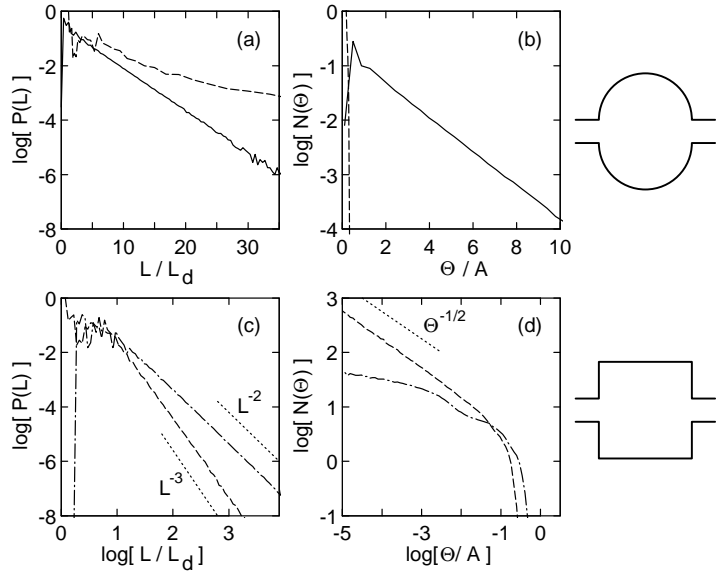}
\end{picture}
\end{center}
\vspace{-6cm}
\caption{Classical distributions of length [(a),(c)] and effective area [(b),(d)]
for the stadium (solid line) and rectangular (dashed line) billiards.
In the stadium both distributions are close to exponential after a short transient 
region and are very different from the distributions for the rectangle, which
show the power-law behavior characteristic of non-chaotic systems.
Different power laws are indicated by
dotted lines. In panel (c), the dash-dotted line is the two-particle
distribution of length differences (Eq.~(\protect\ref{eq:Cdkreg})). In (d) the
dash-dotted line is the two-particle distribution of area differences
for transmitted particles (Eq.~(\protect\ref{eq:CdBreg})).
($A$ is the area of the cavity, $L_{\rm d}$ is the direct length between the
leads. From Ref. \protect\cite{Chaost}.)}
\label{fig:distr}
\end{figure}

In Fig.~\ref{fig:distr}.a we show the length
distribution (which in billiards is equivalent to the length
distribution) for a cavity with the shape of a stadium, and
verify the exponential law (solid line), 
$P(L)=e^{-\gcl L}$ (with $\gcl=\gamma/v$ and
$v$ the constant velocity of the scattering particles). Our curve 
becomes ragged for large lengths due to the finite number of
particles that we are able to simulate. The exponential law
sets in very fast, after a length corresponding to a few bounces.

The appearance of a single scale is not surprising since in 
chaotic scattering the particle moves ergodically over the whole
energy surface while in the scattering region. The value of the
escape rate can be estimated from general arguments of ergodicity
in the case of chaotic cavities with small openings, where the
typical trajectory bounces around many times before it escapes
\cite{Jen91}. Assuming that the instantaneous distribution of
trajectories is uniform on the energy surface, the escape rate
is simply given by $\gamma=F/A$, where $F$ is the flux through
the holes (equal to the size of the holes times $v/\pi$, the factor
of $\pi$ comes from integration over the departing angles), $A$
is the area of the two-dimensional scattering domain. In the case
of small holes this simple estimate reproduces remarkably well 
the escape rates obtained from the numerical determination of
the survival probability using classical trajectories.

Our interest in the escape rate is due to the fact that, as
we will show in the sequel, the energy scale of the conductance
fluctuations is given by this classical quantity. In addition, we
will show that the conductance fluctuations as a function of 
magnetic field are governed by the area distribution. Scattering 
trajectories are open, and therefore do not have a well defined 
enclosed area. Instead, we define the ``effective area" of a 
trajectory $s$ from the circulation of the vector potential along 
the path $\C_s$. We assume a perpendicular magnetic field 
$\bB=B \bf{\hat z}$ generated by a vector potential $\bA$, 
and define the effective area by 

\be
\Theta_s = \frac{2 \pi}{B} \int_{\C_s} \bA \cdot \bf{{\rm d}r} \ ,
\label{eq:ea}
\ee

\nin If $s$ were a closed trajectory, $\Theta_s$ would be equal
to $2\pi$ times the enclosed area.
Unlike the scattering time, $\Theta$ can be positive or
negative. In the chaotic case the distribution of effective
areas, like the distribution of lengths, is expected to depend on a 
single scale. Numerical calculations and analytical arguments 
\cite{Jal90,Jen91,Mac92,RamLech,Ber86,Dor91}
yield a distribution 

\be
N(\Theta) \propto \exp{(-\alc |\Theta|)} \ ,
\label{eq:efardis}
\ee

\nin where the parameter $\alc$ can be interpreted as the inverse of the
typical area enclosed by a scattering trajectory. In 
Fig.~\ref{fig:distr}.$b$ we present the distribution 
$N(\Theta)$ (only for positive $\Theta$, solid line) obtained
from simulating classical trajectories in a stadium cavity,
showing good agreement with the proposed distribution.
Scattering trajectories yield an effective area which is not
gauge-invariant. However, the large (in absolute value) 
effective areas are associated with long trajectories that
bounce many times, which are constituted by many loops
and two extreme ``legs" in and out of the cavity. The dominant
contribution comes from the loops and therefore it is 
gauge-invariant. Changing the choice of the gauge in our 
numerical simulations modifies the distribution for small 
$\Theta$, but not the exponent $\alc$ governing the
distribution of large $\Theta$.

Exploiting the ergodic properties of the chaotic dynamics in 
the scattering domain and assuming that the area is accumulated
in a random-walk fashion, the parameter $\alc$ can be related
to the escape rate and the typical length scale of the cavity
\cite{Jen91,Dor91}. As in the case of the escape rate, these estimations
compare very favorably with the values extracted from numerical
simulations.

In the integrable case the particle moves over only that part
of the energy surface consistent with the conserved quantity.
Therefore, we no longer have a single scale. In situations with
multiple scales we expect to observe power-law distributions
\cite{Bau91,Mac92,Lai92,Lin93}. 
For the case of a rectangular cavity we obtain the results of
Fig.~\ref{fig:distr}$c$ and $d$ with an approximate $L^{-3}$
dependence for the length distribution (dashed) and 
$\Theta^{-1/2}$ for the area distribution (dashed). However,
for integrable cavities the concept of distributions is not
completely rigorous since the survival probabilities depend on 
the initial conditions chosen for the trajectories. For our simulations 
we have used a uniform distribution of $y$ along the entrance lead and 
a $\cos{\theta}$ weighted angular distribution as initial conditions 
(consistently with the classical limit of the quantum problem).

It is important to recall our discussion about the
differences between the traditional Quantum Chaos approach 
and our view of Mesoscopic Physics describing condensed 
matter systems. For instance, the only features of the length 
and area distributions that will be experimentally relevant are 
those happening at scales just a few times larger than the 
typical size and area of the cavity (trajectories shorter than
our physical cut-offs). Therefore, we will not be concerned
about the tails of such distributions \cite{Leg90,Lin93}.

\subsection{The scattering approach to the conductance}

Within the Landauer-B\"uttiker approach, in the 
phase-coherent regime the resistance is not related to an 
intensive resistivity of the type defined in standard condensed 
matter books \cite{Ashcroft} (like, for instance, electron-phonon
interaction), but arises from the elastic scattering 
that electrons suffer while traversing mesoscopic sample between the 
measuring devices. The measuring devices are macroscopic and behave 
as electron reservoirs. They are characterized by an
electrochemical potential $\mu$, which does not vary while
giving and accepting electrons. The role of the reservoirs
is crucial as they render the total system infinite, and
the spectrum continuous. It is only in the reservoirs that the
randomization of electron phases is assumed to take place. 

The simplest 
experimental set up is the two-probe measurement (Fig.~\ref{fig:intro}), 
where the sample is attached between two reservoirs whose electrochemical
potentials differ by the value of the applied voltage $V$, which
is supposed to be very small ($\mu_1\!-\!\mu_2=eV \ll \mu_1$). 
In this work we will concentrate exclusively in ideal two-probe 
experiments. The multiprobe case \cite{But86} does not pose new 
fundamental problems but the theoretical description becomes
more complicated since a matrix of conductance coefficients
must be introduced. These conductances are also expressed in terms of 
transmission coefficients, which can be treated with the 
semiclassical approach \cite{BvH88,Jal90,Bar91,revha,LesHouSt} in 
the same way as in the two-probe case that we discuss here.

The scattering description necessitates a set of 
asymptotic states. In our case such a set is provided
by the propagating channels of the leads connecting
the sample with the reservoirs. Sample, reservoirs
and leads are the three key elements of ballistic
transport. 

Assuming the leads to be disorder-free, with hard walls (of width $W$)
in the $y$-direction and infinite in the $x$-direction, their 
eigenstates with energy $\varepsilon$ are products particle-in-a-box 
wave-functions

\be
\phi_{a}(y)=\sqrt{\frac{2}{W}} \ \sin{\left(\frac{\pi a y}{W}\right)}
\ , 
\label{eq:twf}
\ee

\nin ($a$ integer) in the transverse direction and plane-waves 
propagating in the longitudinal direction, with wave-vectors $k_a$ such
that $\varepsilon=\hbar^2 /(2m)((a \pi/W)^2 + k^{2}_{a})$.
The $N$ transverse momenta which satisfy this relationship
with $k_{a}^2 > 0$ define the $2N$ propagating channels of the leads
with energy $\varepsilon$. The incoming lead-states are 

\be
\label{eq:chann}
\varphi_{1(2),\varepsilon,a}^{(-)}(\br) = \frac{1}{v_{a}^{1/2}} \ e^{\pm i k_{a} x} \
\phi_{a}(y) \ , \hspace{1.2cm} \br=(x,y) \ , \hspace{0.6cm} a=1,\ldots N \ .
\ee

\nin The normalization factor $v_{a}^{-1/2}=(m/\hbar k_{a})^{1/2}$
is chosen in order to have a unit of incoming flux in 
each channel. The subindex 1 (2) corresponds to channels 
propagating from the left (right) reservoir with longitudinal momenta
$k_{a}$ ($-k_{a}$), and $k_{a}$ explicitly positive. 
The outgoing lead-states $\varphi_{1(2),\varepsilon,a}^{(+)}(\br)$ are
defined as in (\ref{eq:chann}), but with the $\pm$ of the exponent inverted.
The time order of outgoing and incoming lead-states is obtained by giving
an infinitesimal positive (negative) imaginary part to $k_a$.

The scattering states corresponding to an electron incoming from lead
1 (2) with energy $\varepsilon$, in the mode $a$ are given, in the 
asymptotic regions, by

\begin{mathletters}
\label{allAFSS}
\begin{eqnarray}
\Psi(\br)_{1,\varepsilon,a}^{(+)} & = & \left\{ \begin{array}{ll}
\varphi_{1,\varepsilon,a}^{(-)}(\br) + \sum_{b=1}^{N} r_{ba} \varphi_{1,\varepsilon,b}^{(+)}(\br)
\ , & \hspace{1cm} \mbox{$x < 0$} \\
\sum_{b=1}^{N} t_{ba} \varphi_{2,\varepsilon,b}^{(+)}(\br) 
\ , & \hspace{1cm} \mbox{$x > 0$}
\end{array} \right. \ .
\label{eq:AFSS0} \\
\displaystyle
\Psi(\br)_{2,\varepsilon,a}^{(+)} & = & \left\{ \begin{array}{ll}
\sum_{b=1}^{N} t^{\prime}_{ba} \varphi_{1,\varepsilon,b}^{(+)}(\br)
\ , & \hspace{1cm} \mbox{$x < 0$} \\
\varphi_{2,\varepsilon,a}^{(-)}(\br) + \sum_{b=1}^{N} r^{\prime}_{ba} \varphi_{2,\varepsilon,b}^{(+)}(\br)
\ , & \hspace{1cm} \mbox{$x > 0$}
\end{array} \right. \ .
\label{eq:AFSS1}
\end{eqnarray}
\end{mathletters}

The $2N \! \times \! 2N$ scattering matrix $S$, relating incoming flux and outgoing fluxes,
can be written in terms of the $N \times N$ reflection and transmission matrices
$r$ and $t$ ($r'$ and $t'$) from the left (right) as

\begin{equation}
S = \left( \begin{array}{lr}
r       & \hspace{0.5cm} t'     \\
t       & \hspace{0.5cm} r'
\end{array} \right) \ .
\label{eq:Smat}
\end{equation}

\nin Current conservation implies that the incoming flux should be equal to 
the outgoing flux, and therefore $S$ is unitary ($S S^{\dagger} = I$). 
In terms of the total transmission ($T=\sum_{a,b}|t_{ba}|^2$) and reflection 
($R=\sum_{a,b}|r_{ba}|^2$) coefficients, the unitarity condition is expressed as $T+R=N$.  
Also, unitarity dictates that $T\!=\!T'$ and $R\!=\!R'$.
Furthermore, in the absence of magnetic field, time reversal invariance
dictates that $S$ is symmetric ($S=S^{\rm T}$). Cavities with geometrical
symmetries (up-down or right-left) are described by scattering matrices
with a block structure \cite{BaMe96}.

The set $\{\Psi_{1(2)\varepsilon,a}^{(+)} \}$ constitutes an
orthogonal (but not orthonormal) basis, \cite{LanAbra,BarSto},

\be
\label{eq:norm}
\int \dif \br \ \Psi_{l,\varepsilon,a}^{(+)*}(\br) \Psi_{l^{\prime},\varepsilon^{\prime},a^{\prime}}^{(+)}(\br)
= \frac{2 \pi}{v_{a}} \ \delta_{aa^{\prime}} \ \delta(k_a-k_{a^{\prime}}) \ \delta_{ll^{\prime}} \ . 
\ee

\nin Using the spectral decomposition of the retarded Green function in this basis and 
taking into account the analytical properties of the transmission amplitudes in the 
complex $k$-plane, we can relate the Green function to the
scattering amplitudes. Alternatively, the formal theory of scattering
(Lippmann-Schinger) can be adapted to wave-guides and obtain \cite{FishLee}

\begin{mathletters}
\label{allTRAMs}
\begin{eqnarray}
\hspace{1.0cm} 
t_{ba} & = & i\hbar(v_{a}v_{b})^{1/2} \ \exp{\left[-i(k_b x^{\prime} - k_a x)\right]}
\int_{\SC_{x'}} dy^{\prime} \int_{\SC_x} dy \ 
\phi_{b}^{*}(y^{\prime}) \ \phi_{a}(y) \ G(\br^{\prime},\br;E)
\label{eq:TRAM0} \\
\hspace{1.0cm} r_{ba} & = & -\delta_{ab}\ \exp{\left[2ik_b x^{\prime}\right]} 
\label{eq:TRAM1} \\
\displaystyle
& + &  i \hbar(v_{a}v_{b})^{1/2}
\ \exp{\left[-i(k_b x^{\prime} + k_a x)\right]} \int_{\SC_{x'}} dy^{\prime} \int_{\SC_x} dy \
\phi_{b}^{*}(y^{\prime}) \ \phi_{a}(y) \ G(\br^{\prime},\br;E) \ ,
\nonumber
\end{eqnarray}
\end{mathletters}

\nin where the integrations take place at the transverse cross sections $\SC_x$ 
on the left lead and $\SC_{x'}$ on the right (left) lead for the transmission
(reflection) amplitudes. The physical observables are obtained from the transmission
and reflection coefficients ($T_{ba}=|t_{ba}|^2$ and $R_{ba}=|r_{ba}|^2$) between modes,
which, by current conservation, do not depend on the choice of the transverse cross sections.
We will use this freedom to take $\SC_x$ and $\SC_{x'}$ at the entrance
and exit of the cavity (or both at the entrance for (\ref{eq:TRAM1})), and we will 
omit the $x$ and $x'$ dependences henceforth.

The intuitive interpretation of the above equations as a particle arriving at the
cavity in mode $a$, propagating inside (through the Green function),
and exiting in mode $b$ is quite straightforward.
Expressing the scattering amplitudes in terms of 
Green functions is extremely useful for analytical
and numerical computations. Diagrammatic perturbation
theory, as well as semiclassical expansions, are built
on Green functions. 

So far, we have presented the scattering theory for 
samples connected to wave-guides. Now, we reproduce
the standard counting argument to relate conductance
with scattering \cite{Landauer,But86,Datta,Imry}. As
stated at the beginning of this section, we assume that
the left reservoir has an electrochemical potential
$\mu_1$ slightly higher than the one of the right reservoir
($\mu_1\!-\!\mu_2=eV$). 

In the energy interval $eV$ between $\mu_2$ and $\mu_1$ 
electrons are injected into right-going states emerging
from reservoir 1, but none are injected into left-going states
emerging from reservoir 2. Consequently, there is a net right-going 
current proportional to the number of states in the interval 
$\mu_1 \! - \! \mu_2$, given by 

\be
 I = \gs e \sum_{a=1}^{N} v_a \ {d{\rm n}_a \over d\varepsilon} \ eV
\sum_{b=1}^{N} T_{ba} = \gs \ \frac{e^2}{h} \ \left(\sum_{a,b=1}^{N} T_{ba}\right)V \ .
\label{Lan1}
\ee

\nin $N$ is the number of propagating channels at the energy
$\mu_1$, the factor $\gs\!=\!2$ takes into account spin degeneracy,
$\sum_{b=1}^{N} T_{ba}$ is the probability for an electron coming in the
mode $a$ to traverse the system, $d{\rm n}_a/d\varepsilon$ quasi-one-dimensional 
density of states (which for non-interacting particles satisfies that 
$d{\rm n}_a /d\varepsilon \! = \! 1/hv_a$). Then, the two-probe conductance is just 
proportional to the total transmission coefficient of the microstructure,

\be
 G = \frac{I}{V} = \gs \ \frac{e^2}{h} \ T = 
\gs \ \frac{e^2}{h} \ {\rm Tr} \{t t^{\dagger}\} \ .
\label{Lan}
\ee

A more rigorous alternative to the counting argument
presented above can be obtained from linear response
(Kubo formula) for the conductivity, within a wave-guide 
geometry \cite{FishLee,Szafer}. The extension to finite
magnetic fields \cite{BarSto,Shepard,NSB}
presents some subtleties, but the final form is 
still the simple looking Eq.~(\ref{Lan}). The magnetic
fields that we consider will always be very
weak, and therefore the zero-field formulation of
the conductance that we presented is sufficient
for our purposes.

\subsection{Quantum interference in ballistic cavities}

The scattering formalism presented in the last chapter is the
base for the semiclassical theory of ballistic transport that
we develop in this work. It is also at the origin of the
numerical studies of elastic scattering due to impurities or
transport through phase-coherent cavities. In this chapter we
describe the results of numerical calculations as a way to
introduce the quantum interference effects that we are 
interested in. These exact computations will later be used
to test the applicability of our analytical results.

In a tight-binding model representing the cavity and the
leads, the Green function can be calculated recursively,
starting from its exact expression in the leads, by matrix
multiplications and inversions \cite{LeeFish,MacK,Bar91}.
The discrete version of Eq.~(\ref{allTRAMs}), together with
(\ref{Lan}), allow us to obtain the transmission coefficient
through the cavity.

We can simulate any potential in the cavity by choosing the
in-site energies of the tight-binding Hamiltonian. In this
work we will concentrate in cavities that are billiards
(defined by hard-walls and with zero potential in the 
classically allowed region). This choice is the simplest
for numerical and analytical calculations, and allows us
to treat the case of a classical dynamics that exhibits hard
chaos. It is clearly a rough approximation to the experimentally
achievable micro-cavities, and its applicability depends on the
fabrication details \cite{Chaos,NRC}.

\begin{figure}
\setlength{\unitlength}{1mm}
\begin{center}
\begin{picture}(110,145)
\leavevmode
\epsfxsize=0.85\textwidth
\epsfbox{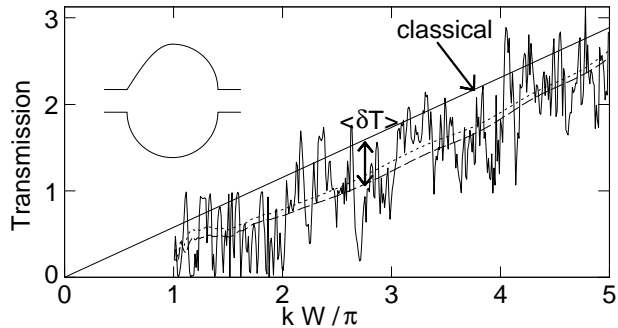}
\end{picture}
\end{center}
\vspace{-8cm}
\caption{
Transmission coefficient for the cavity of the inset as a function of 
wave-vector ($W$ is the width of the leads). The straight solid line 
is the classical transmission, the fluctuating solid line is 
the full quantum transmission $T_{qm}$, and the dashed (dotted) line is 
the smoothed $T_{qm}$ at $B\!=\!0$ ($BA/ \Phi_{0}\!=\!0.25$). 
(From Ref. \protect\cite{Chaost}.)
}
\label{fig:preview}
\end{figure}

In Fig.~\ref{fig:preview} we show the transmission coefficient
of an asymmetric cavity as a function of the incoming flux
$kW/\pi$ ($\mu_1=\hbar^2 k^2/2m$, the integer part of $kW/\pi$
is the number of propagating channels). The overall behavior
of the transmitted flux is a linear increase with $k$. In the next chapters 
we will show that the classical limit of the
semiclassical approximation, corresponding to the neglect of
quantum interference, reproduces the slope of this secular
behavior, which is noted as ``classical". The
term ``classical" should be taken with some caution, since
the incoming flux is still quantized in the modes of the leads.

Superimposed to the secular behavior, we observe fine-structure
fluctuations characteristic of the cavity under study. These
conductance fluctuations, analogous to those of disordered
metals, also appear if we fix the Fermi energy and use the
magnetic field as a tuning parameter. The conductance
fluctuations are characterized by their magnitude, $\langle(\dlT)^2\rangle$,
and the correlation scale as a function of wave-vector $\Delta k_c$ 
(or magnetic field $\Delta B_c$). Our numerical results seem to indicate
that these characteristic scales do not change as we go into
the semiclassical limit of large $k$. However, these results are 
necessarily
limited by the maximum wave-vector that we are able to treat
(scaling with the size of our tight-binding model). In the following 
chapters we will show how semiclassics can be used to extract 
energy-independent values of $\Delta k_c$ and $\Delta B_c$ in chaotic billiards.
The independence of $\langle(\dlT)^2\rangle$ on the incident energy is a necessary
condition if the ballistic conductance fluctuations are to share
some of the universal properties of those of disordered systems.
Semiclassical arguments will be used to support such scale invariance
for chaotic billiards, and we will point out the difficulties concerning 
the calculation of $\langle(\dlT)^2\rangle$.

The secular behavior (dashed line) lies below the classical value
of the transmission coefficient, due to mode effects from confinement 
in the leads. The above defined classical limit only reproduces
the slope of the large-$k$ smoothed transmission coefficient, but
the shift $\langle \delta T \rangle$ (as well as the fluctuations)
does not disappear in the large-$k$ limit. The presence of a 
weak magnetic field tends to decrease the offset, yielding a
secular behavior (dotted line) that runs higher than in the
$B\!=\!0$ case. This is the weak localization effect for 
ballistic cavities \cite{Bar93}. The reason why we chose an asymmetric cavity 
is that the ballistic weak localization effect is strongly dependent on the spatial
symmetries of the cavity \cite{BaMe96}.

Our numerical results show that the conductance fluctuations
and the weak localization effect, first discussed in the
context of disordered mesoscopic conductors, are also present
in ballistic mesoscopic cavities. The distinctions concerning
these two types of mesoscopic systems, that we discussed in
the introduction, force us to rethink the appropriate definition
of averages, as well as the concept of universality.

\subsection{Semiclassical transmission amplitudes}
\label{subsec:sta}

Our goal is to calculate the conductance through a cavity 
(like the ones in Figs.~\ref{fig:intro} and \ref{fig:preview}) by 
using Eqs.~(\ref{Lan})
and (\ref{allTRAMs}) within a semiclassical approach. The Green
function is the Laplace transform of the propagator. The Van Vleck 
expression for the latter, together with a stationary-phase
integration on the time variable, leads to the semiclassical 
approximation for the Green function \cite{gutz_book}

\be
\hspace{0.6cm} G(\brp,\br;E)=\frac{2\pi}{(2\pi i\hbar)^{(d+1)/2}}
\sum_{s(\br,\brp)} \sqrt {D_{s}} \ \exp{\left[\frac{i}{\hbar}S_{s}(\brp,\br;E)-
i \frac{\pi}{2}\nu _{s}\right]} \ .
\label{eq:gfgutz}
\ee

\nin The sum is over classical trajectories $s$, with energy $E$, going
between the initial and final points $\br\!=\!(x,y)$ and $\brp\!=\!(x',y')$. As in the 
previous chapter, we will take $x$ and $x'$ at the junctions between the leads and
the cavity, and therefore we will not write any $x$ dependence. 
$S_{s}=\int_{\C_s} \bp \!\cdot \!\bf{\rm d}\bq$ is the action integral along the 
path $\C_s$. In the case of billiards without magnetic field 
$S_{s}/\hbar=k L_{s}$, where $L_{s}$ is the trajectory length. The factor
$D_{s}$ describing the evolution of the classical probability can be 
expressed as a determinant of second derivatives of the action \cite{gutz_book,brack_book}.
In our geometry, if we denote by
$\theta$ and $\theta^{\prime}$ the incoming and outgoing angles of the trajectory
with the $x$-axis, $D_{s}=(v|\cos{\theta^{\prime}}|/m)^{-1}\mid (\partial
\theta /\partial y^{\prime})_{y}\mid$. We include in the phase $\nu_s$
the Maslov index counting the number of constant-energy conjugate points and
the phase acquired at the bounces with the walls when those are given
by an infinite potential (hard walls). In our calculations we will always take the
spatial dimensionality $d\!=\!2$. 

In the case of hard-wall leads, the transverse wave-functions have the
sinusoidal form of Eq.~(\ref{eq:twf}). Using the semiclassical expression 
(\ref{eq:gfgutz}) of the Green function in Eq.~(\ref{allTRAMs})
we see that, for large integers $a$, the integral over $y$ will be 
dominated by the stationary-phase contribution occurring for trajectories
starting at points $y_0$ defined by

\be
\left(\frac{\partial S}{\partial y} \right)_{y'} = -p_{y} =
- \frac{\bar{a} \hbar \pi}{W} \ , \hspace{1cm} \bar{a}=\pm a \ ,
\label{eq:stphcon}
\ee

\nin The dominant trajectories are those entering the cavity with
the angles $\theta_{\bar{a}}$ such that $\sin\theta_{\bar{a}}\!=\!
\bar{a}\pi/ kW$. Thus, the initial transverse momentum of the 
trajectories equals the momentum of the transverse wave-function. 
As always in this type of reasoning, we have assumed that we could
interchange the order of the integration and the sum over trajectories. 
Integrating the gaussian fluctuations we have

\be
\label{eq:tbaSPA1}
\hspace{0.8cm} t_{ba} = i \ \sqrt{\frac{v_b}{2W}} \int d y' \; \phi_b(y') \sum_{\bar{a}=\pm a}
\sum_{s(\theta_{\bar{a}},y')} {\rm sgn}(\bar{a}) \sqrt{\hat{D}_{s}} \ 
\exp \left[ \frac{i}{\hbar} \hat{S}_{s} (y',\theta_{\bar{a}};E) -
i\frac{\pi}{2}\hat{\nu}_s \right].
\ee

\nin The reduced action is

\be
\label{eq:redact}
\hat{S}(y',\theta_{\bar{a}};E) = S(y',y_0(\theta_{\bar{a}},y');E) +
\frac{\hbar\pi\bar{a}}{W} \ y_0(\theta_{\bar{a}},y') \ ,
\ee

\nin The prefactor is now given by 
$\hat{D}=(v\cos{\theta'})^{-1} |(\partial y /\partial y')_{\theta}|$,
and the new index $\hat{\nu}$ (that we still call Maslov index) is increased by 
one if $(\partial \theta / \partial y)_{y'}$ is positive.
At this intermediate stage we have a mixed representation, with trajectories
starting with fixed angles ($\pm \theta_{a}$) and finishing at points $y'$.
A new stationary-phase over $y'$ calls for. However, this is not possible in
the case where families of trajectories with degenerate action exist and the 
reduced action $\hat{S}$ is a linear function of $y'$. This is the case
of trajectories directly crossing the cavity without suffering collisions or
the case in which the cavity has rectangular shape. In the next section we analyze 
these two cases in detail. Here we will assume that trajectories are isolated and
we can perform the $y'$ integration by stationary phase. The
final points $y^{\prime}_0$ are selected according to

\be
\left(\frac{\partial \hat{S}}{\partial y'} \right)_{\bar{a}} = 
\left(\frac{\partial S}{\partial y'} \right)_{y} = p_{y^{\prime}} =
- \frac{\bar{b} \hbar \pi}{W} \ , \hspace{1cm} \bar{b}=\pm b \ ,
\label{eq:stphcon2}
\ee

\nin implying that the trajectories have an outgoing angle 
$\theta_{\bar{b}}$ such that $\sin\theta_{\bar{b}} =
\bar{b}\pi/ kW$. The semiclassical expression for the transmission
amplitude can then be casted as \cite{Jal90}

\be
t_{ba}=-\frac{\sqrt{2\pi i\hbar}}{2W} \sum_{\bar{a}=\pm a} \sum_{\bar{b}=\pm b}
\sum_{s(\bar{b},\bar{a})} {\rm sgn}(\bar{a}\bar{b}) \ \sqrt
{\tilde{D}_{s}} \ \exp{\left(\frac{i}{\hbar}\tilde{S}_{s}(\bar{b},\bar{a};E)
-i \frac{\pi}{2}{\tilde\nu}_{s}\right)} \
\label{eq:semiclass}
\ee

\nin The reduced action is

\be
\tilde{S}(\bar{b},\bar{a};E)=S(y_{0}^{\prime},y_{0};E)+
\frac{\hbar\pi\bar{a}}{W} \ y_{0} - \frac{\hbar\pi\bar{b}}{W} \ y_{0}^{\prime}  \ .
\label{eq:virial}
\ee

\nin For billiards it can be written as $\tilde{S}=\hbar k \tilde{L}$, with 
$\tilde{L}=L+k y_{0} \sin\theta_{\bar{a}} -k y^{\prime}_{0} \sin\theta_{\bar{b}}$.
The prefactor is now given by

\be
\tilde{D}_{s}=\frac{1}{mv\cos{\theta^{\prime}}} \left|
\left(\frac{\partial y}{\partial \theta^{\prime}}\right)_{\theta}\right| \ ,
\label{eq:dtilt}
\ee

\nin and the Maslov index is

\be
\tilde{\nu} = \nu + H\left(\left(\frac{\partial \theta}{\partial y}\right)_{y'}\right) +
H\left(\left(\frac{\partial \theta^{\prime}}{\partial y'}\right)_{\theta}\right) \ ,
\label{eq:muilt}
\ee

\nin where $H$ is the Heaviside step function \cite{correction}.

Similar arguments can be used to write the semiclassical reflection
amplitude in terms of trajectories leaving and returning to the cross
section at the left entrance with appropriate quantized angles. Note that
there are two kinds of trajectories contributing to
$G(y^{\prime},y;E)$ in the case of reflected paths: those which
penetrate into the cavity and those
which go directly from $y$ to $y'$ staying on the cross section of the lead.
It is only trajectories of the first kind which contribute to
the semiclassical reflection amplitude, as trajectories of the second kind merely cancel
the $\delta_{ba}$ of Eq. (\ref{eq:TRAM1}).

The semiclassical transmission amplitude (\ref{eq:semiclass}) is, for an
open system, the analogous of the Gutzwiller trace formula for the
density of states in a closed system \cite{gutz_book}. Both are expressed as a sum
over isolated classical trajectories, making the connection between
classical and quantum properties transparent. The main difference between the scattering 
and energy-level problems, at the semiclassical level, is that the trace 
formula involves the sum over periodic orbits while the transmission 
amplitude is given by open trajectories that go across the scattering region.
In chaotic systems the number of trajectories connecting two given points
grows exponentially with the trajectory length. In open systems the
trajectories can escape the scattering region, therefore their proliferation
is much weaker than in the close case (although still exponential). Therefore,
the convergence of semiclassical propagators in chaotic 
scattering (even if we ignore the physical cutoffs discussed
in the introduction) will not encounter the difficulties of the trace formula.
From the quantum point of view, since the Gutzwiller trace formula must 
reproduce a delta-function spectrum, it can be conditionally 
convergent at most, while the quantum transmission amplitude is a smooth function of 
the Fermi energy (away from the thresholds at the opening of new modes) 
and so the semiclassical sum can be absolutely convergent (depending on the 
value of the fractal dimension $d$ of the strange repeller governing
the chaotic scattering \cite{Jen93}). 

Chaotic scattering problems have been studied by Miller \cite{Mil74} in the
context of molecular collisions in terms of the semiclassical propagator in the 
momentum representation. In this case the relevant sum in this case is over
classical trajectories with fixed incident and outgoing momenta, similar
but not identical to what we find for the transmission amplitude in our
waveguide geometry.  Our procedure furnishes a more explicit semiclassical
expression to use in the Landauer-B\"uttiker conductance formulas and allows
one to handle more complicated situations like finite magnetic field and
soft walls in the leads \cite{Bar91}, tunneling in the cavities \cite{ingold},
or the presence of families of trajectories \cite{paul}. The last case,
discussed in detail in the next section, leads to an expression of the
semiclassical transmission amplitudes as a sum over {\em families of 
trajectories} (in analogy with the Berry-Tabor formula for the density
of states of integrable systems \cite{ber76}). The simple prescription for 
the Maslov indices makes possible the numerical evaluation 
of the semiclassical transmission amplitude. In this section we will not pursue 
section the explicit summation of Eq.~(\ref{eq:semiclass}) over classical trajectories, 
but in (\ref{subsec:cbdat}) we discuss the work of Lin and Jensen \cite{LinJen} 
addressing such a problem for a circular scattering domain.

\subsection{Transmission coefficients and average values}

Transmission coefficients are obtained from the magnitude squared of
the transmission amplitudes. In a semiclassical approach they are given 
by sums over {\it pairs} of trajectories. 
Since we will be focusing on billiards, it is convenient to
scale out the energy (or momentum) dependence and write

\be
T(k) = \sum_{a,b=1}^{N} T_{ba}(k) = \frac{1}{2} \left(\frac{\pi}{kW}\right) 
\sum_{a,b=1}^{N} \sum_{{\bar a}=\pm a} \sum_{{\bar b}=\pm b}
\sum_{s({\bar a},{\bar b})} \sum_{u({\bar a},{\bar b})} 
F_{{\bar b},{\bar a}}^{s,u}(k) \ ,
\label{eq:scT}
\ee

\be
F_{{\bar b},{\bar a}}^{s,u}(k) = \sqrt{\tilde{A}_{s} \tilde{A}_{u}}
\exp{[i k (\tilde{L}_{s}-\tilde{L}_{u})+i \pi \phi_{s,u} ]} \ .
\label{eq:defF}
\ee

\nin Where $s$ and $u$ label the paths with extreme angles $\theta_{\bar a}$ 
and ${\theta_{\bar b}}$, $\tilde{A}_{s}=(\hbar k/W) \tilde{D}_{s}$, and
$\phi_{s,u}=(\tilde{\nu}_{u}-\tilde{\nu}_{s})/2+\bar{a}+\bar{b}$. Note that 
$\tilde{A}$ is independent of energy, therefore the only energy-dependence 
is in the selection of the injection angles and in the explicit 
$k$-dependence of the actions.

We would like to understand in which sense the semiclassical transmission
coefficients, given by (\ref{eq:scT}) and (\ref{eq:defF}), are able to 
account for the highly structured curve of Fig.~\ref{fig:preview}. We do
not attempt a detailed description or an identification of classical
trajectories (like in Refs.~\cite{ishio95,schwi96}), but a characterization 
of the overall features, like the secular behavior or the statistical 
properties of the fluctuations. For this purpose we need to define an 
average procedure. Since we dispose of a single cavity, our averages must 
be over energy (or wave-vector). Our semiclassical approximations are 
supposed to be valid in the large-$k$ limit, thus for an observable
$O(k)$ we define

\be
\langle O \rangle =
\lim_{q \rightarrow \infty} \frac{1}{q} \int_{q_c}^{q_c+q} dk \ O (k) \ ,
\hspace{1cm}  \frac{q_c W}{\pi} \gg 1 \ .
\label{eq:avgk}
\ee

\nin This average is particularly suited for analytical calculations, 
but not for dealing with experimental or numerical results, where we only 
dispose of a finite $k$-range (and the averages are necessarily local). 
However, if we average over many oscillations, we
expect to get the same results as with (\ref{eq:avgk}). An average over a 
finite energy-range is precisely the effect of finite temperature on the conductance.

The secular behavior is linearly increasing with $k$ (outgoing flux
proportional to the incoming flux). The slope is given by the average of $T(k)/k$, that is,

\be
{\cal T} = \left\langle \frac{\pi}{kW} \ T(k) \right\rangle = \frac{1}{2}
\left\langle \left( \frac{\pi}{kW} \right)^2 
\sum_{a,b=1}^{N} \sum_{{\bar a}=\pm a} \sum_{{\bar b}=\pm b}
\sum_{s,u} F_{{\bar b},{\bar a}}^{s,u}(k) \right\rangle \ .
\label{eq:tbar0}
\ee

In the large $k$-limit the modes are closely spaced in angle, and
the sums over modes can be converted into integrals over angles:
$\sum_{a}^{N} \sum_{{\bar a}=\pm a} \rightarrow (kW)/\pi 
\int_{-1}^{1} d(\sin{\theta})$. After this conversion, the only 
$k$-dependence remains in the phase factors, and we can interchange
the angle-integral with the $k$-average obtaining

\be
{\cal T} = \frac{1}{2} \int_{-1}^{1} d(\sin{\theta})
\int_{-1}^{1} d(\sin{\theta^{\prime}}) \sum_{s(\theta,\theta^{\prime})}
\sqrt{\tilde{A}_{s} \tilde{A}_{u}} \ \langle
\exp{[i k (\tilde{L}_{s}-\tilde{L}_{u})+i \pi \phi_{s,u} ]} \rangle \ .
\label{eq:classT0}
\ee

Our definition of averages immediately yields that 
$k (\tilde{L}_{s}\!-\!\tilde{L}_{u})+ \pi \phi_{s,u} = 0$. In the absence of 
symmetries, such a relation is only possible if $s\!=\!u$. Quantum interference
is therefore absent in the resulting diagonal term. Using the definition
of $\tilde{A}$ and changing variables, from the outgoing angle $\theta^{\prime}$
to the initial position $y$, we have 

\be
{\cal T} = \frac{1}{2} \int_{-1}^{1} d(\sin{\theta})
\int_{0}^{W} \frac{dy}{W} \ f(y,\theta) \ ,
\label{eq:classT}
\ee

\nin where $f(y,\theta)\!=\!1$ if the trajectory with initial conditions
$(y,\theta)$ is transmitted and $f(y,\theta)\!=\!0$ otherwise. The above
expression is purely classical and has the intuitive interpretation of
a probability of transmission. It can also be easily obtained from a
Boltzmann equation approach \cite{Bar91}. The classical form of
the transmission coefficient is relevant when the temperature is high
enough to kill the interference effects, and it has been used to 
understand the early experiments on transport in ballistic junctions at
Helium temperatures \cite{Rouk,Ford,BvH88,BvHra}.

Unlike (\ref{eq:semiclass}), Eq.~(\ref{eq:classT}) is very easy to implement 
numerically, if we have the information of the classical dynamics. We simply have to
sample the space of classical trajectories with random choices of the
initial position and initial angles (with a weight of 
$\cos{\theta}$). Following this procedure yields a value of ${\cal T}$
consistent with the slope of the quantum numerical results (within
the statistical errors with which we can determine them).

The secular behavior of the conductance (dashed line in Fig.~\ref{fig:preview})
is given by the straight line that best fits $T(k)$, defined by its slope 
(equal to the classical transmission probability ${\cal T}$) and the 
shift,

\be
\langle \dlT \rangle = \left\langle \left(T - \left(\frac{\pi}{kW}\right) 
{\cal T} \right) \right\rangle
\label{eq:kappa}
\ee

\nin Operationally, $\dlT$ is well defined. However, it is not
possible to give a simple semiclassical prescription for its calculation.
In (\ref{subsec:wl}) we will address the problem of its magnetic field
dependence, which determines the weak localization effect. 

\subsection{Conductance fluctuations}
\label{subsec:cf}

The most striking feature of the data in Fig.~\ref{fig:preview}
are the conductance fluctuations around the mean value. The shape of
these fluctuations is characterized by their power spectrum. The
semiclassical treatment of the power spectrum \cite{Jal90,Chaost}
was based on the semiclassical approach to $S$-matrix fluctuations 
as a function of energy, introduced by Gutzwiller \cite{Gutz83} and 
extensively developed by Bl\"umel and Smilansky \cite{Blu88} and 
Gaspard and Rice \cite{Gas89}. The main conclusion is that the 
power spectrum is directly related to properties of the classical
phase space.

In order to 
characterize the fluctuations, we introduce the $k$-correlation function  

\be
C_k( \dlk ) = \langle \dlT (k+\dlk) \dlT (k) \rangle \ ,
\label{eq:Cdef}
\ee

\nin with 

\be
\delta T =  T - \left(\frac{kW}{\pi} {\cal T} + \langle \dlT \rangle \right)  \ .
\label{eq:dlTdef}
\ee

The semiclassical calculation of the correlation function involves a
sum over four trajectories,

\be
\hspace{0.3cm} C_k(\dlk) = \frac{1}{4} \left\langle {\left ( \frac{\pi}{kW} \right )}^2
\sum_{a,b}^{N} \sum_{a^{\prime},b^{\prime}}^{N}
\sum_{{\bar a}=\pm a} \sum_{{\bar b}=\pm b}
\sum_{{\bar a}^{\prime}=\pm a^{\prime}} \sum_{{\bar b}^{\prime}=\pm b^{\prime}}
\sum_{s,u}{^{^{\prime}}} \sum_{s^{\prime},u^{\prime}}{^{^{\prime}}}
F_{{\bar b},{\bar a}}^{s,u}(k+\dlk) 
F_{{\bar b}^{\prime},{\bar a}^{\prime}}^{s^{\prime},u^{\prime}}(k) \right\rangle \ .
\label{eq:Cdsc1}
\end{equation}

\nin The ``prime" in the summations over trajectories indicates that the completely 
diagonal terms $s\!=\!u$ (and $s'\!=\!u'$) are excluded, since they contribute to 
${\cal T}$. The semiclassical expression for $C_k(\dlk)$ is fairly complicated since
four trajectories contribute to each summand. We will restrict ourselves to the
diagonal (in modes) component $C_k^{D}(\dlk)$, obtained from (\ref{eq:Cdsc1}) by taking
$a\!=\!a'$ and $b\!=\!b'$.

In taking the $k$-average, we consider the limit when the modes are very dense
and therefore we replace the sums over modes by integrals over angles (as we
did in the previous chapter). The only $k$-dependence is in $\langle \exp[ ik(
\tilde{L}_s - \tilde{L}_u + \tilde{L}_t - \tilde{L}_v )] \rangle$.
The infinite $k$-average implies that the only contribution is for
$\tilde{L}_s - \tilde{L}_u + \tilde{L}_t - \tilde{L}_v = 0$.
Because of the definition of $C_k^D$, all four paths satisfy the same boundary 
conditions on angles, and hence they are all chosen from the same discrete set 
of paths. In the absence of symmetry, the only contribution (excluding 
$s\!=\!u$ and $s'\!=\!u'$) is $s\!=\!u'$ and $u\!=\!s'$. Thus we find 

\be
\hspace{0.6cm} C_k^D(\dlk) = \frac{1}{4}
\int_{-1}^{1} d(\sin\theta)\int_{-1}^{1} d(\sin\theta^{\prime})
\sum_{\sttp} \sum_{\uttp}{^{^{\prime}}} \tilde{A}_s \tilde{A}_u
\exp{\left[i \dlk (\tilde{L}_s - \tilde{L}_u)\right]} \ ,
\label{eq:Cdsc2}
\ee

\nin which is independent of $k$. As previously stated, we characterize
the conductance fluctuations from the Fourier power spectrum

\be
\widehat{C}_k(x) = \int d(\dlk) C_k(\dlk) e^{ix \dlk} \ .
\label{eq:Chatdef}
\ee

\nin The semiclassical approximation to the diagonal term reads

\be
\hspace{0.3cm} \widehat{C}_k^D (x) = \frac{\pi}{2}
\int_{-1}^{1} d(\sin\theta)\int_{-1}^{1} d(\sin\theta^{\prime})
\sum_{\sttp} \sum_{\uttp}{^{^{\prime}}} \tilde{A}_s \tilde{A}_u
\delta(\tilde{L}_s + x - \tilde{L}_u)
\label{eq:Chatsc1}
\ee

In the chaotic case, we can make progress analytically by assuming that 
(1) the trajectories are uniformly distributed in the sine of the angle, 
(2) the angular constraints linking trajectories $u$ and $s$ can be ignored, 
and  (3) the constraint $u \neq s$ can be ignored because of the 
proliferation of long paths. Introducing angular integrations over 
${\tilde \theta}$ and ${\tilde \theta}^{\prime}$ we write

\begin{eqnarray}
\widehat{C}_k^D (x) & = & \frac{\pi}{8}
\int_{-1}^{1} d(\sin\theta)\int_{-1}^{1} d(\sin\theta^{\prime})
\sum_{\sttp}  \tilde{A}_s
\int_{-1}^{1} d(\sin{\tilde \theta})\int_{-1}^{1} d(\sin{\tilde \theta}^{\prime})
\\ \displaystyle
& \times &
\sum_{\uttpt} \tilde{A}_u
\delta(\tilde{L}_s + x - \tilde{L}_u) \propto \int_{0}^{\infty}\! dL \:P(L+x) P(L) \ ,
\nonumber
\label{eq:Chatchaos1}
\end{eqnarray}

\be
P(L) = \frac{1}{4}
\int_{-1}^{1} d(\sin\theta)\int_{-1}^{1} d(\sin\theta^{\prime})
\sum_{\uttp}  \tilde{A}_u \delta(L - \tilde{L}_u) \ ,
\label{eq:Pdef}
\ee

\nin is the classical distribution of lengths ($x$ is taken positive). 
As previously discussed, the distribution 
of lengths is exponential for chaotic billiards for large $L$ 
($P(L) \propto e^{-\gcl L}$), while there may be deviations at small $L$.
Using the exponential form for all lengths, we find that

\be
\widehat{C}_k^D (x) \propto e^{-\gcl |x| } \ ,
\label{eq:Chatchaos2}
\ee

\nin which is valid for all $x$ since $\widehat{C}_D (x)$ must be real and
symmetric. This form for the power spectrum
implies that the wave-vector correlation function is Lorentzian \cite{Blu88}.

\begin{equation}
C_k^D (\dlk) = \frac{C_k^D(0)}{1 + (\dlk/ \gcl)^2} \ .
\label{eq:Chatsc3}
\ee

We have treated the case of billiards, where the energy-dependence is
easily scaled out, and we are lead to the distribution of lengths. But the
result of Ref.~\cite{Blu88} is general for energy-correlation functions. A
semiclassical analysis and an energy-average over intervals small in the
classical scale (unchanged trajectories) but large in the quantum scale
(many oscillations) yields $C_E^D (\Delta E) = C_E^D(0)/[1 + (\Delta E/(\hbar \gamma)^2 ]$.

The conductance fluctuations are on a scale $\Delta E_c=\hbar \gamma$ much larger
than the level spacing $\Delta$, since the openings of the cavity put us in the
regime of overlapping resonances. This regime has been extensively studied in 
Nuclear Physics, in the context of compound nuclei, and the fluctuations we have
discussed are known as Ericson fluctuations \cite{Eric60}. The same physics has also been 
studied in one-dimensional models of chaotic scattering obtained as a variant
of the kicked rotator \cite{borguar}, where the quantum correlation lengths have
been shown to be well described by the numerically computed escape rates.

The derivation of the magnetic field correlation function \cite{Jal90}
follows the same lines as the case treated above. The correlation function 
defined as an average over $k$

\be
C_B( \dlB ) = \langle \dlT (k,B+\dlB) \dlT (k,B) \rangle \ .
\label{eq:CBdef}
\ee

\nin Again, our definition of averages will facilitate the analytical
calculations. In analyzing experimental or numerical results, averages
over finite $k$ or $B$ intervals (small enough not to modify the classical 
dynamics) are generally used. 

Treating the diagonal (in modes) component $C_B^D( \dlB )$, assuming 
$\dlB$ small enough not to change the classical trajectories significatively,
and using the pairing $s\!=\!u'$, $u\!=\!s'$, we are lead to the 
action differences $[ S_s(B+\dlB) - S_u(B+\dlB) + S_u(B) - S_s(B) ]/ \hbar =
(\Theta_s - \Theta_u) \dlB/ \Phi_0$. In analogy with the previous case,
we have

\be
\hspace{0.6cm} C_B^D(\dlB) = \frac{1}{4}
\int_{-1}^{1} d(\sin\theta)\int_{-1}^{1} d(\sin\theta^{\prime})
\sum_{\sttp} \sum_{\uttp}{^{^{\prime}}} \tilde{A}_s \tilde{A}_u
\exp{\left[i \frac{\dlB}{\Phi_0} (\Theta_s - \Theta_u) \right]} ,
\label{eq:CdB1}
\ee

\begin{eqnarray}
\widehat{C}_B^D (\eta) & = & \frac{\Phi_0 \pi}{2}
\int_{-1}^{1} d(\sin\theta)\int_{-1}^{1} d(\sin\theta^{\prime})
\sum_{\sttp} \sum_{\uttp}{^{^{\prime}}} \tilde{A}_s \tilde{A}_u
\delta(\Theta_s + \eta - \Theta_u) \nonumber \\ 
& \propto &
\int_{-\infty}^{\infty}\! d\Theta \:N(\Theta+\eta) N(\Theta) .
\label{eq:CdB3}
\end{eqnarray}

\nin Using the exponential form (\ref{eq:efardis}) of the distribution of 
effective areas $N(\Theta)$, for all values of $\Theta$ yields 
the power spectrum \cite{Jal90}

\be
\widehat{C}_B^D (\eta) \propto e^{-\alc |\eta|} \left(1+ \alc |\eta|\right)
\label{eq:CdB4}
\ee

\nin and the correlation function

\be
C_B^D (\dlB) = \frac{C_B^D(0)}{[1 + (\dlB/ \alpha_{cl} \Phi_0 )^2 ]^2}
\label{eq:CdB5}
\ee

We have only been able to calculate semiclassically the line-shapes
of the diagonal parts of $C_k(\dlk)$ and $C_B( \dlB )$, which are $k$-independent. 
On the other hand, for a chaotic cavity we expect only one characteristic scale 
for each correlation function. We then conjecture that (\ref{eq:Chatchaos2}) and
(\ref{eq:CdB4}) also represent the full power spectra, and that
$C_k$ and $C_B$ are $k$-independent (like $C_k^D$ and $C_B^D$). Thus,
we expect the conductance fluctuations to persist (and remain invariant) in the 
large-$k$ limit.

\begin{figure}
\setlength{\unitlength}{1mm}
\begin{center}
\begin{picture}(110,135)
\leavevmode
\epsfxsize=0.85\textwidth
\epsfbox{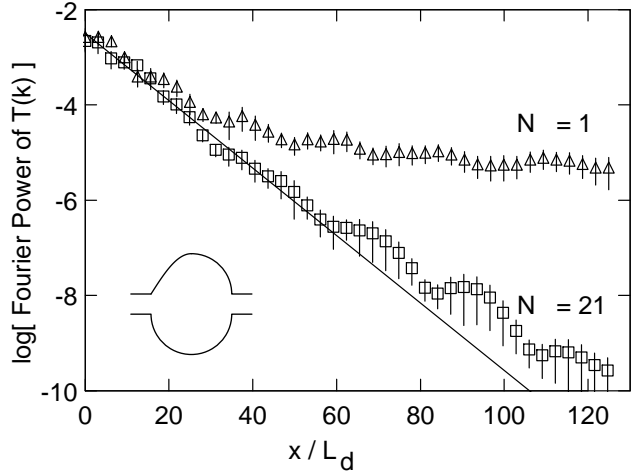}
\end{picture}
\end{center}
\vspace{-7cm}
\caption{
Power spectrum of $T(k)$ for the chaotic structure shown for two
fillings: $N\!=\!21$ (squares) and $N\!=\!1$ (triangles).
In the former case, the agreement with the semiclassical theory is
excellent, in the extreme quantum limit
when only one mode is propagating in the leads, there is more
high-frequency power and the agreement with the semiclassical theory
is poorer. (From Ref. \protect\cite{Chaost}.)}
\label{fig:powsp}
\end{figure}

We need to perform a systematic study of the conductance fluctuations 
in order to test the semiclassical predictions. 
Since there is some arbitrariness in determining the averages of the
numerical data, we directly calculate the power spectra $\widehat{C}_k(x)$
and $\widehat{C}_B(\eta)$ from the Fourier power of the raw data. Once we
smooth them, we verify that they are well represented by Eqs.~(\ref{eq:Chatchaos2}) 
and (\ref{eq:CdB4}) over large ranges of $x$ and $\eta$. In Fig.~\ref{fig:powsp} 
we see (for $\widehat{C}_k$) deviations for small lengths ($x \simeq \LD$) and
for large $x$. Similar agreement is obtained for $\widehat{C}_B$ \cite{Jal90,Chaost}. The 
deviations for small lengths (or areas) are understandable since the chaotic
nature of the dynamics (and the statistical treatment of the trajectories
that we used) cannot give an appropriate description for short trajectories.
The deviations for large lengths arise from the 
limitations of semiclassics and of the diagonal approximation, and they
become less important as we go in the large $k$-limit. 

The analysis of the power spectra has the advantage of separating the 
fluctuations according to their length (or area) scales. This is 
important when we want to make the connection with the experimental
world of Mesoscopic Physics. Only
the features associated with scales smaller than our physical cut-offs
will be relevant. 

\begin{figure}
\setlength{\unitlength}{1mm}
\begin{center}
\begin{picture}(110,145)
\leavevmode
\epsfxsize=0.85\textwidth
\epsfbox{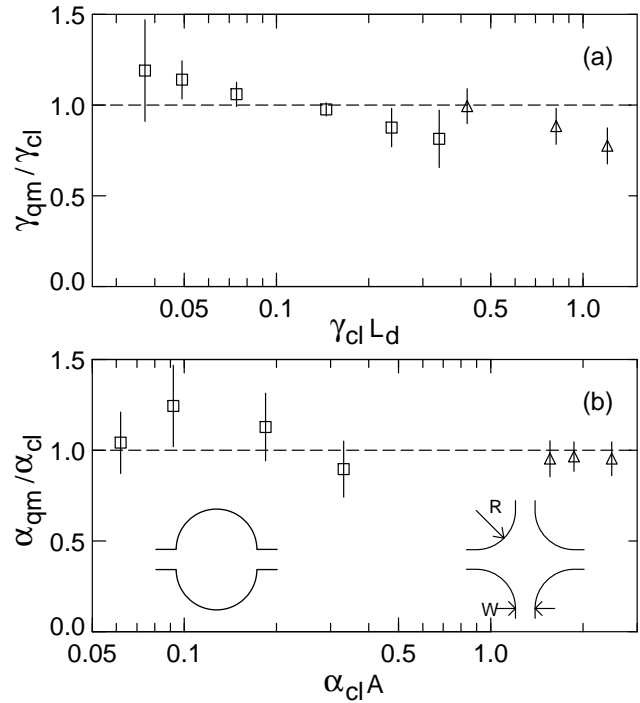}
\end{picture}
\end{center}
\vspace{-4cm}
\caption{
(a) Ratio of the wave-vector correlation length
to the classical escape rate $\gcl$ as a function of $\gcl$ for both
types of structures shown.  4-disc structure (triangles) with $R/W=1,2,4$,
and stadium (squares) with $R/W = 0.5,1,2,4,6,8$.
(b) Ratio of magnetic field correlation length
to $\alc$, the exponent of the distribution of effective areas,
as a function of $\alc$.  4-disc structure with $R/W=1,2,4$,
and open stadium with $R/W = 1,2,4,6$. (From Ref. \protect\cite{Jal90}.)
}
\label{fig:pow3}
\end{figure}

If we fit the power spectra obtained from the {\em quantum numerical
calculations} in an interval where the forms (\ref{eq:Chatchaos2})
and (\ref{eq:CdB4}) are respected we can extract quantum
correlation lengths $\gamma_{qm}$ and $\alpha_{qm}$, that can be
compared with the rates $\gcl$ and $\alc$ obtained from the
simulation of the {\em classical dynamics}. Semiclassics provides
the link, through Eqs.~(\ref{eq:Chatchaos2}) and (\ref{eq:CdB4}),
between quantum and classical parameters. Fig.~\ref{fig:pow3} shows 
that $\gamma_{qm},\alpha_{qm}$ are indeed given by
the classical quantities $\gcl,\alc$ to high accuracy while
they are varied over roughly two orders of magnitude by changing
the size of the structures considered (for the four-probe structure,
we considered the fluctuations of the Hall resistance, which is a
function of the transmission coefficient between leads \cite {But86,
BvH88,Bar91}, for which the previous semiclassical analysis is
applicable). 
The $k$-independence of $\gamma_{qm}$, $C_k(0)$, $\alpha_{qm}$ and
$C_B(0)$ is approximately respected in our numerical simulations
away of the quantum limit of small $N$ \cite{Chaost}.

In an integrable cavity each trajectory belongs to a torus defined by 
two constants of motion. Moreover, if the conserved quantities in the
cavity and in the leads are the same, the scattering trajectories are
organized in families, and as we will see in the next section, we need
to modify the form of the semiclassical transmission amplitudes. Leaving
aside this last case, there are at least three assumptions leading
to Eqs.~(\ref{eq:Chatchaos2}) and (\ref{eq:CdB4}) that break down in the
case of integrable dynamics: the length and area distributions are not 
characterized by a single scale (exhibiting, as we saw in 
(\ref{subsec:qs}), power-laws), there are angular correlations between
incident and outgoing angles, and the constraint of two trajectories
to satisfy the same boundary conditions on angles cannot be ignored.
From Eq.~(\ref{eq:Chatsc1}), the power spectrum $\widehat{C}_k^D(x)$ is
evidently related to the distribution of two distinct paths with a difference
in length of $x$. Thus we conjecture that

\be
\widehat{C}_D (x) \propto \int_{0}^{\infty}\! dL \int_{0}^{1}
d (\sin \theta) \:P_2(L+x,L,\theta)
\label{eq:Cdkreg}
\ee

\nin where $P_2(L+x,L,\theta)$ is the classical distribution for two distinct
trajectories at angle $\theta$, one with length $L$ and the other with length 
$L+x$. Numerical simulations for $P_2$ in rectangular a cavity  (Fig.~\ref{fig:distr}c) 
show that it does not factorize. It is characterized by a decay $1/x^2$ for 
large $x$, while $P(L)$ follows a $1/L^3$ law.

In the case of the power spectrum of $T(B)$, we assume that
a similar expression holds in terms of the classical distribution for two
trajectories at angle $\theta$ with effective areas $\Theta$ and $\Theta+\eta$
\begin{equation}
\widehat{C}_D (\eta) \propto
\int_{-\infty}^{\infty}\! d\Theta \:
\int_{0}^{1} d (\sin \theta) \:N_2(\Theta+\eta,\Theta,\theta) .
\label{eq:CdBreg}
\end{equation}
The result of classical simulation in Fig.\ \ref{fig:distr} shows that
this distribution is roughly constant up to a cutoff which is less
than the area of the rectangle.

\begin{figure}
\setlength{\unitlength}{1mm}
\begin{center}
\begin{picture}(110,155)
\leavevmode
\epsfxsize=0.85\textwidth
\epsfbox{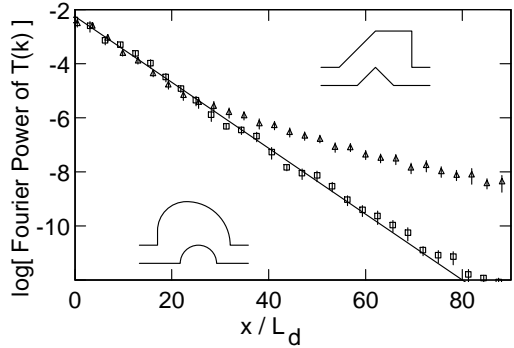}
\end{picture}
\end{center}
\vspace{-11cm}
\caption{
Power spectra of $T(k)$ for the chaotic (squares) and regular
(triangles) structures shown ($N = 33$). The regular
structure has more power at large frequencies because of more trajectories
with large enclosed areas. The line is a fit to Eq.~(\protect\ref{eq:Chatchaos2}).
(From Ref. \protect\cite{Chaost}.)
}
\label{fig:pow4}
\end{figure}

Numerical calculations of integrable cavities yield a more ragged conductance,
as compared with the chaotic case \cite{Mac92,Chaost}. This is reflected by more power strength at
large lengths (Fig.~\ref{fig:pow4}) and areas, consistently with our conjecture
for linking conductance fluctuations and $P_2$. Lai and collaborators \cite{Lai92}
showed that the fine scale energy fluctuations of the $S$ matrix are greatly
enhanced for non-hyperbolic dynamics, as compared with the hyperbolic case, and
predicted for $C_E (\Delta E)$ a cusp-shaped peak at $\Delta E=0$.

\subsection{Weak localization in the ballistic regime}
\label{subsec:wl}

In our discussion of the secular behavior of the conductance
through a ballistic cavity (Fig.~\ref{fig:preview}), we 
pointed that the shift $\langle \delta T \rangle$ is sensitive
to the magnetic field. The presence of a small field increases
the average conductance. By analogy with the disordered case,
we call this effect the ballistic weak localization \cite{Bar93,Chaost}.
It is important to realize that it is an average effect. 
Only after removing the (large) conductance
fluctuations by the $k$-average, we obtain the (small) difference
between the secular behaviors with and without magnetic field. The
two-probe conductance is an even function of the magnetic field,
therefore in a given sample, $g(B)$ may have a {\em maximum or a 
minimum} at $B\!=\!0$. The two possible cases are observed, 
experimentally \cite{Kel94} and in the numerical simulations. 

In our calculation of the slope of the secular behavior (\ref{eq:classT})
we neglected the possibility that unlike paths may have the same
effective action and only considered the diagonal approximation of
pairing each trajectory with itself. Ignoring degeneracies among 
actions is correct in the absence of symmetries, but time-reversal or 
spatial symmetries force us to consider non-diagonal (in trajectories) 
terms. Obtaining exact spatial symmetries of the confining potential
in actual microstructures is quite difficult due to the limitations in
the fabrication procedure, but the time-reversal symmetry is exactly
fulfilled by simply turning off the external magnetic field.

It is easy to identify a set of degenerate trajectories that yields
quantum interference. Let us consider the reflection coefficient and
separate it on its diagonal (in modes) and off diagonal components

\be
R = R^D + R^{OD} = \sum_{a=1}^{N} R_{aa} +  \sum_{b \ne a}^{N} R_{ba} \ ,
\label{eq:decompR}
\ee

\be
R_{aa} = \frac{1}{2} \left(\frac{\pi}{kW}\right)
\sum_{{\bar a}=\pm a} \sum_{{\bar b}=\pm a}
\sum_{s({\bar a},{\bar b})} \sum_{u({\bar a},{\bar b})}
F_{{\bar b},{\bar a}}^{s,u}(k) \ .
\label{eq:scRD}
\ee

\nin The factors $F$ have the same meaning as in Eq.~(\ref{eq:defF}),
but now they refer to reflecting trajectories. Averaging $R(k)/k$ and
doing the diagonal (in trajectories) approximation yields the classical
reflection probability ${\cal R}$ (analogous to Eq.~(\ref{eq:classT}))
verifying ${\cal T}+{\cal R}=1$. In the absence of magnetic field there
are symmetry related trajectories in $R^D$ and in $R^{OD}$. The
time-reversed of a trajectory $s({\bar a},{\bar b})$ is 
$s^{\rm T}({\bar b},{\bar a})$, starting with the angle $\theta_{\bar b}$
and coming back to the lead with angle $\theta_{\bar a}$, therefore it
contributes to $R_{ab}$. If $b \neq a$ there is no interference between
$s$ and $s^{\rm T}$, thus the only diagonal (in trajectories) terms to
be considered are in $R^D$ (trajectories leaving with an angle $\theta$
and returning with $\theta$ or $-\theta$). The contribution of such
trajectories can be treated in a similar way as in the case of ${\cal R}$
and ${\cal T}$ (except that we only have {\em one} sum over modes),

\be
\langle \delta R^D(B\!=\!0) \rangle = \frac{1}{2} \int_{-1}^{1} d(\sin\theta)
\sum_{s( \theta,\pm \theta)} \tilde{A}_s \ .
\label{eq:rd0}
\ee

If we now turn on a magnetic field $B$ weak enough not to modify
appreciably the geometry of the paths, we only need to consider the
phase difference between $s$ and $s^{\rm T}$, that is given by
the effective area $\Theta_s$ according to $(S_s-S_u)/\hbar = 2\Theta_sB/\Phi_0$.
Thus, we obtain

\be
\langle \delta R^D(B) \rangle = \frac{1}{2} \int_{-1}^{1} d(\sin\theta)
\sum_{s( \theta,\pm \theta)} \tilde{A}_s \exp{\left[i \frac{2 B}{\Phi_0} \Theta_s \right]} \ ,
\label{eq:rd1}
\ee

\nin which yields an order unity ($k$-independent) contribution to
the average conductance containing only classical parameters (and $\Phi_0$).

In a chaotic system it is reasonably to assume that there is a
uniform distribution of exiting angles, and that the distribution
(\ref{eq:efardis}) of effective areas is valid  even if we
constraint the initial and final angles of the trajectories to
fixed values of $\theta$ and $\theta^{\prime}$. Therefore we can 
write \cite{Bar93}

\begin{eqnarray}
\langle \delta R^D(B) \rangle & = & \frac{1}{4} \int_{-1}^{1} d(\sin\theta)
\int_{-1}^{1} d(\sin\theta^{\prime}) \sum_{s( \theta,\theta^{\prime})} \tilde{A}_s
\int_{-\infty}^{\infty} d \Theta N( \Theta ) \exp{\left[i \frac{2 B}{\Phi_0} \Theta_s \right]}
\\
& = & \frac{{\cal R}}{1 + (2B/ \alpha_{cl} \Phi_0 )^2} \ .
\nonumber
\label{eq:wlfr}
\end{eqnarray}

\nin The Lorentzian line-shape is governed by the same parameter
$\alc$ of the conductance fluctuations (up to a factor of 2). Notice
that $\langle R^D(B\!=\!0) \rangle = 2 {\cal R}$, and the diagonal reflection
coefficients $R_{aa}$ are on average twice as large as the typical
off diagonal term (${\cal R}/N$). This factor of two enhancement,
known as ``coherent backscattering", is called ``elastic 
enhancement" in Nuclear Physics \cite{Iid90,Wei91}.

\begin{figure}
\setlength{\unitlength}{1mm}
\begin{center}
\begin{picture}(110,135)
\leavevmode
\epsfxsize=0.85\textwidth
\epsfbox{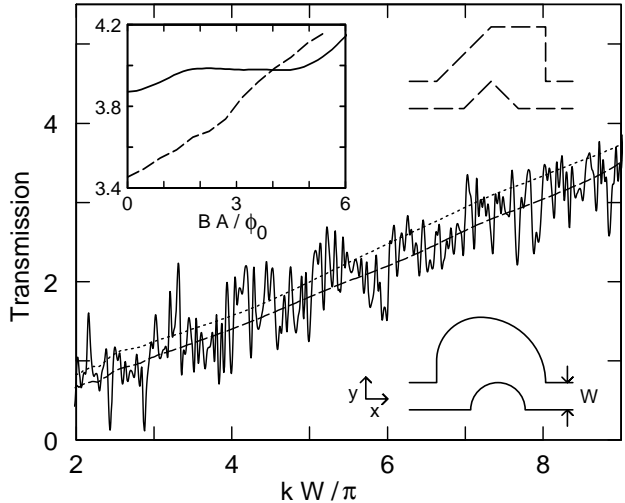}
\end{picture}
\end{center}
\vspace{-6cm}
\caption{
Transmission coefficient as a function of wave-vector for the
half-stadium structure shown in the bottom right.
The $T=0$ fluctuations (solid) are eliminated by
smoothing using a temperature which
corresponds to $20$ correlation lengths. The offset of the resulting
$B= 2 \Phi_{0} /A$ curve (dotted) from that for $B=0$ (dashed)
demonstrates the average magnetoconductance effect.
Inset: smoothed transmission coefficient as
a function of the flux through the cavity ($kW/\pi = 9.5$) showing the
difference between the chaotic structure in the bottom right (solid) and
the regular structure in the top right (dashed).
(From Ref. {\protect{\cite{Bar93}}}.)
}
\label{fig:wl1}
\end{figure}

We have shown that by the application of a magnetic field 
larger than $\alc \Phi_0/2$, the diagonal reflection coefficient
is reduced from $2 {\cal R}$ to ${\cal R}$. By unitarity, this means
that $R^{OD}+T$ must increase by ${\cal R}$. However, there are not
time-symmetric pairs of trajectories producing interference effects
in $R^{OD}$ or $T$, and we are not able to evaluate semiclassically
the corrections (or order 1) to their leading behavior (or order $N$).
The weak localization is the $B$-dependent part of the shift
$\langle \delta T(B) \rangle$ of Fig.~\ref{fig:preview}, but we have
been able to calculate semiclassically only one part of it, the
coherent backscattering $\langle \delta R^D(B) \rangle$.

Our semiclassical result of Eq.~(\ref{eq:wlfr}), predicts two
important parameters of the elastic backscattering: its magnitude
and its magnetic field scale. Due to the off-diagonal contribution, 
we expect the magnitude of the weak localization effect to be 
different than the coherent backscattering. On the other hand,
as with the conductance fluctuations, we expect the weak
localization of a chaotic cavity to be characterized by a single
field scale, that should necessarily be the one of the coherent 
backscattering. We confirm such a conjecture with the numerical
results of Fig~\ref{fig:wl1}, where the magnetoconductance of a
chaotic cavity is shown to have a Lorentzian line-shape, with
the width given by the classical parameter $\alc$.

Our numerical studies allow to address the problem of the 
off-diagonal contributions. For the two structures of 
Fig.~\ref{fig:wl2} we split the field dependent part of
the (smoothed) total reflection coefficient $R(B\!=\!0)-R(B)$
(solid) in its diagonal (dashed) and off-diagonal (dotted)
parts. For the structure with the stopper $\langle \delta
R^D \rangle$ is approximately independent of $k$, its 
magnitude is within  $30 \% $ of ${\cal R}$, and the elastic
enhancement factor goes approximately from 2 to 1 when we
turn on the field. But there is an important off-diagonal 
contribution, of opposite sign. Therefore, the weak localization
effect is reduced respect to the coherent backscattering. 
The structure without stoppers exhibits similar features,
but has a very reduced weak localization effect. Also,
the magnitude of the coherent backscattering differs
considerably from ${\cal R}$ and there is an important
net variation as a function of $k$. Such effects are
probably due to the presence of short paths and the 
approximate nature of the uniformity assumption used to
obtain Eq.~(\ref{eq:wlfr}). 

\begin{figure}
\setlength{\unitlength}{1mm}
\begin{center}
\begin{picture}(110,135)
\leavevmode
\epsfxsize=0.85\textwidth
\epsfbox{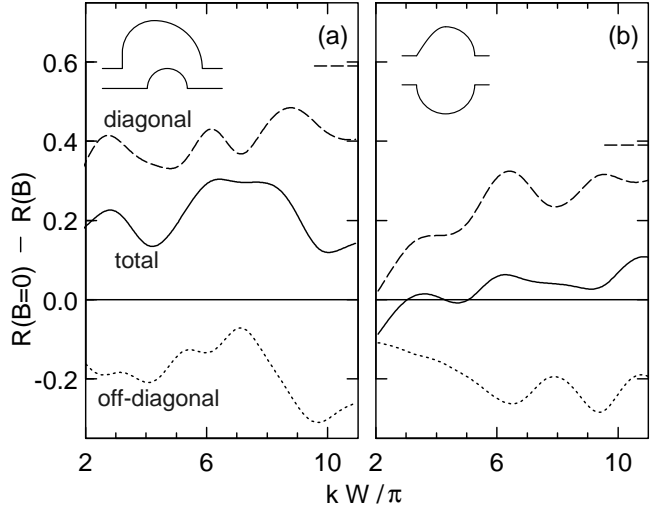}
\end{picture}
\end{center}
\vspace{-6cm}
\caption{
Change in the total (smoothed) reflection coefficient (solid), as well as  the diagonal
(dashed) and off-diagonal (dotted) parts, upon changing $B$ from $0$
to $2 \alpha_{cl} \Phi_{0}$. The dashed ticks on the right mark the classical
value of ${\cal R}$. (From Ref. \protect\cite{Bar93}.)
}
\label{fig:wl2}
\end{figure}

In (\ref{subsec:svrmt}) we will discuss some proposals to 
cope with this limitation of the semiclassical approximation
and the predictions of the theory of random matrices applied
to ballistic transport. With the random matrix approach we
are able to calculate averages of diagonal, as well as 
non-diagonal, coefficients.

In the case of integrable cavities similar considerations to those
invoked in the discussion of conductance fluctuations are pertinent.
Lacking a rigorous theory, we conjecture how to modify our formalism
taking into account the assumptions that break down in the integrable
case. For instance, the uniformity assumption leading to Eq.~(\ref{eq:wlfr})
is no longer valid, and we should work with an angle-dependent
distribution function $N(\Theta,\theta)$. Evaluating (\ref{eq:rd1})
numerically for circular and polygonal shapes leads [approximately] to
$\langle \delta R^D(B) \rangle \propto |B|$ for small $B$ \cite{Bar93,revha}. 
Such linear behavior can be obtained by taking the Fourier transform
of a distribution $N(\Theta) \propto 1/\Theta^2$
In of Fig.~\ref{fig:wl1} (inset) we see for the quantum mechanical
calculations in a polygonal cavity the triangular shape of the weak 
localization peak, together with the Lorentzian one of the chaotic case.
We thus verify that the difference in the area distributions of 
chaotic and integrable cavities leads to a qualitative difference
in the shape of the weak localization peak.

%% file: var3.tex
%=============================================================================

%
% SECTION VII: SCATTERING AND INTEGRABILITY IN QUANTUM TRANSPORT
%
% file: var3.tex
%
% last version: 06/10
%

\section{Scattering and integrability in quantum transport}
\label{sec:saiiqt}

In the previous section we analyzed quantum interference effects
in ballistic cavities using a semiclassical approach. In the 
discussion leading to the semiclassical from of the transmission
amplitude, we stressed the fact that the simple form of 
(\ref{eq:semiclass}), expressed as a sum over classical trajectories,
was only valid in the case in which trajectories were isolated, as
typically happens for chaotic systems. In
this chapter we treat the case of trajectories that appear in 
{\em families} and we obtain a different form of the semiclassical
transmission amplitudes than in the previous case. This difference,
arising from the classical dynamics, has important consequences at
the level of the quantum observables. We first discuss in this section
the example of the families of trajectories that directly cross
a ballistic cavity \cite{Chaost,LinJen}, and then the case of
the scattering through a rectangular billiard by using a continuous
fraction approach \cite{paul}. Transport through a square cavity has 
also been discussed by Wirtz and collaborators \cite{Wirtz}; an 
alternative form of the transmission coefficients was proposed and 
the correspondence between quantum results and classical trajectories 
was established. At the end of the section we present another integrable 
case: circular cavities, for which the semiclassical form (\ref{eq:semiclass})
is applicable, and can be adapted in order to incorporate effects of 
diffraction and tunneling.

\subsection{Direct trajectories} 

A typical structure, like that of
Fig.~\ref{fig:intro} will have a family of trajectories that cross
directly from one lead to the other, without suffering any collision.
This is the simplest case to illustrate the modifications that we
need to include in the semiclassical approach to transport due
to the presence of families of trajectories. As discussed in 
(\ref{subsec:sta}), we can always do the first integration
of (\ref{eq:TRAM0}) by stationary-phase. The resulting 
Eq.~(\ref{eq:tbaSPA1}) for the contribution of direct trajectories 
to the transmission amplitude $t_{ba}$ writes as 

\begin{eqnarray}
t^{\rm d}_{ba}= \frac{i}{W} \sqrt{\frac{\cos{\theta_{b}}}{\cos{\theta}}} \ 
\left\{ \int^{W}_{\LD \tan{\theta}} \ dy' \ \sin{\left[\frac{\pi b y^{\prime}}{W}\right]} 
\ \exp{\left[i \left(\frac{k }{\cos{\theta}} + \frac{\pi a y_{0}}{W} \right) \right]} 
\right. \nonumber \\ \left.
- \int_{0}^{W-\LD \tan{\theta}} \ dy'\ \sin{\left[\frac{\pi b y^{\prime}}{W}\right]}
\ \exp{\left[i \left(\frac{k \LD}{\cos{\theta}} - \frac{\pi a  y_{0}}{W} \right) \right]} \right\} 
\ . 
\label{tdjj}
\end{eqnarray}

\nin $W$ is the width of the leads and $\LD$ the distance between them. The two
terms correspond to the two families, with injection angles $\pm \theta$, verifying
$\sin{\theta}\!=\! a\pi/ kW$. The length of all the trajectories is $L=\LD/\cos{\theta}$, and
$y_{0}=y' \mp \LD \tan{\theta}$ is the initial point of the trajectory reaching
the exiting cross section at $y'$ in the first (second) term. Of course, we
cannot perform as before the integral over $y'$ by stationary-phase as there
is no quadratic term in the action. However we can calculate the
integral in a closed form. For the diagonal term we obtain \cite{Chaost}

\be
t^{\rm d}_{aa}= - \exp{\left[i k L\right)} \left\{\left(1-\rho\tan{\theta}\right]
\exp{\left(-i \pi a \rho \right)}
 + \frac{1}{\pi a} \sin{\left(\pi a \rho \right)}\right\} \ ,
\label{tdjj2}
\ee

\nin with $\rho=(\LD/W)\tan{\theta}$. For large quantum numbers $a$ we can 
neglect the last (small and rapidly oscillating) term. The off-diagonal
terms ($a \neq b$) vanish if $a$ and $b$ have different parity. For modes 
with the same parity we have

\begin{eqnarray}
\hspace{0.6cm} t^{\rm d}_{ba} & = &  - \frac{2}{\pi} \ \sqrt{\frac{\cos{\theta_{b}}} 
{\cos{\theta_{a}}}} \ \exp{i\left[k L - \pi a \rho\right]} 
\left\{\frac{1}{b+a} \ \exp{\left[-i \frac{(b+a)\pi \rho}{2} \right]} 
\ \sin{\left[\frac{(b+a)\pi \rho}{2}\right]} \right.  \\ 
\displaystyle
&  & \hspace{1cm} -
\left.
\frac{1}{b-a} \ \exp{\left[-i \frac{(b-a)\pi \rho}{2} \right]} 
\ \sin{\left[\frac{(b-a)\pi \rho}{2}\right]} \right\} \ .
\nonumber
\label{tndjj2}
\end{eqnarray}

\nin As expected, the result is strongly peaked for $b\!=\!a$ (where
Eq. (\ref{tdjj2}) should be used), but shows
nevertheless an important off-diagonal contribution for close
quantum numbers $b$ and $a$. In Ref.~\cite{LinJen}, Lin generalized
these results to the case in which the leads are not collinear.

It is important to notice the different dependence on $\hbar$ (or $k$) for
the contribution from the family of direct trajectories and the isolated
trajectories. The number of modes that support direct trajectories is
$N(W/\LD)$, and therefore the effect of direct trajectories does not
disappear in the semiclassical limit. The existence of direct trajectories 
complicates the comparison between experiment (both numerical and 
laboratory) and the semiclassical theory. Thus in many of our numerical 
simulations we have introduced ``stoppers'' in the billiards which 
eliminate this effect; the experimentalists have also tried to avoid this 
problem by displacing the leads \cite{Kel94}, having an angle smaller 
than $\pi$ between them \cite{Mar92,marcusgroup3}, or using stoppers
inside the cavity \cite{Kel94,lee97}. 

\subsection{Scattering through a rectangular cavity}
\label{subsec:starc}

\begin{figure}[tbh]
\setlength{\unitlength}{1mm}
\begin{picture}(100,110)
\put(-10,-40)
{\epsfxsize=17cm\epsfbox{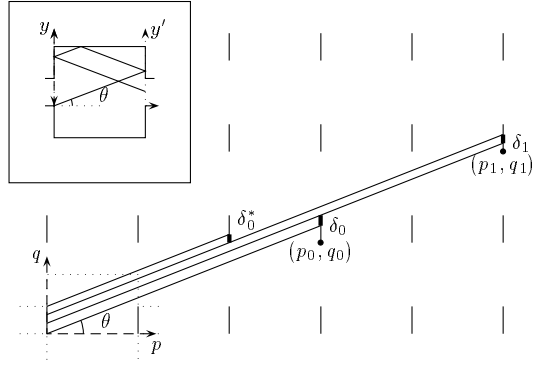}}
\end{picture}
\vspace{-5cm}
\caption{Unfolded space for the dynamics in a square cavity (inset).
The trajectory entering at the lowest point ($y\!=\!0$) of the left
lead with an angle $\theta$ is shown in the original and extended
space. It belongs to the family that leaves the cavity through the
exiting lead $(p_0,q_0)$ and has a weight $\delta_0$. For this
angle $\theta$ there are two families contributing
to the transmission amplitude (0 and 1) and one family contributing
to the reflection amplitude ($0^{*}$).
(From Ref. \protect\cite{paul}.)}
\label{fig:unfold}
\end{figure}

The dynamics in a rectangular cavity is integrable, and the conserved
quantities are the absolute value of the projections of the momentum
of the particle along the two axis. The motion of the particle inside 
the cavity can be represented by a straight line in the extended (or
unfolded) space where the cavity is periodically repeated in both directions
(Fig.~\ref{fig:unfold}). In order to simplify the notation, we work in this
section with a square cavity  of length side $1$ (without loss of
generality since we can always scale the sides of a rectangular cavity
to the square of unit length).

For a trajectory starting at the point $y_0$ of the entrance lead
with and angle $\theta$, we would like to know after how many bounces
with the walls, the particle is going to leave the cavity. In the 
unfolded space, this problem translates into finding the first 
window ({\em i.e.} image of the lead connections) encountered by the straight
line $D(y_0)$ defined by $y=y_0\!+\!x\tan{\theta}$. If we now change
$y_0$, keeping $\theta$ fixed, the nearby trajectories will exit by
the same window until a certain critical value $y_0^{c}$, where we
pass to another family of trajectories associated with a different 
window. Evaluating the families of trajectories with a a given injection
angle leads to a diophantic problem \cite{paul}.
The coordinates $(p,q)$ of the exiting windows are part of the
intermediate fractions (or Farey series) appearing in the continued fraction
representation of $1/\tan{\theta}$. Our interest in fixed injection angles
stems from the semiclassical approximation to the transmission amplitude,
that according to Eq.~(\ref{eq:tbaSPA1}) selects the angles $\pm \theta$,
with $\sin{\theta}\!=\!a\pi/ kW$.

Let us start with the trajectory entering the 
cavity at the lowest point ($y_0\!=\!0$), whose exiting lead is defined by the 
segment $[(p_0,q_0);(p_0,q_0+\!W)]$ which intersects $D(0)$ (see Fig.~\ref{fig:unfold}). 
As we increase $y_0$, we will remain within a family of degenerate trajectories 
(that we note by $n\!=\!0$) until the exiting point $y_0\!+\!p_0 \tan{\theta}$ hits the 
uppermost point of the segment. The pair $(p_0,q_0)$ must verify the conditions:

\begin{enumerate}
\item  [{\em a.}] \ $0<p_0\tan\theta -q_0<W$ \ ,

\item  [{\em b.}] \ $\forall (p,q) \ {\rm such} \ {\rm that} \ 0 < p\tan\theta -q < p_0\tan\theta -q_0 
\Rightarrow p > p_0 $ .
\end{enumerate}
According to $a$, the first $y$-interval is $[0,\delta_0]$ (or equivalently,
the first $y'$-interval is $[W-\delta_0,W]$), while condition $b$ means that
$(p_0,q_0)$ is the first lattice point verifying $a$. The uppermost family will be
associated with an interval $[(p^*_0,q^*_0);(p^*_0,q^*_0\!+\!W)]$, where the pair
$(p^*_0,q^*_0)$ verifies similar conditions as $(p_0,q_0)$:
\begin{enumerate} \setcounter{enumi}{2}
\item  [{\em c.}] \ $0<W+p^*_0\tan\theta-q^*_0<W$ \ ,
\item  [{\em d.}] \ $\forall (p,q) \ {\rm such} \ {\rm that} \ W\!+\!p^*_0\tan\theta\!-\!q^*_0<
W\!+\!p\tan\theta\!-\!q<W \Rightarrow 
p>p^*_0$ \ .
\end{enumerate}
According to $c$, the first $y$-interval of the uppermost family is
$[W\!-\!\delta_0^*,W]$ (the first $y'$-interval is $[0,\delta_0^*]$),
while $d$ implies that $(p^*_0,q^*_0)$ is the first lattice point 
verifying $c$.

Now that we have determined the lowest and the uppermost families for the
$[0,W]$ $y$-interval, the following sequences of families can be obtained
by reducing ourselves to the $y$-interval $[\delta_0,W\!-\!\delta_0^*]$, and with
the changes of $(p_0,q_0)$, $(p^*_0,q^*_0)$ by $(p_1,q_1)$, $(p^*_1,q^*_1)$
the conditions $a-d$ define the next two families. Continuing this procedure
until the two sequences of families begin to overlap each other, we obtain
the sequence of lower families defined by their initial and final points, 

\begin{mathletters}
\label{allEXL}
\begin{eqnarray}
y'_{i}(p_l,q_l) & = & W+(p_l-p_{l-1})\tan\theta-(q_l-q_{l-1}) \ ,
\label{eq:EXL0} \\
\displaystyle
y'_{f}(p_l,q_l) & = & W \ ,
\label{eq:EXL1} 
\end{eqnarray}
\end{mathletters}

\nin The width (or ``weight") of the family is 
$\delta_l=y'_{f}-y'_{i}=q_l-q_{l-1}-(p_l-p_{l-1})\tan\theta$.
For the upper-families we have

\begin{mathletters}
\label{allEXU}
\begin{eqnarray}
y'_{i}(p_u^*,q_u^*) & = & 0 \ ,
\label{eq:EXU0} \\
\displaystyle
y'_{f}(p_l^*,q_l^*)& = & (p_u^*-p_{u-1}^*)\tan\theta-(q_u^*-q_{u-1}^*) \ ,  
\label{eq:EXU1}
\end{eqnarray}
\end{mathletters}

\nin and the width $\delta_u^*=(p_u^*-p_{u-1}^*)\tan\theta-(q_u^*-q_{u-1}^*)$
The very last family is simultaneously shadowed by lower and upper 
families, therefore has $y'_{i}$ given by (\ref{eq:EXL0}) and $y'_{f}$ by
(\ref{eq:EXU1}).

We can now establish the relationship with the continued fraction representation of 
$\Theta\!=\!1/\tan\theta$, that is defined by the sequences $(\alpha_m)$ and 
$(a_m)$ as follows \cite{khinchin}:
\begin{equation}
\label{eq:khin1}
\begin{array}{cc}
  \alpha_0 = \Theta & a_0 = [\alpha_0] \\
  \alpha_{m+1} =  \frac1{\alpha_m - a_m} &  \quad a_{m+1} =
  [\alpha_{m+1}] \ .
\end{array}
\end{equation}

\nin $[\alpha]$ denotes the integer part of $\alpha$. The best rational
approximations to $\Theta$ are the fractions $P_{m}/Q_{m}$, called
convergents, and obtained from the recurrence relations
\begin{equation}
\label{eq:khin2}
\left\{\begin{array}{ll}
      P_{m}=P_{m-2}+a_m P_{m-1} \\
      Q_{m}=Q_{m-2}+a_m Q_{m-1} \ , 
    \end{array}
\right.
\end{equation}

\nin with the initial choice of $(P_{-1},Q_{-1})=(1,0)$ and
$(P_{0},Q_{0})=([\Theta],1)$. Since $(P_m)$ and $(Q_m)$ are sequences of
integers, we can represent the convergents as lattice points
$(P_m,Q_m)$ that approach the straight line $D(0)$ as $m$ increases from
above (even $m$) and below (odd $m$).

The $m$-th Farey sequence (or intermediate fractions) are the intermediate 
lattice points on the segment $[(P_{m-2},Q_{m-2}),(P_{m},Q_{m})])$ defined by
\begin{equation}
\label{eq:deffar}
\left\{
  \begin{array}[c]{l}
    p_{m}^k = P_{m-2} + k  P_{m-1}\\
    q_{m}^k = Q_{m-2} + k  Q_{m-1}
  \end{array}
\right. \quad 0\leq k \leq a_m \ .
\end{equation}

\nin This sequence sequence of equally spaced lattice points $(p_{m}^k,q_{m}^k)$
starts at the convergent $(P_{m-2},Q_{m-2})$ (for $k\!=\!0$) and finishes at
$(P_{m},Q_{m})$ (for $k\!=\!a_m$). The translation vector is given by the
coordinates of the convergent $(P_{m-1},Q_{m-1})$. The intermediate
fractions with odd (even) $m$ verify the properties
the conditions $b$ ($d$), thus determining the lattice points 
associated with the exiting leads. The two types
of families are given by the Farey sequences of even and odd order convergents, with the
additional requirements that the exiting points are closer than 
$W$ to the straight line $D(0)$. Odd (even) values of $p_n$ correspond to 
transmission (reflection) trajectories. The initialization $(p_{-1},q_{-1})=(1,0)$
and $(p_{0},q_{0})=([\Theta],1)$ is an arbitrary choice, and some special care
has to be taken in the case in which $\tan{\theta}<1/2$ where they may exist lattice
points between the initial ones.

The continued fraction representation of a generic (irrational) real number 
results in an infinite sequence of convergents. However, the fact that 
we have defined a finite distance $W$ to approach $D(0)$ implies that once the
lower and upper families begin to overlap we do not need to consider higher
convergents. Therefore, each angle $\theta_a$ is associated to a finite
number of families. In addition, it can be shown that at most three 
sequences of intermediate fractions are relevant \cite{paul}.

\subsection{Semiclassical transmission amplitudes for square cavities}

In the extended space, the Green function between points at the entrance and
exit leads equals to that of the free space taken between the initial
point and the image of the final one (with the adjustment of the phases at
the reflections on the hard walls).

As in the previously studied cases, the semiclassical approach for the
transmission amplitudes $t_{ba}$ selects the incoming angle $\theta$
corresponding to the mode $a$ (Eqs.~(\ref{eq:stphcon})-(\ref{eq:redact})).
Symmetry arguments for a rectangular cavity dictate that $t_{ba}\!=\!0$ if
$a\!+\!b$ is odd. For even $a\!+\!b$ we perform the $y'$-integration for
each family of trajectories $n$, obtaining a transmission amplitude 

\be
  \label{eq:tabinter}
  t_{ba}= -\frac{i}{W} \ \sqrt\frac{\cos\theta_b}{\cos\theta_a}
  \sum_{n}  \varepsilon_n \exp[i k \tilde{L}_n]
  \left\{I_{n}(a\!+\!b)-I_{n}(a\!-\!b)\right\} \ ,
\ee

\nin where

\be
  \label{eq:idef}
I_{n}(x) = \frac{W}{\pi x} \left(\exp{\left[i \frac{\pi x}{W}
y_{f}^{\prime(n)}\right]} - \exp{\left[i \frac{\pi x}{W} y_{i}^{\prime(n)}\right]}
\right) \ .
\ee

\nin $(p_n,q_n)$ are the coordinates of the exiting lead in the extended space,
$y_{i}^{\prime(n)}$ and $y_{f}^{\prime(n)}$ are the extreme points of
the exiting interval. We have defined the phase 
$\varepsilon_n\!=\!\exp[i\pi(a\!+\!1) \varepsilon(q_n)]$, with 
$\varepsilon(q_n)\!=\!0$ ($1$) for even (odd) $q_n$. The parity of $q_n$ 
appears in the phase due to mirror symmetries involved in going to 
the extended space. The phase gained at the hard-walls is given by the index
$\tilde{\nu}_n=2(p_n\!+\!q_n\!-\!1)$.
The trajectories of the $n$-th family have a length
$L_n = p_n/\cos{\theta_a}$ (all lengths are expressed in units of the
side of the square), we have defined the reduced length 
$\tilde{L}_n = p_n \cos{\theta_a} + q_n \sin{\theta_{a}}$.
The trajectories that contribute to the transmission amplitude are those
going from the left to the right lead, therefore only the values $n$ with
odd $p_n$ are considered. Similarly to the contribution of direct trajectories, the 
case $\bar{a}\!+\!\bar{b}=0$ has to be treated separately for the $y'$-integration.
However, the corresponding result is included in Eq.~(\ref{eq:tabinter}) by taking the
limit $x \! \rightarrow \! 0$. Obviously, $\bar{a}\!+\!\bar{b}=0$ corresponds to the
maximum transmission since this is the case where the classical trajectory arrives 
to the exiting lead with the quantized angle of mode $b$. 

When the $y'$-intervals are of the form $[W\!-\!\delta_l,W]$ or $[0,\delta_u^*]$
(Eqs.~(\ref{allEXL})-(\ref{allEXU})) we can separate the semiclassical sum
into upper ($u$) and lower ($l$) families and write

\be
  \label{eq:tab}
  t_{ba}= \frac{1}{W} \ \sqrt\frac{\cos\theta_b}{\cos\theta_a}
  \sum_{n=l,u} \varepsilon_n \varepsilon_n^{\prime} \exp{[ik \tilde{L}_n]}
  \left\{\Delta_n(a+b)-\Delta_n(a-b)\right\} \delta_n \ ,
\ee

\nin with the family-dependent function $\Delta_n$ defined by

\be
  \label{eq:Delta}
\Delta_n(x) = \frac{2W}{\pi x \delta_n} 
\exp \left[{i\varepsilon_n^{\prime}\frac{\pi x}{2W}\delta_n}\right] \ 
\sin \left[\frac{\pi x}{2W}\delta_n\right] \ ,
\ee

\nin $\varepsilon_n^{\prime}\!=\!1$ if $n\!=u$ and 
$\varepsilon_n^{\prime}\!=\!-\!1$ if $n\!=l$ \cite{correction2}.  
As previously stated, only odd $p_n$ should be considered in the
expansion for the transmission amplitude. The ``last" family
does not have a $y'$-interval with the simple form of the previous ones.
Therefore, when this family contributes to the transmission amplitude, 
we cannot use for it the simple form (\ref{eq:tab}) depending only on the
weights $\delta_n$, but Eq.~(\ref{eq:tabinter}) together with
the limits (\ref{eq:EXL0}) and (\ref{eq:EXU1}). 

The forms (\ref{eq:tabinter}) and (\ref{eq:tab}) of the transmission amplitude 
provide a very
powerful method to numerically compute the conductance through a rectangular
cavity. For each $a\!=\!1\ldots N$, we only need to calculate a finite number of 
convergents and intermediate fractions of the continued fraction representation 
of $1/\tan{\theta_a}$ and recursively obtain the weighting intervals 
$\delta_n$. The advantage over quantum methods based on discretization
(recursive Green function, wave function matching, etc) is that it can be
used for large wave-vectors $k$. The method actually gets more exact with
increasing $k$ since it involves a semiclassical approximation. 

\begin{figure}
\setlength{\unitlength}{1mm}
\begin{picture}(100,110)
\put(-80,20){\epsfxsize=9cm\epsfbox{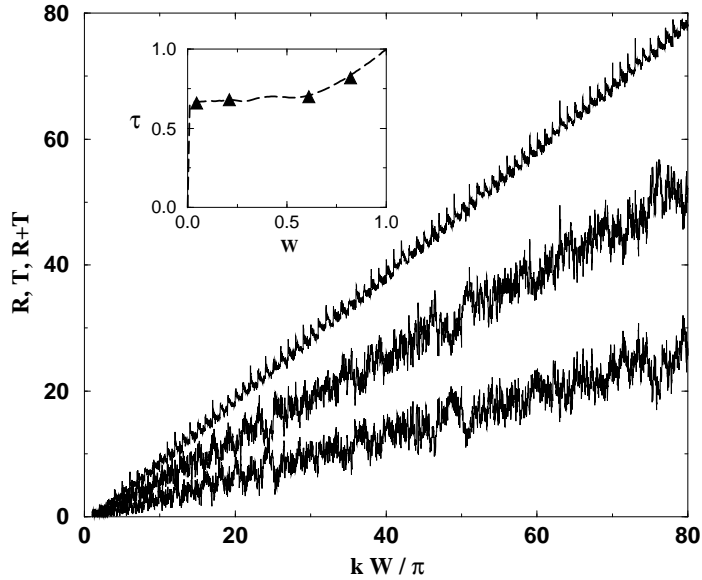}}
\end{picture}
\vspace{-2cm}
\caption{Reflection $R$ (lower curve), transmission $T$ (medium), and $R+T$
(upper) for a square cavity as a function of wave-vector for an opening of
$W=0.05$ (in units of the size of the square). Inset: classical transmission
probability as a function of $W$ from the slope of $T(k)$
(triangles) and from Eq.~(\protect\ref{eq:classT}) (dashed). 
}
\label{rtarpt}
\end{figure}

In Fig.~\ref{rtarpt} we present the conductance of a square cavity calculated
from Eq.~(\ref{eq:tab}) for an small opening ($W\!=\!0.05$) in a large 
$k$-interval. We remark two salient features:
the linear increase of the mean conductance with $N=kW/\pi$ and the fluctuations
around the mean, which become larger as $k$ increases. The linear
increase of $\langle T \rangle$ with a slope given by the classical 
transmission coefficient agrees with the classical behavior discussed in 
the previous section. The 
increase of the fluctuations obtained within the semiclassical approach is
consistent with previous quantum computations \cite{Chaost}. 
The quantum mechanical calculations of Wirtz {\em et al.} \cite{Wirtz} for
a square cavity allowed to identified the peaks of the Fourier transform of
the transmission amplitude with the families (or bundles)
of classical trajectories contributing in the semiclassical expansion.

Unitarity,
the mathematical translation of charge conservation, is a critical test for 
semiclassical approximations. Fig.~\ref{rtarpt} shows that it is relatively well respected,
except at the opening of new modes, where we obtain diffraction peaks. The curve 
$R\!+\!T$ has a slope of $N$, within an error of $5\%$. The
departures from unitarity are smaller than the fluctuations, therefore it is
justified to approach the latter with our semiclassical methods.

\subsection {Mean conductance in a square cavity}

Our semiclassical approach can be further simplified in order to render the
calculations analytically tractable.
The function $\Delta_n(x)$ defined in Eq.~(\ref{eq:Delta}) is peaked for $x\!=\!0$
(when the quantized angles of the incoming and outgoing modes coincide with the
angle of the trajectory) and decays on the scale of $W/\delta_n$, therefore it
can be approximated by the rectangular function

\be
\Pi_n(x) = \left\{
 \begin{array}[c]{l}
    \exp{\left[i\varepsilon_n^{\prime}\frac{\pi x}{2W}\delta_n \right]} \ , \qquad
    {\rm if} \quad |x|<\frac{W}{2\delta_n} \ , \\
    0 \qquad {\rm otherwise \ .}
  \end{array}
\right.
\ee

Thus, the semiclassical $t_{ba}$ simplifies to:

\be
\label{eq:tab2}
t_{ba} = \frac{1}{W}\sqrt\frac{\cos\theta_b}{\cos\theta_a}
\sum_{n} \Pi_n(a-b) \varepsilon_n \varepsilon_n^{\prime} \delta_n
\ \exp[ik \tilde{L}_n] \ .
\ee

Within this approximation, the total transmission coefficient
is be expressed as a sum over pairs of families of trajectories 
(with odd $p_n$ and $p_{n'}$).

\be
  \label{eq:simp}
  \hspace{0.6cm} T= \frac{1}{W^2} \sum_{a,b=1}^{N}
  \frac{\cos\theta_b}{\cos\theta_a} \sum_{n,n'} \varepsilon_n
  \overline{\varepsilon}_{n'}
  \varepsilon_n^{\prime} {\varepsilon}_{n'}^{\prime}  \delta_n
  \delta_{n'} 
  \exp{[ik(\tilde{L}_n\!-\!\tilde{L}_{n'})]} \
  \Pi_n(a\!-\!b) \overline{\Pi}_{n'}(a\!-\!b) \ .
\ee

The secular behavior of the transmission coefficient can be obtained 
from the $k$-average of $T(k)/k$, leading to the diagonal approximation 
between the families of trajectories ($n\!=\!n'$). In the
case of isolated trajectories, the diagonal approximation
yielded the classical probability of transmission by
pairing individual trajectories. In the present case, the concept of
families of trajectories replaces the role of individual trajectories.
Inserting the definition of the function $\Pi$, we find (to leading
order in $kW/\pi$)

\be
  \langle T\rangle \simeq \frac{1}{W^2} \ \sum_{a=1}^{N} \ \sum_{n} \delta_n^2
  \left( \sum_{b=b_{min}}^{b_{max}}
  \frac{\cos\theta_b}{\cos\theta_a} \right) \ ,
\ee

\nin $b_{min}=\max{\{a\!-\!W/2\delta_n,1\}}$ and
$b_{max}=\min{\{a\!+\!W/2\delta_n,N\}}$. In the classical limit of 
$N\!=\!kW/\pi\gg 1$ the sum over $b$ can be approximated by an integral
leading to

\be
\label{eq:simpspl2}
  \langle T \rangle \simeq \sum_{a=1}^{N} \sum_{n} \frac{\delta_n}{W} \ .
\ee

For each mode $a$ we have simply obtained the total weight of
trajectories contributing to transmission. Converting also the sum over 
$a$ into an integral we write

\be
\langle T \rangle \simeq \frac{kW}{\pi} \ {\cal T}
\qquad  ,  \qquad 
{\cal T} = \int_{0}^{\pi/2} d\theta \cos\theta \ \sum_{n} \frac{\delta_n}{W} \ .
\label{tracin}
\ee

\nin Thus, the total transmission coefficient is proportional to the
number of modes, and the constant ${\cal T}$ is a purely geometric factor.
Breaking the contribution of families into that of individual trajectories
we are left with the usual classical transmission probability (\ref{eq:classT}).

In the inset of Fig.~\ref{rtarpt} we compare the mean slope in the numerical
(semiclassical) results (triangles) with the classical transmission probability
${\cal T}$ obtained from Eq.~(\ref{eq:classT}) by sampling the space of classical
trajectories with random choices of initial conditions $\theta$ and $y$ (dashed). 
We verify the good agreement between the two approaches and we see that ${\cal T}$ 
remains almost constant over a large interval of variation of the opening.
A more efficient path to ${\cal T}$ than Eq.~(\ref{eq:classT}) is to sample the angles $\theta$ and to incorporate
the weights $\delta_n$ emerging from the intermediate fractions of $1/\tan\theta$, as
suggested by Eq.~(\ref{tracin}). We then see
that the continued fraction approach is not only useful for evaluating
semiclassical effects, but also for classical properties like the
transmission coefficient (and also the length distribution). Random sampling
of classical trajectories is an appropriate procedure for chaotic
structures, where the ergodicity of phase space results in an
exponential distribution of lengths. On the other hand, integrable
cavities exhibit power-law distributions, which are more difficult
to obtain by trajectory sampling. In this case, the continued fraction
approach is very efficient since, for a given angle, only a finite number
of terms are relevant, and the contributing families are incorporated
at once according to their weight.

\subsection {Conductance fluctuations in a square cavity}

As visible from Fig.~\ref{rtarpt}, the oscillations around the
mean transmission coefficient grow with larger $N$.
We will now evaluate the local fluctuations 
$\langle (\delta T)^2 \rangle = \langle (T-[(kW/\pi){\cal T} + \langle \dlT \rangle])^2 \rangle$. 
We begin with the simplified expression (\ref{eq:simp}) of $T$, and write

\begin{eqnarray}
  \hspace{1.0cm}
  T^2 & = & \frac{1}{W^4} \sum_{a,b,a',b'=1}^{N} \frac{\cos \theta_b}{\cos
  \theta_a}\frac{\cos \theta_{b'}}{\cos \theta_{a'}}
  \sum_{n,n^{\prime},n^{\prime\prime},n^{\prime\prime\prime}}
\varepsilon_n \overline{\varepsilon}_{n^{\prime}} \varepsilon_{n^{\prime\prime}}
\overline{\varepsilon}_{n^{\prime\prime\prime}} 
\varepsilon_n^{\prime} \overline{\varepsilon}_{n^{\prime}}^{\prime}
\varepsilon_{n^{\prime\prime}}^{\prime}
\overline{\varepsilon}_{n^{\prime\prime\prime}}^{\prime}
\delta_n \delta_{n^{\prime}} \delta_{n^{\prime\prime}} \delta_{n^{\prime\prime\prime}}
\displaystyle
\\ 
&  & \exp{\left[ ik (\tilde{L}_n\!-\!\tilde{L}_{n'}\!+\!\tilde{L}_{n^{\prime\prime}}\!-\!
\tilde{L}_{n^{\prime\prime\prime}})\right]}
\ \Pi_n(a\!-\!b) \overline{\Pi}_{n^{\prime}}(a\!-\!b) \Pi_{n^{\prime\prime}}(a'\!-\!b')
  \overline{\Pi}_{n^{\prime\prime\prime}}(a'\!-\!b') \ .
\nonumber
\label{eq:g2}
\end{eqnarray}

\nin As before, we only consider the terms having a null phase, for which

\be
  \label{eq:ph0}
 \tilde{L}_n-\tilde{L}_{n'}+\tilde{L}_{n^{\prime\prime}}-
\tilde{L}_{n^{\prime\prime\prime}} = 0 \ .
\ee

\nin This condition is satisfied with the pairing  $\tilde{L}_n=\tilde{L}_{n'}$
and $\tilde{L}_{n^{\prime\prime}}=\tilde{L}_{n^{\prime\prime\prime}}$, but the
resulting term cancels against the square of the average transmission coefficient.
A non trivial pairing is obtained when $\tilde{L}_n=\tilde{L}_{n^{\prime\prime\prime}}$
and $\tilde{L}_{n'}=\tilde{L}_{n^{\prime\prime}}$ with $\tilde{L}_n \ne
\tilde{L}_{n'}$, which implies $a\!=\!a'$. The contribution of this pairing to the
local fluctuations is

\begin{eqnarray}
 \langle (\delta T)^2 \rangle_{I} =
  \frac{1}{W^4}\sum_{a=1}^{N} \sum_{b,b'=1}^{N}
  \frac{\cos \theta_b \cos \theta_{b'}}{\cos^2 \theta_a} 
  \sum_{n,n'} \Pi_n(a\!-\!b)\overline{\Pi}_{n'}(a\!-\!b)\Pi_{n'}(a\!-\!b')
  \overline{\Pi}_n(a\!-\!b') \delta_n^2 \delta_{n'}^2 \nonumber \\
  = \frac{1}{W^4}\sum_{a=1}^{N} \sum_{n,n'} \left(\min{\left\{\frac{W}{\delta_n},
\frac{W}{\delta_{n'}}\right\}}
\right)^2 \delta_n^2 \delta_{n'}^2 \ .
\label{eq:fluflu}
\end{eqnarray}

\nin In the semiclassical limit the sum over $a$ can be converted into an integral
dictating a linear behavior of $\langle (\delta T)^2 \rangle_{I}$ with respect
to $k$.

\begin{figure}
\setlength{\unitlength}{1mm}
\begin{picture}(100,110)
\put(-80,20){\epsfxsize=9cm\epsfbox{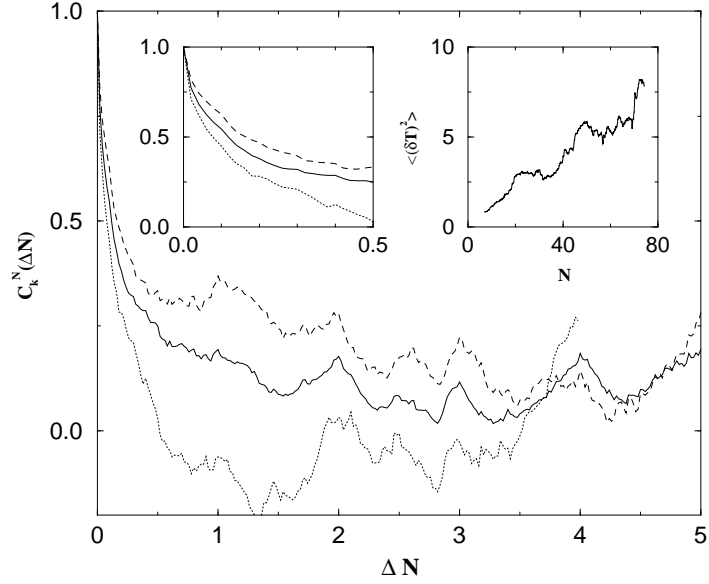}}
\end{picture}
\vspace{-2cm}
\caption{Correlation function locally normalized according to
Eq.~(\ref{eq:corrfun}) for three $N$-intervals: 20-40 (dotted),
40-60 (solid), and 60-80 (dashed) with an opening $W=0.05$. Left inset: blow up of the small 
$\Delta N$ region showing a cusp at the origin. Right inset: 
local variance $\protect \langle (\delta T)^2(k) \rangle$ exhibiting
a linear increase with $N=kW/\pi$.
}
\label{fig:corrfunc}
\end{figure}

The two pairings above described are those usually considered in dealing with chaotic
cavities, except that in such cases we take individual trajectories instead
families. In the integrable system we are studying there is another non trivial
way of satisfying Eq.~(\ref{eq:ph0}), that is, $\tilde{L}_n\!-\!\tilde{L}_{n'}=
\tilde{L}_{n^{\prime\prime}}\!-\!\tilde{L}_{n^{\prime\prime\prime}}$, with
$\tilde{L}_n \ne \tilde{L}_{n'}$ and $\tilde{L}_n \ne \tilde{L}_{n^{\prime\prime}}$.
This typically happens when $n,n',n^{\prime\prime}$, and $n^{\prime\prime\prime}$
belong to the same ($m$-th) Farey sequence and they are respectively associated
with the intermediate functions $(p_m^{k+j},q_m^{k+j})$, $(p_m^{k},q_m^{k})$,
$(p_m^{k'},q_m^{k'})$, and $(p_m^{k'+j},q_m^{k'+j})$, with $\!k\ne\!k'$, $j\!\ne\!0$
and $k,k',k\!+\!j,k'\!+\!j \in (0,a_{m})$. Also, since we are dealing with
transmission coefficients, we need $p_m^{k}$, $p_m^{k'}$, $p_m^{k+j}$ and $p_m^{k'+j}$
to be odd, which implies that if $j$ is odd, $P_{m-1}$ must be even. Under such conditions,
we have
$\tilde{L}_n-\tilde{L}_{n'} = j(P_{m-1} \cos \theta_a + Q_{m-1} \sin \theta_a)$.
If neither $n$ nor $n'$ correspond to the first family of the sequence
($k,k' \ne 0$) the four exiting intervals have the same length
$\delta_m= \left|P_{m-1}\tan \theta - Q_{m-1}\right|$.

This last contribution to the local fluctuations can be expressed as a sum
over the convergents

\be
  \langle (\delta T)^2 \rangle_{II} = \sum_{a=1}^{N} \sum_{m}
  \sum_{k,k'=0}^{a_m} \sum_{j} \frac{\delta_m^2}{W^2} \ ,
\ee

\nin with the above specified restrictions for $k,k'$ and $j$. As in the
previous case, converting the sum over $a$ into an integral yields a
contribution $\langle (\delta T)^2 \rangle_{II}$ to the local fluctuations
that is linear in $k$, with a purely geometrical coefficient given by the
continued fraction representation of $1/\tan \theta$.

The linearity of $\langle (\delta T)^2 \rangle=\langle (\delta T)^2 \rangle_{I}+
\langle (\delta T)^2 \rangle_{II}$ with $k$ that we have demonstrated is 
consistent with the numerical results of Fig.~\ref{fig:corrfunc} (right inset). 

The conductance fluctuations of chaotic cavities are characterized by
two $k$-independent parameters: the correlation length $\gcl$ and 
the variance $\langle (\delta T)^2 \rangle$. In our integrable cavity
we have seen that $\langle (\delta T)^2 \rangle$ is not universal, but energy
dependent. Also, there does not exist a characteristic time for exiting the
cavity. Therefore it is not obvious that a correlation function depending
only on the energy (or momentum) increment can be defined. That is why we 
consider the normalized correlation function

\be
\label{eq:corrfun}
  C_k^N(\dlk) = \frac{\langle \delta T(k\!+\!\Delta k)\delta
  T(k)\rangle}{\langle (\delta  T)^2(k) \rangle} \ ,
\ee

\nin where $k$ varies on an interval much larger than $\Delta k$, but
small enough to neglect secular variations.

From Eqs.~(\ref{eq:corrfun})
and (\ref{eq:tab}) we obtain numerically the correlation functions
shown in Fig.~\ref{fig:corrfunc} for three $N$-intervals: 20-40 (dotted),
40-60 (solid), and 60-80 (dashed). A singularity for small $\Delta k$
appears in all the $N$-intervals in the form of a cusp around the origin (inset).
The linear behavior of $C(\Delta k)$ is to be contrasted to the Lorentzian
correlation function expected for a chaotic cavity. It agrees with the prediction
of Ref.~\cite{Lai92} and it is consistent with quantum calculations yielding 
decay of the power spectrum as $x^{-2}$ at large $x$ (see Fig.~\ref{fig:pow4}
and Ref.~\cite{Chaost}).

We have seen an example of an open system with integrable dynamics exhibiting
larger fluctuations than those of the chaotic case. The situation is analogous to the
density of states of closed systems, which is characterized by stronger
fluctuations in integrable than in chaotic geometries. The augmented
fluctuations in {\em integrable closed and open} geometries can be
traced to the same origin: {\em the bunching of trajectories into
families} in the semiclassical expansions, the Berry-Tabor formula and
Eq.~(\ref{eq:tab}) respectively.

The unbounded fluctuations we have found are unlikely to be experimentally
detected in Mesoscopic Physics. It is a very small effect that necessitates
a range of variation of $N$ much larger than what is normally achieved 
\cite{Kel94,lee97,zozou97}. Also, the cusp of the correlation function at the
origin is related with very long trajectories, that may be longer than our
physical cut offs. The case of the square is rather special amont integrable
systems since the conserved quantities of the cavity are the same as in 
the leads. Therefore, he geometry of the leads plays a very important role 
\cite{bird,zozou97} and renders the quantum signatures of integrability in 
open systems quite involved.

\subsection{Circular billiards, diffraction and tunneling}
\label{subsec:cbdat}

The circular billiard is particularly interesting because it has been
realized experimentally \cite{Mar92,Chaos,chang94,persson,lee97}, and it is an integrable
geometry where the semiclassical transmission amplitude (\ref{eq:semiclass})
is applicable since the contributing trajectories are isolated. Also, the
proliferation of trajectories with the maximum number of bounces considered 
is much weaker than for the chaotic case, making the explicit summation of 
Eq.~(\ref{eq:semiclass}) much easier. Lin and Jensen \cite{LinJen} undertook
such a calculation considering trajectories up to 100 bounces. As expected,
going into the semiclassical limit by increasing the number of modes $N$
results in a better fulfillment of the unitarity condition $T\!+\!R=N$ (only
a 1\% deviation for $N\!=\!20$). Moreover, they demonstrated that the coherent
backscattering is significantly reduced by off-diagonal contributions to the 
total reflection, and they obtained conductance fluctuations as a function of 
energy with high degree of regularity.

The signature of classical trajectories in the numerically obtained
quantum transmission amplitudes has been established for circular 
billiards \cite{ishio95,schwi96,ingold}. In particular,
the Fourier transform of the transmission amplitudes shows strong
peaks for lengths corresponding to the classical trajectories
contributing in the semiclassical expansion (\ref{eq:semiclass}).
Since the injection angle depends on $k$, a given trajectory contributes
only over a limited energy range. This is why in geometries with stable
trajectories, like the circle, the Fourier peaks are more pronounced than
for the stadium billiard. It has also been shown in Ref.~\cite{ishio95} 
that $\langle(\dlT)^2\rangle$ increases with $k$, consistently with the behavior found in
the last chapter for another integrable case (the square).

\begin{figure}[tbh]
\setlength{\unitlength}{1mm}
\begin{picture}(110,145)
\put(10,40){\epsfxsize=8cm\epsfbox{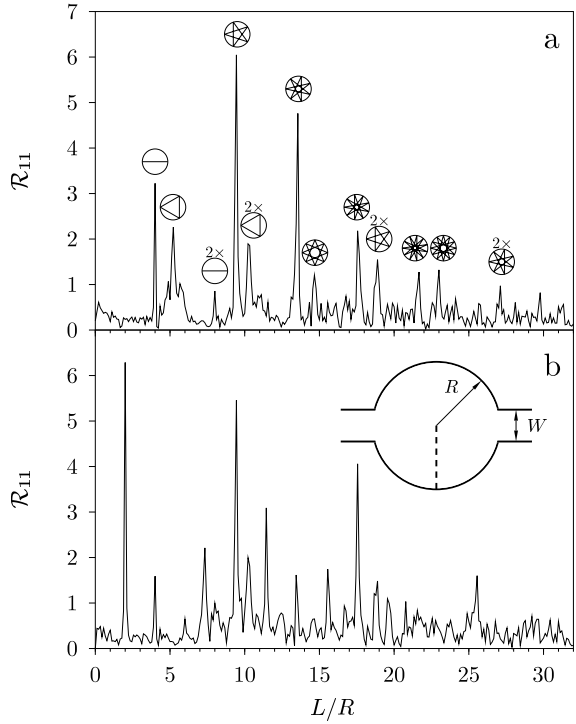}}
\end{picture}
\vspace{-4cm}
\caption{Length spectrum ${\cal R}_{11}$ (magnitude squared of the Fourier transform
of $r_{11}(k)$) for the circular billiard of the inset. 
(a) the case without a barrier, (b) with a thin and infinitely high barrier
(shown as the dashed line reaching from the edge to the center of the circle). Lengths 
are scaled to the radius $R$ of the circle. In the absence of a barrier we identify the 
peaks in the length spectrum with periodic trajectories.
(From Ref. \protect\cite{ingold}.)}
\label{fig:pacfigure}
\end{figure}

Diffraction effects become important when the number of incoming
modes $N$ is small. The semiclassical approach we have presented can be
generalized to include diffraction effects in the transmission and
reflection amplitudes \cite{vattay}. In the extreme limit of $N\!=1\!$,
the wave-front impinging into the cavity is approximately circular
and has the entering lead as the source. The semiclassical propagation
of this wave-front can be built from classical trajectories launched
from the center of the lead in all directions \cite{schwi96}. The
resulting expansion reproduces, at the level of Fourier transforms,
rather well the exact quantum mechanical data in the case of a
circular billiard. Some of the harmonics of the quantum reflection
amplitude do not correspond to classical trajectories, but to ``ghost
paths", or diffractive trajectories (like reflections off the mouth
of an exiting lead), that can be incorporated in an approximate way
in this previous semiclassical formalism.

In Fig.~\ref{fig:pacfigure} we present the correspondence between 
peaks in the squared modulus ${\cal R}_{11}(L)$ of the Fourier transform
of $r_{11}(k)$ and the classical trajectories (including their repetitions)
contributing to (\ref{eq:semiclass}), established in Refs.~\cite{ishio95,schwi96,ingold},
for the circular billiard of the inset. The first peak is not a classical
trajectory contributing to reflection, but corresponds to diffraction off 
the lead mouths \cite{ishio95,schwi96}. This effect can
be interpreted in terms of a trajectory that gets reflected back at the right
lead (ghost path). For larger lengths $L$, we can identify a triangular path,
a five-star path, a seven-star path, and so on. Since we are considering $r_{11}$,
and since the angular momentum is conserved, the outgoing angle is opposite to the
incoming one, and the transport trajectories tend to coincide with periodic orbits. 

Placing a sufficiently high barrier into the cavity considerably changes the
reflection amplitude and ${\cal R}_{11}$ (panel b); some of the peaks are reduced,
other augmented, and new length scales appear. Some of the new peaks are related
with diffractions induced by the barrier. Interestingly enough, the dependence of 
the peak amplitudes is not monotonic with the strength
of the barrier. This behavior is well described by a simple modification of 
Eq.~(\ref{eq:semiclass}) where each contribution is affected by the tunneling
amplitude of a two-dimensional plane-wave encountering a barrier of infinite length \cite{ingold}.

It is important to note that the straightforward identification of classical
trajectories in the quantum calculations is always at the level of transmission 
and reflection amplitudes. For the transmission and reflection coefficients,
the identification is more problematic since we deal with pairs of trajectories,
and the Fourier transforms yield differences of lengths. 

%% file: var4.tex
%=============================================================================

%
% SECTION VII: EXPERIMENTS ON BALLISTIC TRANSPORT AND OTHER ASPECTS
%              OF THE THEORY
%
% file: var4.tex
%
% last version: 07/10
%

\section{Experiments on ballistic transport and other aspects of the theory}
\label{sec:eobtaoaott}

In the two previous sections we developed the semiclassical theory of
ballistic transport, studying conductance fluctuations and weak
localization in chaotic and integrable cavities. Our emphasis was
set in the signatures of the underlying classical mechanics, but we
did not discussed in detail the actual experimental work. We intend to
provide here a brief description of the experimental results relevant
in the development and testing of the theory. (For reviews on 
experiments in ballistic transport see Refs.~\cite{Chaos,csf,Westervelt}.)
The initial application of Quantum Chaos ideas in transport through
ballistic cavities was followed by work in different regimes,
other systems, and also the development of new theoretical tools. 
Without attempting to review such developments, we briefly discuss
in this section the regime of mixed dynamics, the semiclassical
approach of Kubo conductivity applied to the antidot lattice, and
the connection between semiclassics and the random matrix theory of
ballistic transport.

\subsection{Conductance fluctuations and weak localization in ballistic
microstructures}
\label{subsec:cfawlibm}

\begin{figure}
\setlength{\unitlength}{1mm}
\begin{picture}(100,110)
\put(-10,140){\epsfxsize=12cm\epsfbox{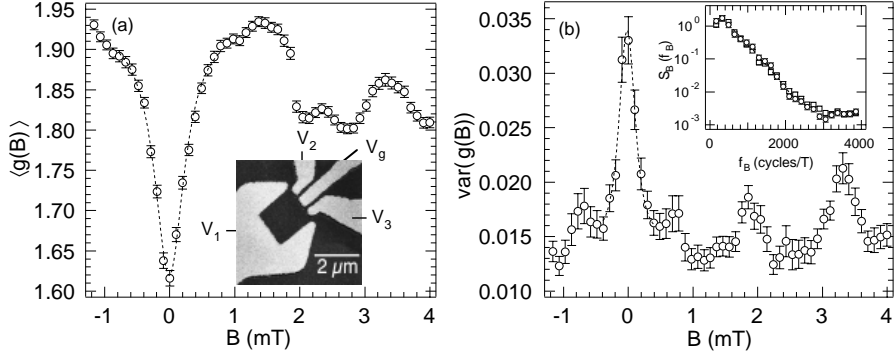}}
\end{picture}
\vspace{-5.5cm}
\caption{
(a) Shape-averaged conductance showing a weak localization peak fitted
to a Lorentzian (dashed). Inset: electron micrograph of device (the gate
voltage $V_g$ is
used to produce shape distortion. (b) Variance of shape-distortion
conductance fluctuations (in units of $(e^2/h)^2$) Inset: power spectral
density in magnetic field fitted by Eq.(\protect\ref{eq:CdB4}). 
(Adapted from Ref. \protect\cite{marcusgroup2}.)}
\label{fig:charly}
\end{figure}

The statistical analysis of the low-temperature magnetoconductance in 
ballistic $GaAs/AlGaAs$ quantum dots was first performed by Marcus and
collaborators \cite{Mar92,Chaos}. Different shapes (stadium and circle)
with steep-walled confining potentials were achieved and the leads were
oriented at right angles to reduce transmission via direct trajectories.
The transport mean free path was estimated to be several times the size
of the structures ($a \simeq 0.5 \mu m$) indicating that the ballistic
regime was achieved, and the number of conducting channels was $N\!=\!3$
(that is, not quite in the semiclassical limit of the theory). Analyzing
the magnetoconductance as we did in (\ref{subsec:cf}) gave for the 
stadium cavities a power spectrum in agreement with Eq.~(\ref{eq:CdB4})
over three orders of magnitude, and a deviation for large areas was
observed (as in the inset of Fig.~\ref{fig:charly}.b). The conductance fluctuations
in the circular billiard were more structured (more weight in the 
high harmonics of the power spectrum) compared with the case of the
stadium, demonstrating experimentally the possibility of distinguishing
ballistic cavities according to their classical dynamics. 

The magnetic field scale of the fluctuations was found to be consistent
with the semiclassical prediction, and increasing with the mean
conductance through the dot \cite{marcusgroup1}. In the completely
coherent picture of Sec.~\ref{sec:qttccc}, the parameter $\alc$ 
governing the area distribution in a chaotic cavity was given by
the geometry and the escape rate. A phenomenological way of 
introducing decoherence \cite{But88} is by attaching a virtual
lead that draws no current but provides a channel for phase breaking.
Taking into account the virtual lead in our description will increase
the escape rate (proportionally to $L_{\Phi}$), and then $\alc$.
The measurement of the conductance fluctuations allowed to estimate
the temperature dependence of $L_{\Phi}$, thus providing an example
where the theoretical ideas of Quantum Chaos were used to test
fundamental properties of condensed matter systems.

The systematic study of conductance fluctuations, and the measurement
of the ballistic weak localization, require a considerable amount of
averaging. A given magnetoconductance curve offers only a limited interval
for averaging since once the cyclotron radius becomes comparable to the
size of the structure the nature of the classical dynamics may change.
In order to cope with this problem, alternative types of averages have
been developed: by tuning the Fermi energy \cite{Kel94,zozou97}, by thermal
cycling the sample \cite{berry94,bird}, by different realizations of the
residual disorder in an array of identical cavities \cite{chang94}, and
by small distortions in the shape of the cavity \cite{marcusgroup2,lee97}.

Keller {\em et al.} \cite{Kel94} fabricated microstructures where the electron density
(and hence $\kf$) was tunable while maintaining the geometry 
approximately fixed. Different shapes were considered: a stadium with
displaced leads, the ``stomach" and a polygonal shape with stoppers
(the last two shapes are shown in the insets of Fig.~\ref{fig:pow4}).
In the chaotic cavities a good quantitative agreement with the semiclassical
theories is obtained, with the scale of the fluctuations depending on the
cavity size. The polygonal geometry did not show qualitative differences with
the chaotic case, illustrating the difficulties of observing experimentally
the signatures of integrable dynamics. The energy average yielded a weak-localization
peak, and the conductance of a given sample did not always show a minimum at
$B\!=\!0$, consistent with the lack of self average of ballistic cavities. 

The use of sub-micron stadium-shaped quantum dots (with $N$ up to 7) cycled at room 
temperature allowed Berry and collaborators \cite{berry94} to obtain average values
and separate the weak localization peak from the conductance fluctuations. The 
line-shape of the peak was found to be Lorentzian, in agreement with the semiclassical
prediction. Moreover, the field scales of the weak localization and conductance
fluctuations were found to be related by the factor of 2 that we discussed in
(\ref{subsec:wl}), giving strong support to the applicability of the semiclassical
approach.

Chang and collaborators \cite{chang94} fabricated arrays of microstructures
connected in parallel and considered two shapes: stadium and circle. In each case,
the 48 cavities were nominally identical but actually slightly different due
to uncontrollable shape distortions and residual disorder. Thus, the conductance 
fluctuations were averaged out. The weak localization peak was found to be 
Lorentzian for the stadium cavities and triangular for the circular ones, in
agreement with the semiclassical prediction and detailed numerical calculations
\cite{revha}. Rectangular cavities, however, failed to yield a cusp of the 
magnetoresistance at $B\!=\!0$.

Microstructures admiting small shape distortions (less than 5 \% in the area) by 
tuning the voltage of lateral gates (inset of Fig.~\ref{fig:charly}.a) were
developed by Chan, Marcus collaborators \cite{marcusgroup2}. The lithographic shape of the
cavity is clearly non-chaotic, however it is expected that dot-specific features
tend to average away by the effect of the shape distortions. Also, these are
relatively large structures, where disorder definitely affects the long trajectories.
$N$ was not in the semiclassical regime, as it was tuned to a value of 2.
Conductance can be studied as a function of magnetic field and shape distortion, 
allowing to gather very good statistics. The fluctuations as a function of magnetic 
field show very good agreement with Eq.~(\ref{eq:CdB4}) (inset of Fig.~\ref{fig:charly}.b). 
The shape distortion fluctuations yield an exponential power spectrum, in agreement with the
calculations of Bruus and Stone showing that the semiclassical formalism of Sec.~\ref{sec:qttccc}
can be extended to this case \cite{Bruus94}.  A Lorentzian shape for the weak localization peak 
was obtained, with a width related to the characteristic field of the conductance fluctuations, as predicted by
semiclassical theory. The magnitude of the shape fluctuations at non-zero field
had a factor of 2 reduction with respect to the zero field value, and the line shape
of $\langle(\dlT)^2\rangle$ was found to be a Lorentzian squared \cite{efetov}. The rich statistics that this
type of structures allows to gather can be used to generate not only the moments of the
conductance, but its whole distribution, and compare with the random matrix theory
predictions \cite{BarMel,JPB}. As we will see in the next chapter, the detailled 
study of the conductance distribution allows to estimate the coherence length
\cite{marcusgroup3}. Interestingly enough, $L_{\Phi}$ seems to saturate below certain
temperature (of de order of 100 $mK$), in analogy with similar observations in 
disordered metals \cite{Moha}

Bird and collaborators used thermal cycling on rectangular
cavities to measure a weak localization peak that changes its
shape from Lorentzian to triangular as the quantum
point contacts at the entrance of the cavity are closed \cite{bird}. The
transition occurs for $N \simeq 2$ and demonstrates the
non-trivial role played by the leads. Measurements and numerical analysis 
by Zozoulenko {\em et al.} \cite{zozou97} on square cavities suggested that, 
depending on the geometry of the leads, transport through the cavity is
effectively mediated by just a few resonant levels, illustrating the importance 
of the injection conditions in the integrable case.

Lee, Faini and Mailly used shape and energy averages to
extract the weak-localization peak of chaotic (stadia and
stomach) and integrable (circles and rectanges) \cite{lee97}.
The former exhibited a Lorenztian line-shape, consistently
with the theoretical predictions. However, among the integrable
cavities, only the rectangle showed the expected triangular
shape, while the circle yielded a Lorentzian. Chan has proposed
\cite{chang94} that this apparently discrepancy with his results is
due to the shorter physical cut offs of Ref. \cite{lee97},
hindering long trajectories to exhibit the signatures of the
integrable dynamics.

\subsection{Semiclassics vs. random matrix theory}
\label{subsec:svrmt}

In our discussion of ballistic weak localization 
(\ref{subsec:wl}) we stressed the fact that our
semiclassical approach allows to calculate only one 
part of the effect, the elastic backscattering, 
corresponding to the interference of time-reversed
trajectories in the diagonal reflection coefficients
$R_{aa}$. Numerical calculations, as well as elementary
considerations based on current conservation, show that
the off-diagonal terms $R_{ba}$ are also sensitive
to a weak field. However, this $B$-dependence is not
easily calculable within the diagonal semiclassical
approximation of only keeping terms where we pair one 
trajectory with itself or with its time-symmetric.
Moreover, the semiclassical results for the weak
localization may depend on how we organize the sums.
If instead of going through the approach developed
in Sec.~\ref{sec:qttccc} of calculating transmission amplitudes
and then the coherent backscattering, we calculate the
average magnetoconductance effect directly from the real-space
Green function, we obtain a vanishing result \cite{Ullmo}.

The problem that we face in calculating the magnitude of 
weak localization (or conductance fluctuations) arises from 
the fact that those are effects of order $N^0$, while 
semiclassics only assures the compliance of unitarity to
leading order ($N$). A complementary approach to semiclassics
is that of random matrix theory (RMT), where we describe the
ballistic transport through a chaotic cavity by its 
scattering matrix $S$ (Eq.~\ref{eq:Smat})), and unitarity
is automatically preserved by the condition $S S^{\dagger}=I$.

Random matrix approaches have been applied to a variety of
physical problems, ranging from Nuclear Physics to level
statistics in small quantum systems and conductance fluctuations
in disordered mesoscopic conductors \cite{Haake,Bohigas,ALWrev,BeenRMP,WeiPR}.
The basic assumption is that the relevant matrix is the most
random among those verifying the required symmetries of the problem.
The appropriate ensembles for unitary matrices are Dyson's
COE and CUE (circular orthogonal and circular unitary) ensembles
describing respectively the cases with ($\beta=1$) and without
($\beta=2$) time-reversal symmetry. For unitary matrices the 
eigenvalues are pure phases, and the random assumptions on
the distribution of the matrices translate into the statistical
properties of the eigenphases. B\"umel and Smilansky proposed
that chaotic scattering is represented by COE and CUE $S$-matrices,
and derived statistical properties of the eigenphases from a 
semiclassical analysis \cite{Blu90}. 

Eigenphases are not directly related with transport properties,
and therefore a parameterization of Dyson matrices in terms of 
conductance parameters needs to be established \cite{JP}. From
that, the statistical properties of the conductance can be
established. Without going through the derivations, we quote
the results for the average conductance and the second moment
\cite{BarMel,JPB}

\be
\label{eq:RMT}
\langle T \rangle = \frac{1}{2} N + \langle \delta T \rangle
\qquad  ,  \qquad \langle \delta T \rangle = \frac{\beta-2}{4\beta} 
\qquad  ,  \qquad 
\langle(\dlT)^2\rangle = \frac{1}{8\beta} \ .
\ee

\nin The weak localization effect $\Delta \langle T \rangle = \langle T \rangle_{\beta=1}-
\langle T \rangle_{\beta=2} = -1/4$ can be shown to be the sum of a diagonal contribution
($\sum_{a} \Delta \langle R_{aa} \rangle = 1/2$) and an off-diagonal contribution
($\sum_{b \ne a} \Delta \langle R_{ba} \rangle = -1/4$), consistently with our results 
of (\ref{subsec:wl}).

RMT is able to predict not only the moments of the conductance, but the whole distribution 
in the case of small $N$. In particular, for $N\!=\!1$ the distribution is highly non-gaussian. The
experimental quest for such peculiar distribution has lead to the consideration of 
virtual leads, discussed in (\ref{subsec:cfawlibm}), and which can also be incorporated
in a random matrix description \cite{BaMeFL,BrBeFL}. Determining experimentally the number of
channels in the virtual leads has become a very useful way of estimating $L_{\Phi}$
\cite{marcusgroup3}.

The random matrix hypothesis for the Hamiltonian of a chaotic dot (or the zero-dimensional
non-linear sigma model) coupled to leads
yields equivalent results to those of Eq.~(\ref{eq:RMT}) \cite{Iid90,Wei91} and 
allows to calculate the crossover $\beta=1$ to $\beta=2$ as a function of magnetic 
field \cite{efetov,Weid94,WeiPR}.

Unlike semiclassics, RMT is able to produce numerical values for the 
conductance fluctuations and weak localization. However, its applicability to
ballistic cavities necessitates an ergodic behavior of classical trajectories. 
For instance, the existence of direct trajectories dramatically changes the
previous values, and that is why the quantitative agreement of RMT and
numerical simulations is only obtained in structures like the one at the
bottom of Fig.~\ref{fig:pow4}, where the short-time scales were eliminated from the
problem by introducing stoppers. Short-time process can be incorporated into the
RMT approach \cite{DoroSm,MeBa}, but the formalism and its use becomes 
considerably less simple. Obviously, RMT is not of any help when dealing
with cavities with integrable or mixed dynamics.

We therefore see that RMT and semiclassics are complementary techniques. The former
immediately yields the universal behavior in the case of ergodic dynamics. The 
semiclassical diagonal approximation is applicable to a much wider range of classical
behaviors, but it is not able to produce universal numbers. In addressing this failure we
have to consider the possible breakdown of our two main assumptions: semiclassics and diagonal
approximation. Similar questions arise in the study of the
density-density correlator of closed systems, where
standard semiclassics cannot recover the small energy
(long time) quantum behavior. It has been suggested in this
context that taking into account action correlations between
very long trajectories may correct this failure of the diagonal
semiclassical approximation \cite{Argetal}. In the context of
ballistic transport, Argaman \cite{Argaman} pointed to the
failure of the diagonal approximation for special pairs of long trajectories where
the difference of actions does not vary enough in the integration interval.
Similar arguments were presented by Aleiner and Larkin
\cite{AlLa} separating the dynamics in scales shorter and longer than the 
Ehrenfest time. In the former the correlations of the disorder potential have to be
taken into account, in the latter the classical diffusion equation can be used. Both
approaches reproduce (under certain conditions of ergodicity and invoking the presence of
a small amount of disorder) the universal values of Eq.~(\ref{eq:RMT}), but they are
not genuinely semiclassical in the sense of giving an operational prescription to
handle a given geometry from the knowledge of the classical trajectories. 

\subsection{Mixed dynamics}
\label{subsec:md}

In our study of ballistic transport we have considered, so far, the extreme cases of
hyperbolic and regular classical dynamics. Semiclassical theory predicts qualitative
differences between these two cases, and the experiments seem to support these results.
It is natural to ask about the behavior with a mixed phase space containing both
chaotic and regular regions. This is the most generic situation for a dynamical system,
and since the microstructures do not have hard-walls potentials, it is experimentally
relevant too.

Ketzmerick considered the problem of a dynamical system with mixed phase space \cite{roland},
where the trapping generated by the infinite hierarchy of cantori leads to a power-law
for the escape rate of the cavity $P(\tau) \propto t^{-\beta}$, with the exponent
$\beta > 1$. From a semiclassical diagonal approximation to the conductance,
Ketzmerick proposed that the graph of $g$ {\em vs.} $E$ has (in the case $\beta < 2$)
the statistical properties of fractional Brownian motion with fractal dimension
$D=2-\beta/2$.

Numerical simulations by Huckestein and collaborators \cite{bodo}
in cavities with mixed dynamics connected to leads yielded the power-law distribution
of the classical escape rate, but the quantum curve $g(E)$ failed to exhibit fractal
behavior. On the other hand, Casati and and collaborators \cite{casati} obtained
a fractal structure for the survival probability in an open quantum system classically
described by a map with mixed phase space.

These unsettled theoretical issues illustrate the difficulties of describing generic
Hamiltonian systems, and are particularly interesting given the recent experiments of
Sachrajda {\em et al} \cite{NRC}. The conductance fluctuations  measured in soft-wall
stadium and Sinai billiards exhibited fractal behavior in the curves $g(B)$ over two
orders of magnitude in magnetic field. Normally, we would have expected to be more
difficult to observe the fractal behavior in experiment than in the simulations, since
the physical cut-offs discussed in the introduction will limit the scale that can
be resolved in phase space.

\subsection{Semiclassical approach to bulk conductivity}
\label{subsec:satbc}

In Sec.~\ref{sec:qttccc} we mentioned that the Landauer formulation of the conductance
could be obtained from linear response (Kubo formula) theory. The latter is given in
terms of matrix elements of the current operator. The semiclassical approximation for
matrix elements \cite{Wilk} and the consistent use of stationary-phase integrations
can be then used to yield a semiclassical expression for the longitudinal conductivity 
as the sum over classical (Drude) component \cite{Fleisch} and an oscillatory 
component \cite{KlausEPL,Greg}. The latter is given by a periodic orbit expansion
where the coefficients of the trace formula are affected by the classical correlator
of the longitudinal component of the velocity along the trajectory. This approach has
been very helpful to understand the classical and quantum oscillations of the 
magnetoconductivity in antidot lattices \cite{Weiss93}. This formulation needs the
incorporation of a small amount of disorder to produce a finite conductivity and has
the drawback of not being able to yield a weak localization effect (since it is 
given as an expansion over {\em single} orbits). Even if the Kubo and Landauer
approaches for the conductance give equivalent results at the quantum mechanically, 
such a connection has not been established at the semiclassical level. As discussed in 
(\ref{subsec:svrmt}) the difficulty arises from the lack of unitarity of the semiclassical 
approximations that we have described.

%% file: var5.tex
%
% SECTION V: ORBITAL MAGNETISM IN CLEAN SYSTEMS
%
% file: var5.tex
%
% last version: 07/10
%

\section{Orbital magnetism in clean systems}
\label{sec:clean}

The problem of orbital magnetism in an electron gas has a long 
history, going back to the pioneering work of Landau demonstrating the
existence of a small diamagnetic response at weak fields $H$ and
low temperatures $T$ (such that $\kb T$ exceeds the typical spacing
$\hbar \omega$, \ $\omega=eH/mc$) \cite{Land,Peierls}. The orbital
response of a non-interacting 2DEG has a small diamagnetic value 

        \be
        \label{eq:chilandau}
        -\cl = -\frac{\gs e^2}{24 \pi m c^2}
        \ee

\nin ($\gs\!=\!2$ is the spin degeneracy factor). The constriction
of the electron gas (2D or 3D) to a finite region (area of volume) 
introduces a new energy scale in the problem (the typical level spacing 
$\Delta$) leading to a modification of the Landau susceptibility. 
The finite-size corrections to the Landau susceptibility have 
received considerable attention; various geometries and physical
regimes have been studied \cite{RUJ95,schmid}. We are interested in the
orbital response of ballistic microstructures, as measured in the 
experiments of Refs.~\cite{levy93,BenMailly}. We will show that semiclassical
methods, similar to those used in the previous sections, allow us
to undertake such studies \cite{vO94,URJ95}. Moreover, the
differences arising from the nature of the classical dynamics turn
out to be more important than in the case of transport.

The magnetic susceptibility is a thermodynamical property 
and it is sensitive to the effect disorder and interactions. In this 
section we start with the simplest 
model of non-interacting electrons in a clean cavity. In the following
sections we will incorporate the effect of residual disorder and
electron-electron interaction.

%%%%%%%%%%%%%%%%%%%%%%%%%%%%%%%%%%%%%%%%%%%%%%%%%%%%%%%%%%%%%%%%%%%%%%%%% 

\subsection{Thermodynamic Formalism}
\label{subsec:thermo}

For a system of electrons in an area $A$, connected to a reservoir of 
chemical potential $\mu$, the magnetic susceptibility is defined by

\be
\cgc =  - \frac{1}{A}
\left(\frac{\partial^{2}\Omega}{\partial H^{2}} \right)_{T,\mu} \ .
\ee

The notation with the superscript GC is used in order to emphasize the
fact that we are working in the grand canonical ensemble. 

\be
\label{eq:gcpott}
\Omega(T,H,\mu) = \Omega_0(T,H,\mu) + \Omega_i(T,H,\mu)
\ee

\nin is the thermodynamic potential, decomposed in a non-interacting part 

\be
\label{eq:gcpot}
\Omega_0(T,H,\mu) = -\frac{1}{\beta} \int\,  {\rm d}E \, d(E) \ln[1 +
\exp(\beta(\mu-E))]
\ee

\nin depending on the one-particle density of states $d(E)$ ($\beta=1/\kb T$)
and the term $\Omega_i$ arising from electron-electron interactions 
\cite{AGD,Fetter}. We will not consider this last contribution until 
Sec.~\ref{sec:eeiitbr}.

The choice of the ensemble in the macroscopic limit of the area $A$
and the number of particles ${\bf N}$ going to infinity is a matter of
convenience. As it was recognized in the context of persistent currents
in disordered rings \cite{BM,Imry,ensemble}, the equivalence 
between the ensembles may break down in the mesoscopic regime.
Although the number of electrons can be large for a mesoscopic system,
the fact that $\bf N$ is fixed must be taken into account
(by working in the canonical formalism) if a disorder or energy averaged 
susceptibility of an {\em ensemble} of isolated micro-systems is examined. 

The magnetic susceptibility of a system of $\bf N$ electrons is

\be 
\label{eq:sus}
\chi = - \frac{1}{A}
\left(\frac{\partial^{2}F}{\partial H^{2}} \right)_{T,{\bf N}} \ ,
\ee

\nin with the free energy $F$ and the thermodynamical potential 
$\Omega$ are related by means of the Legendre transform

\be
\label{eq:free}
F(T,H,{\bf N}) = \mu {\bf N} + \Omega(T,H,\mu) \ .
\ee

In Condensed Matter Physics the simple dependence of the thermodynamical
potential on the density of states in (\ref{eq:gcpot}) makes the grand 
canonical ensemble the easiest to work with. In the case of disordered
metals, in which the density of states (DOS) can be separated in a smooth and 
a (small) fluctuating part, Imry proposed convenient representation for 
the canonical free energy in terms of grand canonical quantities 
\cite{Imry}. In our clean case we use the decomposition

\be 
\label{eq:dosc1}
d(E) = \bar d(E) + \dosc(E) \; 
\ee

\nin in the mean (or Weyl) part $\bar d$ and the periodic orbit
contribution $\dosc$. (Rigorously speaking, $\dosc$ is not small
since it is the sum of delta functions, therefore the expansion
(\ref{eq:dosc1}) has to be used after some thermal broadening
\cite{RUJ95}.) We define a mean chemical potential $\bar{\mu}$
as the one that accommodates the $\bf N$ particles with
the mean DOS $\bar d$. 

\be
{\bf N} = N(\mu) = \bar{N}(\bar{\mu})\ .
\label{eq:mGCE}
\ee

\nin Here

\be
N(\mu) = \int_0^{\infty} \dif E \ d(E) \ f(E\!-\!\mu)  
\label{eq:NOS}
\ee

\nin with the Fermi distribution function
\be 
\label{eq:fermi}
f(E-\mu) = \frac{1}{1 + \exp [\beta (E-\mu)]} \ .
\ee

\nin $\bar{N}$ is then obtained in Eq.~(\ref{eq:NOS}) by replacing 
$d(E)$ by $\bar{d}(E)$. Expanding (\ref{eq:free}) to second order 
in $\mu-\bar{\mu}$ leads to an expansion of the free energy in
terms of grand canonical quantities \cite{ensemble,Imry}

\be
\label{eq:fd}
F({\bf N}) \simeq F^{0} + \Delta F^{(1)} +\Delta F^{(2)} \ ,
\ee

\nin with

\begin{mathletters}
\label{allDF}
	\begin{eqnarray}
	     F^{0} & = & \bar \mu {\bf N} + \bar \Omega(\bar \mu) \; ,
	      \label{eq:DF0}  \\
	     \Delta F^{(1)} & = & \Oosc (\bar \mu)  \; ,
	        \label{eq:DF1} \\
	\displaystyle 
	     \Delta F^{(2)} & = & \frac{1}{2 \bar d (\bar \mu)} \
				\left( \Nosc (\bar \mu) \right)^{2} \ .
	     \label{eq:DF2}
        \end{eqnarray}
\end{mathletters}

\nin The functions $\Oosc$ and $\Nosc$ are expressed by means of 
Eqs.~(\ref{eq:gcpot}) and (\ref{eq:NOS}), respectively, upon inserting the 
oscillating part $\dosc$ of the density of states. The leading order contribution 
to $F$ is given by the first two terms $F^{0} + \Delta F^{(1)}$ yielding 
the susceptibility calculated in the {\em grand} canonical case with the
chemical potential $\bar \mu$. $F^0$ gives rise to the
diamagnetic Landau--susceptibility ($-\cl$) independently
of the confining geometry \cite{Kubo64,Prado,RUJ95}. We will show that,
as in the case of persistent currents, the average value of $\Delta F^{(1)}$
vanishes and the additional term $\Delta F^{(2)}$ becomes the dominant one.

\subsection{Semiclassical treatment of susceptibilities}
\label{subsec:sus}

For a semiclassical computation of $\Delta F^{(1)}$ and $\Delta F^{(2)}$ 
and their derivatives respect to $H$ we calculate $\dosc(E,H)$ from 
the trace 

	\be
	\label{eq:trace}
	d(E,H) = -\frac{\gs}{\pi} \ {\rm Im} \int \dif \br \ G(\br,\br;E)
	\ee

\nin of the semiclassical one--particle Green function (Eq.~(\ref{eq:gfgutz})). 
Its contribution to $\dosc(E)$ is given by all classical paths $s$
(of non--zero length) joining $\br$ to $\br'$ at energy $E$ 
\cite{gutz_book,brack_book}.

The evaluation of the trace integral (\ref{eq:trace}) for chaotic
and integrable systems leads to the Gutzwiller \cite{gutz_book}
and Berry--Tabor \cite{ber76} periodic--orbit 
trace formulas, respectively. In order to calculate the magnetic 
susceptibility at small fields one has to carefully distinguish 
\cite{RUJ95} between the three possibilities: a chaotic billiard, 
the special case of an integrable billiard remaining integrable upon 
inclusion of the $H$--field, and the more general case where the field 
acts as a perturbation breaking the integrability of a regular structure. 
We start this section with the last situation, focusing on the experimentally
relevant case of ballistic squares \cite{levy93} and we later discuss
chaotic and circular cavities.

For a generic integrable system (a {\em regular} geometry) any 
perturbation breaks the integrability of the dynamics. The 
Poincar\'e-Birkhoff theorem \cite{arnold:book} states that as soon as 
the magnetic field is turned on, all resonant tori ({\it i.e.} all
families of periodic orbits) are instantaneously broken, leaving only two
isolated periodic orbits (one stable and one unstable). Therefore, neither 
Gutzwiller nor Berry--Tabor trace formulas are directly applicable and 
a uniform treatment of the perturbing $H$--field is necessary 
\cite{ozor:book}. In the integrable
zero--field limit each closed trajectory belongs to a torus
$I_\bM$ and we can replace $\br$ in the trace integral (\ref{eq:trace})
by angle coordinates $\Theta_1$ specifying the trajectory within
the (one--parameter) family and by the position $\Theta_2$ on the
trajectory. For small magnetic field the classical orbits can be 
treated as essentially unaffected while the field acts merely on the 
phases in the Green function in terms of the magnetic 
flux through the area ${\cal A}_\bM
(\Theta_1)$ enclosed by each orbit of family $\bM$. Evaluating the trace
integral along $\Theta_2$ for the semiclassical Green 
function of an integrable system leads in this approximation
to a factorization  of the density of states

\be 
\label{eq:uniform_d}
\dosc(E) = \sum_{\bsM \neq 0} \C_\bsM(H) \ d^0_\bsM(E)  
\ee

\nin into the contribution from the integrable zero--field limit

\be
\label{eq:d0}
d^0_\bsM(E) = B_\bM \, \cos\left( \kf L_\bM - 
\nu_\bM \frac{\pi}{2} - \frac{\pi}{4} \right)
\ee

\nin ($L_\bM$ and $\nu_\bM$ are respectively the length and Maslov index of the 
orbits of family $\bM$ and $B_\bM$ is the semiclassical weight \cite{ber76}) 
and the function
	
\be 
\label{eq:thetatrace} 
\C_\bsM(H) = \frac{1}{2\pi} \int_0^{2\pi} \dif \Theta_1 \,
\cos \left[ 2 \pi \frac{H \A_\bsM(\Theta_1) }{ \Phi_0} \right] \;
\ee

\nin containing the $H$--field dependence ($\Phi_0\!=\!hc/e$). Calculating 
$\Delta F^{(1)}$ from Eq.~(\ref{eq:DF1}) and taking the derivatives 
respect to $H$ gives the grand canonical contribution to the 
susceptibility at small magnetic field
	
\be 
\label{eq:chi1}
\frac{\chi^{(1)}}{\cl} =- \frac{24\pi}{\gs}  \ m A 
\left( \frac{\Phi_0}{2\pi A} \right)^2 \, 
\sum_{\bM} \ \frac{R_T(\tau_\bsM)}{\tau_\bsM^2} \ 
d^0_\bsM(\bar \mu) \ 
\frac{d^2 \C_\bsM}{d H^2} \, .
\ee

\nin Here, $\tau_\bM$ is the period of a closed orbit of family $\bM$ and

\be
\label{eq:R_T}
R_T(\tau) = \frac{\tau/\taut}{\sinh(\tau/\taut)} 
\hspace{1cm} ; \hspace{1cm} \taut = \frac{\hbar \beta}{\pi} = \frac{L_T}{\vf} \ .
\ee

\nin is a damping factor which arises from the convolution integral in 
Eq.~(\ref{eq:gcpot}) and gives an exponential suppression of 
long orbits according to the temperature dependent cut off $\taut$
(or $L_T$). As discussed in the introduction, this physical cutoff is 
important from a physical as well as computational point
of view, as conceptual difficulties associated with the questions
of absolute convergence of semiclassical expansions at zero temperature 
do not arise.

Eq.~(\ref{eq:chi1}) is the basic equation for the susceptibility
of an individual microstructure. When considering ensembles of 
ballistic microstructures  however, an average $(\overline{~\cdot~})$ 
over Fermi energy ($\mu=\hbar^2\kf^2/2m$) or over the system size $a$
has to be performed since there will usually be a dispersion of 
$\kf$ and $a$ among the members of the experimental ensemble. These 
averages lead to variations in the phases 
$\kf L_\bM$ of the DOS (\ref{eq:d0})
which are much larger than $2\pi$. Therefore,
$\chi^{(1)}$ vanishes upon ensemble average. In order to
characterize the orbital magnetism of ensembles we introduce the
{\em typical} susceptibility $\chit = (\overline{\chi^2})^{1/2}$
(the width of the distribution) and the ensemble averaged 
$\overline{\chi}$ (its mean value, which is non--zero
because of the {\em positive} term $\Delta F^{(2)}$ in the expansion 
(\ref{eq:fd})). 

If we assume that there are no degeneracies in the lengths of 
orbits from different families $\bM$ we obtain for $\chit$
	
\be 
\label{eq:chit}
\left( \frac{\chit}{\cl} \right)^2 = 
\left(\frac{24 \pi}{\gs} m A\right)^2 
\left( \frac{\Phi_0}{2\pi A} \right)^4 \, 
\sum_{\bM} \frac{R_T^2(\tau_\bsM)}{\tau_\bsM^4} \ 
\overline{d^0_\bsM(\bar \mu)^2}  \, 
\left(\frac{d^2 \C_\bsM}{d H^2}\right)^2 \, .
\ee

In calculating $\overline{\chi}$, the grand canonical 
contribution $\chi^{(1)}$ from $\Delta F^{(1)}$ 
vanishes under energy average and the semiclassical approximation to the 
canonical correction $\Delta F^{(2)}$ (Eq.~(\ref{eq:DF2}) yields

	\be 
        \label{eq:chi2}
	\frac{\overline{ \chi}}{\cl}   \simeq
	\frac{\overline{ \chi^{(2)} }}{\cl}  = 
	- \frac{24 \pi^2}{\gs^2} \hbar^2
	 \left( \frac{\Phi_0}{2\pi A} \right)^2\, 
	\, \sum_{\bM} \frac{R^2_T(\tau_\bsM)}{\tau_\bsM^2} \ 
        \overline{ d^0_\bsM(\bar \mu)^2} \ 
	\frac{d^2 \C^2_\bsM}{d H^2} \ .
        \ee

Eqs.~(\ref{eq:chi1})--(\ref{eq:chi2}) provide the
general starting point for a computation of the susceptibility
of integrable billiards at small fields.

\subsection{Square billiards}

In a square billiard each family of periodic orbits can be labeled by
$\bM = (M_x, M_y)$ where $M_x$ and $M_y$ are the number of bounces 
occurring on the bottom and left side of the billiard. The length of 
the periodic orbits for all members of a family is
$L_\bM = 2 a (M_x^2 + M_y^2)^{1/2}$ and $B_\bM = m/\sqrt{\hbar k L_\bM}$.
Since $\nu_\bM = 4 (M_x + M_y)$ the Maslov index will be 
omitted from now on. The advantage of the square geometry, from the
calculational point of view, is that as in (\ref{subsec:starc}), we can
work in the extended space and use the free form of the Green function.

\begin{figure}
\setlength{\unitlength}{1mm}
\begin{picture}(100,110)
\put(15,160)
{\epsfxsize=12cm\epsfbox{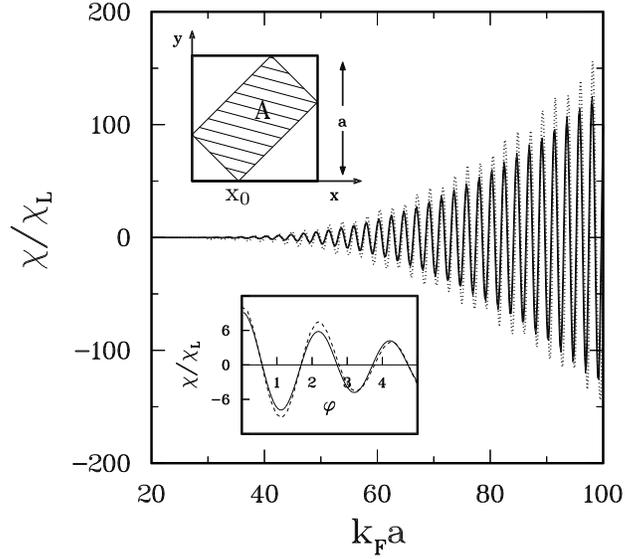}}
\put(55,76){{\bf ${\rm x}_0$}}
\end{picture}
\vspace{-2cm}
\caption{
Magnetic susceptibility of a square as a function of $\protect
k_{\scriptscriptstyle F} a$ from numerical calculations (dotted line)
at zero field and at a temperature equal to 10 level-spacings. The 
solid line shows our semiclassical approximation
(Eq.~(\protect\ref{eq:chiH0}))
taking into account only the family $(1\!,\!1)$ of shortest orbits.
The period $\pi/ \protect\sqrt{2}$ of the quantum result indicates the
dominance of the shortest periodic orbits enclosing non-zero area
with length $L_{11}=2\protect\sqrt{2} a$ (upper inset). Lower inset: 
amplitude of the oscillations (in $k_{\scriptscriptstyle F} L_{11}$)
of $\chi$ as a function of the flux through the sample from 
Eq.~(\protect\ref{eq:chi0}) (solid) and numerics (dashed).
(From Ref. \protect\cite{URJ95}.)
}
\label{fig:chi1}
\end{figure}

In the inset of Fig.~\ref{fig:chi1} we represent a trajectory of
the family ${\bf M}=(1,\!1)$ of the shortest flux--enclosing periodic 
orbits. Instead of $\Theta_1$ we use the lower reflection point
$x_0$ as orbit parameterization within the family.
The orbits $(1,\!1)$ have the unique length $L_{11} = 2 \sqrt{2} a$ and
enclose a normalized area $\A(x_0)=4 \pi \epsilon x_0(a-x_0)/a^2$,
where the index $\epsilon = \pm 1$ specifies the direction in which 
the trajectory is traversed. This set is particularly relevant in view
of the cut off (\ref{eq:R_T}) discussed above, killing exponentially
the long trajectories (for the usual experimental conditions
$L_T \simeq 2a$). The (small) contribution of the long trajectories
can be calculated along the same lines that in the case of the family
$(1,\!1)$ \cite{RUJ95}.

The computation of $d_{11}^0 (\mu)$ for the square geometry gives
for $\chi^{(1)}$ (Eq.~(\ref{eq:chi1})) yields

\be 
\label{eq:chi1_sq}
\frac{\chi^{(1)}}{\chi^0} = \sin{\left(\kf L_{11}+\frac{\pi}{4}\right)}
\int_{0}^{a} \frac {{\rm d}x_0}{a} \A^2(x_0) \cos (\varphi \A(x_0))
\ee

\nin as a function of the  total flux $\varphi = H a^2/\Phi_0$. The 
prefactor is defined by

\be
\label{eq:chi0}
\chi^0 = \chi_{\scriptscriptstyle L} \ 
\frac{3}{(\sqrt{2}\pi)^{5/2}} \ (\kf a)^{3/2} R_T(L_{11}) \  .
\ee

In the case of a square geometry, the integrals can be evaluated analytically,

\be
\label{Csimple}
{\cal C}_{11}(\varphi) = \frac{1}{\sqrt{2 \varphi}}
          \left[ \cos(\pi \varphi) {\rm C}(\sqrt{\pi \varphi}) +
                 \sin(\pi \varphi) {\rm S}(\sqrt{\pi \varphi}) \right] \ ,
\ee

\nin where $\rm C$ and $\rm S$ respectively denote the cosine and sine
Fresnel integrals, and the zero-field susceptibility of a clean cavity is 
\cite{vO94,URJ95}

\be
\label{eq:chiH0}
\chi^{(1)} = \cl \ \frac{4}{5 (\sqrt{2}\pi)^{1/2}} \ (\kf a)^{3/2} 
\ \sin{\left(\kf L_{11}+\frac{\pi}{4}\right)} \ R_T(L_{11}) \  .
\ee

The results (\ref{eq:chi1_sq}) and (\ref{eq:chiH0}) show the 
$(\kf a)^{3/2}$-dependence typical for regular systems, with rapid 
oscillations as a function of $\kf$ (that make $\chi^{(1)}$ to 
vanish when averaged over variations of $\kf$ or $a$). 
In Fig~\ref{fig:chi1} we see the good agreement between
Eq.~(\ref{eq:chiH0}) (solid) and the numerically obtained quantum
mechanical susceptibility (dotted). The small deviations between
both traces is due to higher repetitions (trajectories 
$\bM = (M,M)$) that can easily be incorporated in our formalism
\cite{RUJ95}). The oscillations as a function of the flux at a given
number of electrons in the square (solid) are in agreement with the
numerical calculations (dashed).

The typical and average susceptibilities for a square geometry where only 
the dominant contributions of the family $(1,\!1)$ are considered, can be written as

\be  
\label{eq:chit_sq} 
	\frac{\chi^{(t)}}{\chi^0}   \simeq  
	\frac{\sqrt{\overline{{\chi^{(1)}}^2}}}{\chi^0}
         = \frac{1}{\sqrt{2}} \  
        \int_{0}^{a} \frac{{\rm d}x_0}{a} 
	\A^2(x_0) 
	\cos \left( \varphi\A(x_0) \right) \ ,
\ee

\be 
\label{eq:chi2_sq}
\frac{\overline{ \chi}}{{\overline{\chi}^0}} = \frac{1}{2}
        \int_{0}^{a} \frac{{\rm d}x_0}{a} \int_{0}^{a} 
	\frac{{\rm d}x_0'}{a}  \left[
	\A^{2}_{-} \cos(\varphi \A_{-}) +
	\A^{2}_{+} \cos(\varphi \A_{+})\right] \ ,
\ee

\nin with 

\be
\label{eq:chi0bar}
        \frac{\overline{\chi}^0}{\cl} = 
       \frac{3}{(\sqrt{2}\pi)^3} \ (\kf a) \  R^2_T(L_{11})
\ee

\nin and $\A_{\pm} = \A(x_0)\!\pm\!\A(x_0')$. As in the previous case,
the integrals (\ref{eq:chit_sq}) and (\ref{eq:chi2_sq}) can be
calculated in terms of Fresnel functions. In particular, the average zero-field
susceptibility is paramagnetic and attains the value \cite{vO94,URJ95}

\be
\label{square:chizf}
        \overline{\chi^{(2)}}({H\!=\!0}) =
        \frac{4\sqrt{2}}{5\pi} \ \kf a \ \chi_{\scriptscriptstyle L} \ R^2_T(L_{11}) \ .
\ee
 
Since $\kf a \gg 1$ we have that $\chi^{(t)} \gg \overline{ \chi} \gg \cl$, and then
we obtain a large enhancement of the Landau susceptibility by the effect of
confinement in an integrable geometry.

\begin{figure}
\setlength{\unitlength}{1mm}
\begin{picture}(100,110)
\put(-5,160)
{\epsfxsize=11cm\epsfbox{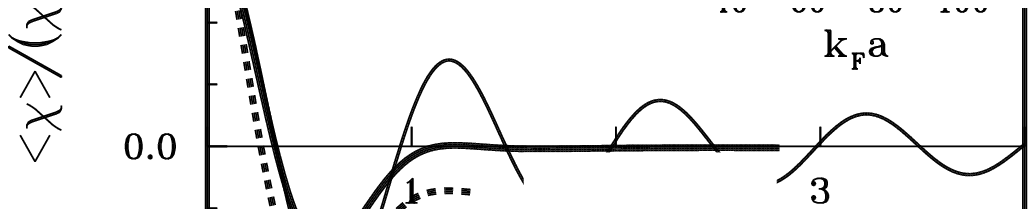}}
\end{picture}
\vspace{-4cm}
\caption{
Thin solid curve: average magnetic susceptibility of an ensemble of
squares from Eq.~(\protect\ref{eq:chi2_sq}). Thick solid line:
average over an ensemble with large dispersion of sizes. Thick dashed
curve: average from numerics. The shift
of the numerical with respect to the semiclassical results
reflects the Landau susceptibility (due to $\protect F^0$
in Eq.~(\protect\ref{eq:fd})) not included in the latter. Inset: average
susceptibility as a function of $\protect k_{\scriptscriptstyle F} a$ for various
temperatures (8, 6 and 4 level spacings, from below) and a flux 
$\protect\varphi=0.15$ and numerics (dashed).
(From Ref. \protect\cite{URJ95}.)
}
\label{fig:chi2}
\end{figure}

The average susceptibility from (\ref{eq:chi2_sq}) (thin solid line in 
Fig.~\ref{fig:chi2}) oscillates in the scale of one flux through the sample.
For ensembles with a wide dispersion of lengths (like in the experiment of
Ref.~\cite{levy93}) we have to also consider the effect on ${\cal C}$ of the average 
over $a$ (on a classical scale), which will suppresses the $B$-oscillations (thick solid
line).
 
\subsection{Integrable versus chaotic behavior}
\label{subsec:ivsb}

Squares constitute a generic example of an integrable system perturbed by a 
magnetic field. It is interesting to compare our results with two extreme
cases: circles (which remain integrable under the perturbation) and completely
chaotic systems. The periodic orbits of the circular billiard are labeled by
the topology $\bM = (M_1,M_2)$, where $M_1$ is the number of turns around the 
circle until coming to the initial point after $M_2$ bounces. We can therefore
write the field-dependent DOS as in (\ref{eq:uniform_d}), and we easily see
that the susceptibility for circular billiard (of radius $R$) has the same 
parametric dependence of the square: $(\kf R)^{3/2}$ for $\chi^{(1)}$ and $\chi^{(t)}$, 
and $(\kf R)$ for $\overline{\chi}$ \cite{RUJ95}. The orbital response of small
rings is usually expressed in terms of the persistent current 
$I=-c(\partial F/\partial \Phi)_{T,{\bf N}}$ which can be calculated along the
same lines than the susceptibility of the circle \cite{sursci} and shows the
previous parametric dependence on $(\kf R)$. The semiclassical results are in 
good agreement with the measurments of Ref.~\cite{BenMailly}. The generic behavior of 
integrable systems can be traced to the $\kf^{-1/2}$ dependence of 
$\dosc(E)$ in Eq.~(\ref{eq:uniform_d}).

For chaotic systems (of typical length $a$) with hyperbolic periodic orbits, 
the Gutzwiller trace formula provides the appropriate path to calculate 
$\dosc(E,H)$. For temperatures at which only a few short periodic orbits are
important, $\chi$ can have any sign, and its magnitude is of the order
of $(\kf a)\cl$ \cite{Sha,aga94}. For an ensemble of chaotic billiards
$\chi \propto \cl$. Therefore both susceptibilities are considerably reduced 
respect to the integrable case. The magnetization line-shape can be calculated
along similar lines than those used in (\ref{subsec:wl}) for the 
coherent backscattering \cite{RUJ95,KlaBer}.

%% file: var6.tex
%=============================================================================

%
% SECTION VI: SEMICLASSICAL APPROACH TO WEAK DISORDER
%
% file: var6.tex
%
% last version: 07/10
%

\section {Semiclassical approach to weak disorder}
\label{sec:disord}

In the previous section we described important differences in the
orbital response of clean cavities according to their underlying
classical mechanics. On the other hand, we know that any
perturbing potential, such as the one provided by the disorder,
immediately breaks the integrable character of the classical 
dynamics \cite{arnold:book}. A natural question to pose is whether 
the difference chaotic versus integrable will survive when we go 
from the {\em clean} to the {\em ballistic} regimes and we 
consider the residual disorder that is always present in actual 
microstructures.

Disorder is usually studied in terms of the ensemble average over
impurity realizations, since it is a perturbation of a electrostatic
potential whose detailed nature is unknown. Typically, quantum perturbation
theory is followed by the average over the strengths and positions of the
impurities \cite{AGD,LeeRam}. This approach is suited for macroscopic 
metallic samples (which are self-averaging) or ensembles of mesoscopic 
samples (where different microstructures present different impurity 
configurations). The possibility of measuring a single disordered 
mesoscopic sample posses a conceptual difficulty since there is not an 
average process involved. In this section we first develop a general formalism of
disordered Green functions for the ballistic regime, applicable to a
wide range of physical problems, and then use this formalism to
calculate the various averages of the susceptibility that we encounter 
in different experimental situations \cite{RapComm,JMP}.

\subsection{Disorder models}

We assume that the disorder is generated by means of a given
realization of a two-dimensional Gaussian potential of
the form

\be 
\label{eq:garapo}
V({\bf r}) = \sum_{j}^{N_i} \frac{u_j}
{2 \pi \xi^2} \
\exp{\left[-\frac{({\bf r}\!-\!{\bf R}_j)^2}{2 \xi^2} \right]} \ ,
\ee

\nin
provided by $N_i$ {\em independent} impurities located at 
points ${\bf R}_j$ with uniform probability on an area 
${\sf V}$ ($n_i = N_i/ {\sf V}$). The strengths $u_j$ obey
$\langle u_j u_{j \prime} \rangle = u^2 \delta_{jj^{\prime}}$.
This model yields, in the limit $\xi \rightarrow 0$, the white 
noise disorder of $\delta$-function scatterers
$V({\bf r})=\sum_{j}^{N_i} u_j \delta({\bf r}\!-\! {\bf R}_j)$.

\nin The disorder potential $V(\br)$ is characterized by its 
correlation function

\be 
\label{eq:corr}
C(|\br-\br'|)  = \langle V(\br) V(\br') \rangle = \frac{u^2 n_i}{ 4\pi \xi^2} 
\ \exp{\left[-\frac{(\br-\br')^2}{4\xi^2}\right]} \ .
\ee

Our model is quite simple, we do not aim to describe disorder 
within a microscopic model with realistic distributions of residual 
impurities in the semiconductor heterostructure, as done in 
Refs.~\cite{Davies,Stopa}. The  model of Gaussian disorder
is particularly appropriated for analytical calculations. On the other hand,
many of the results that we will give only depend on the correlation
function $C(|{\bf r-r'}|)$, and are therefore valid for a wide class of 
disorder models. 

Disorder effects depend on several length scales: the 
Fermi--wavelength $\lf$ of the electrons, the disorder correlation 
length $\xi$ and the size $a$ of the microstructure.
In the bulk case of an unconstrained (2DEG) we distinguish 
between {\em short range} (SR, $\xi\!<\!\lf$) and {\em finite range} 
(FR, $\xi\!>\!\lf$) disorder potentials. In the case of a 
microstructure a third, {\em long range} (LR) regime for 
$\xi\!>\!a\!>\!\lf$ has to be considered. The cleanest samples used 
in today experiments are in the finite range regime $a > \xi > \lf$.

\subsection{Single-particle Green function}
\label{subsec:spgf}

If we assume a microstructure with size $a\gg\lf$ and work in the 
FR or LR regimes, where the disorder potential is 
smooth on the scale of $\lf$, the use of the semiclassical expression
(\ref{eq:gfgutz}) for the single--particle Green function 
$G(\brp,\br;E)$ is well justified. The classical mechanics of 
trajectories with length $L_s \ll \lt$ is essentially 
unaffected by disorder. Therefore the dominant effect on 
the Green function results from the shifts in the phases due to the 
modification of the actions (while the amplitudes $D_s$ and topological 
indices $\nu_s$ remain nearly unchanged). The first--order approximation
to the action along a path $\C_s$ in a system with weak 
disorder potential is

\be 
\label{eq:action_approx}
S_s^d 
\simeq S^c_s + \delta S_s \, ,
\ee

\nin here the clean action $S^c_s$ is obtained by integrating along
the {\em unperturbed} trajectory $\C_s^{\rm c}$ without disorder
({\it i.e.} $S^c_s = \kf L_s$ in the case of billiards without magnetic field)
instead of the actual path $\C_s$. The correction term
$\delta S_s$ is obtained, after expanding ${\bf p} = \sqrt{2 m 
[E-V({\bf q})]}$ for small $V/E$, from the integral 
	
\be 
\label{eq:dis_action}
\delta S_s \ = \ - \ \frac{1}{\vf} \ 
\int_{\C_s^{\rm c}} V(\bq) \ {\rm d}q \ .
\ee

In this approximation an impurity 
average $\langle \ldots \rangle$ acts only on $\delta S_s$
and the disorder averaged Green function reads
	
\be 
\label{eq:avegre}
\langle G(\brp,\br;E)\rangle  =  \sum_s G_{s}^{c}(\brp,\br;E) \ 
\left\langle \exp{\left[\frac{i}{\hbar} \delta S_s
\right]} \right\rangle \ .
\ee

\nin $G_{s}^{c}$ is the contribution of the 
trajectory $s$ to the zero-disorder (clean) Green function $G^{c}$. 

For trajectories of length $L_s \gg \xi$, the contributions to $\delta S$ 
from segments of the trajectory separated more than $\xi$, are uncorrelated. 
The stochastic accumulation of action along the path can be 
therefore interpreted as determined by a random-walk process, 
resulting in a Gaussian distribution of $\delta S_s(L_s)$. 
For larger $\xi$ or shorter trajectories ($L_t \not \gg \xi$), 
one can still think of a Gaussian distribution of the de-phasing $\delta S_s$
provided $V(\br)$ is generated by 
a sum of a large number of independent impurity potentials.
As a consequence of the Gaussian character of the distribution 
of $\delta S_s(L_s)$, the characteristic function involved in 
Eq.~(\ref{eq:avegre}) is given by
	\be
	\label{eq:gauss_av}
	\langle \exp{\left[\frac{i}{\hbar}\delta S_s\right]} \rangle =
	\exp{\left[-\frac{\langle \delta S^2_s \rangle}{2\hbar^2} 
	\right]}
	\ee
and therefore entirely specified by the variance 
	\be 
	\label{eq:dS2}
	\langle \delta S_s^2 \rangle = 
	\frac{1}{\vf^2} \int_{\C_s^c} {\rm d}q \ 
	  \int_{\C_s^c} {\rm d}q' \langle V(\bq) V(\bq') \rangle \ .
	\ee

For an unconstrained 2DEG the sum in Eq.~(\ref{eq:avegre}) is reduced 
to the direct trajectory joining $\br$ and $\brp$.
If $L=|\br-\br'| \gg \xi$ the inner integral in Eq.~(\ref{eq:dS2}) 
can be extended to infinity and we obtain 
	\be
        \label{eq:dS2fr}
	\langle \delta S^2 \rangle = 
	   \frac{L}{\vf^2} \ \int {\rm d}q \ C(\bq) \; .
	\ee

\nin The semiclassical average Green function for the bulk 
exhibits therefore an exponential behavior \cite{RapComm,Mirlin96}
(on a length scale $\lt > L \gg \xi$)
	
\be 
\label{eq:avegrebulk}
\langle G(\brp,\br;E)\rangle \ = \ G^{c}(\brp,\br;E) \ 
\exp{\left[-\frac{L}{2l}\right]} \, ,
\ee 

\nin with the damping governed by an inverse elastic mean free path

\be
\label{eq:mfp}
\frac{1}{l} = \frac{1}{\hbar^2\vf^2} \ \int {\rm d}q \ C(\bq) =
\frac{u^2 n_i}{4\sqrt{\pi}\hbar^2 \vf^2\xi } \ , 
\ee 

\nin for the Gaussian potential (\ref{eq:garapo}).

Quantum diagrammatic perturbation theory for the potential 
(\ref{eq:garapo}) shows that the damping (\ref{eq:avegrebulk}) 
of the one--particle Green function is also valid in the SR regime
(where our semiclassical approach is no longer applicable). The
quantum elastic MFP agrees with the semiclassical 
result (\ref{eq:mfp}) in the limit $\kf \xi \gg 1$, and it is
equal to $l_\delta=(\vf \hbar^3)/(m n_i u^2)$ for short range.
The quantum results for the transport MFP \cite{RapComm} are
$\lt=l_\delta$ in the SR regime and $\lt=4(\kf \xi)^2 l$ for
$\kf \xi \gg 1$ (and therefore the transport MFP may be 
significantly larger than $l$ \cite{DSS}).

We now turn from the semiclassical treatment of the bulk to that of a
confined system. We treat the ballistic regime $\lt\! >\! a$ where both,
the confinement {\em and} the impurities have to be considered.
Confinement implies that the clean $G^c(\brp,\br;E)$ is
given as a sum over all direct and multiply reflected 
paths connecting $\br$ and $\br'$; disorder modifies the
corresponding actions according to Eq.~(\ref{eq:dis_action}). 

In the SR and FR regimes the 
damping of each contribution $\langle G_{s}\rangle$ to 
$\langle G \rangle$ acquires a damping $\exp(-L_s/2l)$ according
to its length $L_s$,

\be 
\label{eq:avegrefr}
\langle G(\brp,\br;E)\rangle  =  \sum_s G_{s}^{c}(\brp,\br;E) \ 
\exp{\left[-\frac{L_s}{2l} \right]}   \ .
\ee

In the long range regime and for $\xi \sim a$ the correlation integral 
(\ref{eq:dS2}) can no longer be approximated 
by $L_s\int_{-\infty}^{+\infty} dq C(\bq)$ due to correlations 
across different sectors of an orbit (with distance smaller $\xi$).
Therefore the orbit--geometry enters into the correlation integral.
For $\xi \gg a$ we can expand $C(|\br-\br'|)$ (up to first order in
$(|\brp-\br|/\xi)^{2}$) and obtain a damping exponent that depends
quadratically on $L_s$ (in contrast to linear behavior in the finite 
range case), with a length scale given by the geometrical mean of the 
bulk MFP $l$ and $\xi$ \cite{JMP}.

\subsection{Two-particle Green function}

The typical susceptibility (Eq.~(\ref{eq:chit})), the  ensemble
averaged susceptibility (Eq.~(\ref{eq:chi2})), or in general
a density-density correlation function, involve squares of the density 
of states. Writing the latter (Eq.~(\ref{eq:trace})), in terms of the 
difference between advanced and retarded Green functions $(G^{(+)}-G^{(-)})$
we are left with products of one-particle Green functions. The
most interesting terms are the cross products 
$G^{(+)}(\brp,\br;E) \ G^{(-)}(\br,\brp;E) = G^{(+)}(\brp,\br;E) {G^{(+)}}^*(\brp,\br;E^{*})$, 
because they survive the energy average and are sensitive to changes 
in the magnetic field. 

In the non-interacting approach that we have used so far in this work, the 
two-particle Green function factorizes into a product of one-particle Green 
functions \cite{Fetter}, therefore we will use in this section the former 
as a synonym for the latter. Unless specified, the Green functions will
be retarded ones. Let us consider the product 
$G(\brp_1,\br_1;E) G^*(\brp_2,\br_2;E)$. The effect of the disorder 
potential can be taken into account perturbatively, for each realization 
of the disorder, in the same way as before by using 
Eqs.~(\ref{eq:action_approx}) and (\ref{eq:dis_action}).
We can therefore write the disorder average as a double sum
over trajectories $s$ and $u$

        \begin{eqnarray}
        \langle G \ G^\ast \rangle & = & \sum_s \sum_u \
        \langle G_s \ G^\ast_u \rangle
        = \sum_s \sum_u \  G^c_s \ G^{c\ast}_u \langle
        \exp{\left[\frac{i}{\hbar} (\delta S_s-\delta S_u)\right]} \rangle
        \label{eq:Gproduct}\\
        & = &
         \sum_s \sum_u \  G^c_s \ G^{c\ast}_u
        \exp \left[-\frac{\langle (\delta S_s-\delta S_u)^2\rangle}{
        2\hbar^2} \right] \; . \nonumber
          \end{eqnarray}

Here it is necessary to take into account the correlation
of the disorder potential between points on trajectories $s$ and
$u$.  One limiting case is that where
$s$ and $u$ are, either the same trajectory, or the time reversal
one of each other.  In these cases their associated contribution acquires
exactly the same phase shift and
 $\langle G_s \ G^\ast_s \rangle = |G^c_s|^2$.
Within our approximation, the diagonal contributions $s=u$,
which {\em e.g.} are responsible for the classical part of the
conductivity, remain disorder-unaffected, since we
assume the trajectories have a length much smaller
than $\lt$. (A semiclassical consideration of these effects
for trajectories of length of the order of $\lt$ or
larger was performed in Ref.~\cite{Mirlin96} for the bulk, 
giving a damping of the two--point Green function on the scale of $\lt$.)
At the opposite extreme,  if trajectories $s$ and $u$ are completely
uncorrelated, {\em i.e.}, for long trajectories in classical chaotic 
systems or trajectories in integrable systems with a spatial distance 
larger than $\xi$, the average in Eq.~(\ref{eq:Gproduct}) factorizes:
$\langle G_s \ G^\ast_u \rangle  =
\langle G_s  \rangle \langle  G^\ast_u \rangle$
and lead to single-particle damping behavior.

The double sum Eq.~(\ref{eq:Gproduct}) may also 
involve pairs of trajectories which stay within a distance of the 
order of  $\xi$ (as for nearby paths on a torus of an integrable 
system).  In this case the behavior of
$\langle G_s \ G^\ast_u \rangle$ is more complicated and
depends of the confinement geometry of the system under
consideration.  As a simple illustration of the interplay between
disorder correlation and families of orbits, let us consider
for the case of two trajectories $s$ and $u$ joining respectively
the points $\br_1 = (0,0)$ with $\brp_1 = (L,0)$ and $\br_2 = (0,y)$ 
with $\brp_2 = (L,y)$. We assume $L\gg\xi$ and ignore the confinement
effects. The separation $y$ between the trajectories may be of the order 
of $\xi$. The variance of the relative phase between $s$ and $u$ is

        \be
        \langle  (\delta S_s-\delta S_u)^2\rangle =
        2 L \ \frac{(K(0) - K(y))}{\vf^2}
        \ee

\nin with  $K(y) = \int_{-\infty}^{+\infty} C(x,y) \ {\rm d} x = 
(\hbar^2 \vf^2/l) \exp(-y^2/(4\xi^2))$ for Gaussian potentials.
Therefore $\langle G G^\ast \rangle =   G^c \ G^{c\ast} \tilde{f}(y)$ with

\be
\label{eq:ftilde}
\tilde{f}(y) = \exp\left[-\frac{L}{l}\left(1 - \frac{K(y)}{K(0)}\right)
\right] \ .
\ee

\nin This result is general for two trajectories of length $L$ running
parallel a distance $y$ apart. The function $\tilde{f}(y)$ expresses in a 
very simple way that as $y \to 0$, the effect of disorder disappears 
($\tilde{f}(0) = 1$) while for $y\gg\xi$ the function $\tilde{f}(y)$ 
behaves as the square of single particle Green function damping.

%%%%%%%%%%%%%%%%%%%%%%%%%%%%%%%%%%%%%%%%%%%%%%%%%%%%%%%%%%%%%%%%%%%%%%%%% 

\subsection{Fixed-size impurity average of the magnetic susceptibility}
\label{subsec:applic1}

We consider here a disorder average (which will henceforth be called a
fixed--size impurity average)
of an ensemble of structures for which the parameters of the
corresponding clean system (geometry, size, and chemical potential)
remain fixed under the change of impurity realizations.
As  shown in the previous section, averages over weak disorder exponentially 
damp, but do not completely suppress, oscillatory contributions 
(with phase $\kf L_s$) to the single--particle Green function 
arising from the paths of a confined system.

We treat regular billiards at zero or small magnetic fields, where
the density of states has the $H$--dependence of the formulae 
(\ref{eq:uniform_d})--(\ref{eq:thetatrace}).
The general result for $\chi^{(1)}$, Eq.~(\ref{eq:chi1}), formally
persists in the presence of smooth disorder with the replacement of 
$\C_{\rm M}$ by
   
\be 
\label{eq:thetatr_dis} 
\langle \C^d_\bsM(H) \rangle  = \frac{1}{2\pi} \int_0^{2\pi} 
\dif \Theta_1 \, \cos \left[ 2 \pi \frac{H \A_\bsM(\Theta_1) 
}{ \Phi_0} \right] \ \exp\left[-\frac{\langle(\delta S_{\rm M}
(\Theta_1))^2 \rangle}{2\hbar^2}\right] \; ,
\ee

\nin where $\langle\delta S_{\rm M}^2(\Theta_1)\rangle$ 
is given by Eq.~(\ref{eq:dS2})
with the integrals performed along the orbits of the family $\bf M$
parameterized by $\Theta_1$. 

In the case of square billiards, where the dominant contribution 
arises from the family $(1,1)$, $\delta S(x_0)$ is independent of $x_0$ 
and in the FR regime we have

\be 
\label{eq:chi1_disfr}
\langle	\chi \rangle  \simeq 
\langle	\chi^{(1)}\rangle  =  \chi^{(1)}_{\rm c}   
\exp{\left[-\frac{L_{11}}{2l}\right]}  \ ,
\ee

\nin where $\chi^{(1)}_{\rm c}$ denotes the susceptibility of the
system without disorder (Eq.~(\ref{eq:chi1_sq})). 

\begin{figure}
\setlength{\unitlength}{1mm}
\begin{picture}(100,110)
\put(0,160){\epsfxsize=9cm\epsfbox{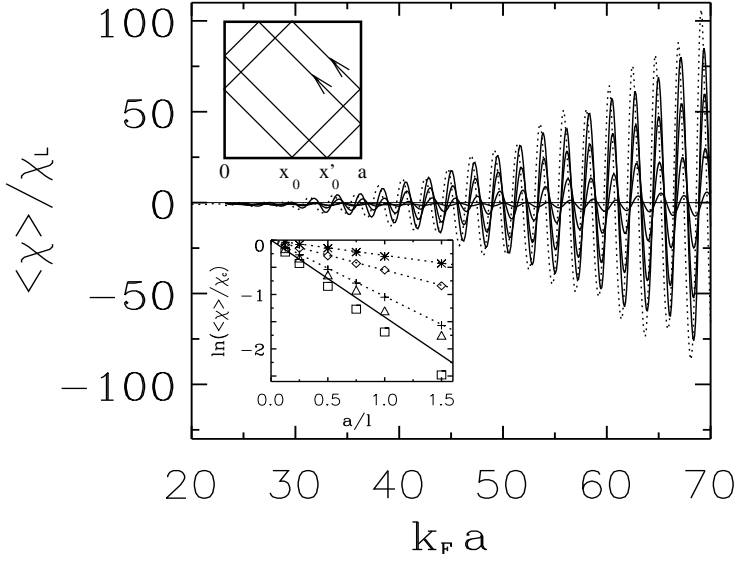}}
\end{picture}
\vspace{-2cm}
\caption{Susceptibility $\protect \langle \chi \rangle $ of a square
billiard (at zero magnetic field and at a temperature equal to
6 level spacings) as a function of $\protect k_{\scriptstyle F}
a$ for the clean case (dotted) and for increasing Gaussian
disorder ($\protect \xi/a = 0.1$) with elastic
MFP $\protect l/a = 4, 2, 1, 0.5$ (solid lines in the order of
decreasing amplitude). Upper inset: Two representative
periodic orbits belonging to the family (1,1) of a square billiard.
Lower inset: ratio $\protect \langle \chi \rangle / \chi_{\rm c}$ as a function
of the inverse elastic MFP $a/l$ for $\protect \xi/a = 4, 2, 1, 0, 0.5$
(from the top). The symbols indicate the numerical quantum results
for $\xi = 0$ (triangles), $\xi/a= 0.5$ (squares), $\xi/a=1$ (crosses),
$\xi/a= 2$ (diamonds), $\xi/a=4$ (stars). The solid line is the
$\protect \xi = 0$ result. The the dotted lines going through the 
different sets of symbols for show the semiclassical 
dampings in various regimes.
}
\label{fig1rapcom}
\end{figure}

Fig.~\ref{fig:chi1} shows results of the numerical quantum 
simulations for the average susceptibility $\langle \chi \rangle$ 
of an ensemble of squares with fixed size but different disorder 
realizations at a temperature $\kb T = 3 \gs \Delta$, where $\Delta$ 
is the mean level spacing. The characteristic oscillations on a scale 
$\kf L_{11}$ are the signature of the $(1,1)$ family and persist upon
inclusion of disorder such that the elastic MFP $l$ is of the order of
the system size. The clean susceptibility (dotted line) is increasingly
damped (solid lines) with decreasing elastic MFP ($l/a=4,2,1,0.5$) for fixed
$\xi/a=0.1$ (which represents a typical disorder correlation length in 
experimental realizations). For details of the numerical simulations
see Ref.~\cite{RapComm}. The lower inset depicts the ratio
$\langle \chi \rangle/\chi_{c}$ for various correlation
lengths $\xi$ and $l$. The damping is exponential
as predicted by Eq.~(\ref{eq:chi1_disfr}) and the decays are
well reproduced by the analytical expressions for the SR (solid line), FR and
LR (dotted lines).

%%%%%%%%%%%%%%%%%%%%%%%%%%%%%%%%%%%%%%%%%%%%%%%%%%%%%%%%%%%%%%%%%%%%%%%%% 

\subsection{Combined impurity and energy average of the susceptibility}
\label{subsec:applic2}

The previously treated fixed-size impurity average is not realistic since 
in experimentally realizable structures disorder averages 
cannot be performed independently from size-averages. The
detailed features of the confining potential do not remain unchanged for
different impurity configurations. Therefore we have to consider
{\em energy and disorder} averages. The typical susceptibility is
now defined by $ \chi^{(t)} = \langle \overline{\chi^2} \rangle^{1/2}$.
It applies to the case of repeated measurements on a given microstructure
when different impurity realizations (and simultaneous changes in $\kf$)
are obtained by some kind of perturbation ({\em e.g.} cycling to room temperature).
From now on we will reserve the term $ \chi^{(t)}_{\rm cl}$ for the clean
typical susceptibility $ (\overline{\chi^2})^{1/2}$. The energy and impurity
averaged susceptibility $\langle \overline{\chi} \rangle$ describes the
magnetic response of an ensemble of a large number of microstructures with
different impurity realizations and variations in size. This is the relevant
quantity for the interpretation of the experiment of Ref.~\cite{levy93}.

The semiclassical results for $\chi^{(t)}$ and $\langle \overline{\chi} 
\rangle$ for a system of  integrable geometry are obtained in an
analogous way as we proceeded in the previous chapter. That is,
by including in the integral (\ref{eq:thetatrace})
for ${\cal C}_{\bf M}$ a $\Theta_1$-dependent disorder-induced
phase $\exp{[i\delta S(\Theta_1)/\hbar]}$. 
However, now we have to take the square of ${\cal C_{\bf M}}$ 
before the impurity average, and cross correlations between different paths
on a torus {\bf M}, or between different tori, have to be considered. 
We discuss this effect, typical of integrable systems, for the case of a 
square billiard. For sake of clarity we moreover assume a temperature range such
that only the contribution of the shortest closed orbit has to be taken into 
account. The contribution of orbits $(1,1)$ for the typical susceptibility reads:

\be  
\label{eq:chit_dis} 
\hspace{0.7cm} \left( \frac{\chi^{(t)}}{\chi^0} \right)^2  = \frac{1}{2} 
\int_{0}^{a} \frac{{\rm d}x_0}{a} 
\int_{0}^{a} \frac{{\rm d}x_0'}{a}  \
\A^2(x_0) \A^2(x_0')
\cos \left( \varphi\A(x_0) \right)
\cos \left( \varphi\A(x_0') \right)
f(x_0,x_0')   \, ,
\ee

\nin with $\chi^0$ defined as in Eq.~(\ref{eq:chi0}). The function

\begin{eqnarray}
\label{eq:aveexpoad}
f(x_0,x_0') & = &
 \left\langle \exp{\left\{\frac{i}{\hbar}
\left(\delta S(x_0)-\delta S(x_0') \right) \right\}} \right\rangle \\
& = &  \exp{\left\{-\frac{1}{2\hbar^2} \left[\langle
\delta S^2(x_0)\rangle+\langle\delta S^2(x_0')\rangle - 2
\langle \delta S(x_0)\delta S(x_0')\rangle \right]
 \right\}}
\nonumber
\end{eqnarray}

\nin accounts for the effect of disorder on pairs of orbits
$x_0$ and $x_0'$. For the magnetic response of an 
energy and disorder averaged ensemble we find correspondingly:

\be
\label{eq:chia_dis}
\frac{\langle\overline{ \chi} \rangle}{{\overline{\chi}^0}} =
\frac{1}{2} \ \int_{0}^{a} \frac{{\rm d}x_0}{a} \int_{0}^{a} 
\frac{{\rm d}x_0'}{a}  \left[\A^{2}_{-} \cos(\varphi \A_{-}) +
\A^{2}_{+} \cos(\varphi \A_{+})\right] f(x_0,x_0') 
\ee

\nin with $\overline{\chi}^0$ and $\A_{\pm}$ defined as in Eq.~(\ref{eq:chi0bar}).

In the case of short range we reach the border of applicability of 
our semiclassical approximation. If we nevertheless take 
$\xi \longrightarrow 0$, we see that orbits with $x_0 \neq x_0'$ 
are disorder-uncorrelated and all such pair contributions are 
exponentially damped. Using exclusively the family (1,1), one obtains
an overall suppression of the typical and average susceptibility 
	
\be
\label{eq:xi0t_dis}
\lim_{\xi \rightarrow 0} \chi^{(t)} = \chi^{(t)}_{\rm cl} \ e^{-L_{11}/2l_\delta}  
\qquad  ,  \qquad \lim_{\xi \rightarrow 0} \langle\overline{\chi} \rangle
 =  \overline{\chi} \ e^{-L_{11}/l_\delta}   \ .
\ee

\nin Note that the exponent for $ \langle\overline{\chi} \rangle $
differs by a factor $1/2$ from that for $\langle \chi \rangle$. 

In the finite range case of $\lf \! < \! \xi \! \ll \! a$, the phase shifts 
$\delta S(x_0)$ and
$\delta S(x_0')$ in $f(x_0,x_0')$ are accumulated in a correlated way,
if the spatial distance of two orbits $x_0$ and $x_0'$ is smaller than 
$\xi$. To evaluate the product term 
$2 \langle \delta S(x_0)\delta S(x_0')\rangle $ in the exponent
of $f(x_0,x_0')$ in this regime the
integrations are performed as in Eq.~(\ref{eq:dS2}) but with
$\bf q$ and $\bf q'$ running along paths starting at $x_0$,
respectively $x_0'$. 
Ignoring the additional correlations
occurring near the bounces off the boundaries of the billiard, the 
trajectories $x_0$ and $x'_0$ (see inset Fig.~\ref{fig1rapcom})
can be regarded as straight lines
remaining at a constant distance $y=|x_0 - x'_0|/\sqrt{2}$
from another. We can therefore approximate $f(x_0,x'_0)$ by
$\tilde f(|x_0 - x'_0|/\sqrt{2})$ with the function $\tilde f$
given by Eq.~(\ref{eq:ftilde}).  For Gaussian correlation we thus have
	
\be
\label{eq:f_gauss}
f(x_0,x_0')  =   \exp\left\{-\frac{L_{11}}{l}
\left[ 1 - \exp\left(-\frac{(x_0-x_0')^2}{8\xi^2}\right)\right]
\right\}  \ .
\ee

\nin Orbits separated by $|x_0-x_0'| \gg \xi$ are disorder--uncorrelated
and exponentially suppressed: 
$ f(x_0,x_0')  \simeq \exp(-L_{11}/l)$. In contrast to that, disorder 
only weakly affects trajectories separated by $|x_0-x_0'| < \xi$.

The disorder averages in the finite range regime lead,
 by means of the function $f$, to a non--exponential damping of
the susceptibilities for systems with families of periodic orbits. 
This behavior becomes obvious for the case of square billiards
where at $H\!=\!0$ the integrals (\ref{eq:chit_dis}) and (\ref{eq:chia_dis}) 
can be evaluated analytically in the limits of $L_{11} \ll l$ 
(extreme ballistic) and $L_{11} \gg l$ (deep ballistic).
We find for the typical and average susceptibility at $H=0$ in 
the finite range case for $L_{11} \ll l$ \cite{RapComm,JMP}
        
\be \
\label{ebt}
\left(\frac{ \chi^{(t)}}{\chi^{(t)}_{\rm c}} \right)^2 \simeq
1-\frac{L_{11}}{l} \left(1-c_t\ \frac{\xi}{a} \right) 
\qquad  ,  \qquad 
\frac{\langle \overline{\chi}\rangle}{\overline{\chi}} \simeq
1-\frac{L_{11}}{l} \left(1-c_a\ \frac{\xi}{a} \right)  \ ,
\ee

\nin and for $L_{11} \gg l$ (by steepest descent):

\be 
\label{dbt}
\left(\frac{ \chi^{(t)}}{ \chi^{(t)}_{\rm c}} \right)^2  \simeq 
c_t \ \left(\frac{\xi}{a}\right) \ \left(\frac{l}{L_{11}}\right)^{1/2} 
\qquad  ,  \qquad
\frac{\langle \overline{\chi}\rangle}{\overline{\chi}}  \simeq 
c_a \ \left(\frac{\xi}{a}\right) \ \left(\frac{l}{L_{11}}\right)^{1/2} \ .
\ee

\nin The constants in the above equations are $c_t = (20/7) \sqrt{2 \pi}$ 
and $c_a = 2 \sqrt{2\pi}$ \cite{RapComm}. Eq.~(\ref{ebt}) expresses the limit of very
weak disorder, showing that the small disorder effect is further reduced 
due to the correlation of the disorder potential. The other limit, 
Eq.~(\ref{dbt}), is more interesting since it shows that disorder correlation 
effects lead to a replacement of the exponential disorder damping by a power law.

\begin{figure}
\setlength{\unitlength}{1mm}
\begin{picture}(100,110)
\put(-85,40){\epsfxsize=6.5cm\epsfbox{fv17a.ps}}
\put(-10,40){\epsfxsize=6.5cm\epsfbox{fv17b.ps}}
\put(5,57){\Large (a)}
\put(80,57){\Large (b)}
\end{picture}
\vspace{-4cm}
\caption{Ratio between disorder averaged and zero--disorder results
for (a) the typical $\chi^{(\rm t)}$ (thin curves)
and (b) the ensemble averaged $\langle \overline{\chi} \rangle $
(thick curves) susceptibilities as a function of
decreasing elastic MFP $l$ for different values of $\xi/a$.
The symbols denote the numerical quantum results, the solid lines
(for $ \xi > 0$) the semiclassical integrals
(\protect\ref{eq:chit_dis}) (a) and (\protect\ref{eq:chia_dis}) (b) and the dashed lines
asymptotic expansions of the integrals for large $a/l$.  }
\label{fig2rapcom}
\end{figure}

Fig.~\ref{fig2rapcom} depicts in logarithmic representation
our collected results for the disorder averaged typical (a) 
and averaged (b) susceptibility for square
billiards (at $H=0$ and $\kb T = 2\gs\Delta$) as a function of
the inverse elastic MFP for different disorder correlation
lengths. The symbols denote results from numerical quantum simulations and 
the full curves semiclassical results from numerical integration
of the Eqs.~(\ref{eq:chit_dis}) and (\ref{eq:chia_dis}). For the
short range case $\xi = 0$ they reduce to Eq.~(\ref{eq:xi0t_dis})
predicting an exponential decrease with exponent $L_{11}/l_{\delta}$
which is in line with the quantum calculations (circles). The 
semiclassical results for the finite range are on the whole
in agreement with the numerical results for $\xi/a = 0.1$ (diamonds),
$\xi/a = 0.2$ (triangles) and $\xi/a=0.5$ (squares). The
semiclassical curves seem to overestimate the 
damping of the typical susceptibility. The dotted curves (shown for
$a/l \geq 1$) depict the limiting expressions (\ref{dbt}) 
in the regime $L_{11} > l$.  

\subsection{Relation to experiment and other theories}
\label{subsec:exp}

In the experiment of L\'evy {\em et al.} \cite{levy93}, the magnetic
susceptibility was measured for an array of about $10^5$ ballistic square-like
cavities. The size of the squares is on average $a\!=\!4.5 \mu m$, with
a large dispersion (estimated between 10 and 30\%) along the array.
Each individual square is a mesoscopic ballistic system since
the phase-coherence length is estimated to be $L_{\Phi} = 15$--$40$ $\mu m$ 
and the elastic MFP $l = 4.5$--$10$ $\mu m$.
The potential correlation length can be estimated 
to be of the order of $\xi/a \simeq 0.1$. Taking the most
unfavorable case of $l \simeq a$ we obtain,
with respect to the clean case, a disorder reduction
for the averaged susceptibility of
$\langle \overline{\chi} \rangle /\overline{\chi} \simeq 0.37$,
showing that the features of the clean integrable systems (strong
paramagnetic susceptibility at $H=0$) persist upon inclusion
of disorder. Since $\overline{\chi} \simeq 100 \ \cl$
\cite{vO94,URJ95,RUJ95}, our calculations
for the paramagnetic response of the
ballistic squares agree quantitatively with the experimental
findings (given the experimental uncertainties). However,
the temperature dependence obtained in the experiment is much 
slower than the exponential suppression implied by Eqs.~(\ref{dbt}),
(\ref{square:chizf}) and (\ref{eq:R_T}). This is one of the
motivations to consider the interaction effects that we discuss
in the next section.

In a related theoretical work, Gefen and collaborators\ \cite{braun} followed a 
complementary approach
to ours and calculated the disorder--averaged susceptibility
for an ensemble of ballistic squares based on long trajectories
[strongly] affected by scattering from $\delta$--like impurities.
They found that the average susceptibility does not depend
on the elastic MFP. At the temperatures of experimental relevance, these
very long trajectories are irrelevant, and this is the reason why our
numerical results do not present the proposed effect. McCann and Richter
combined a diagrammatic approach with semiclassical techniques and
studied the transition between the clean and diffusive regimes for
$\delta$-like potentials \cite{McCann}. They separated the contributions from 
trajectories persisting in the clean limit (as we have considered in
this section) form scattered paths (as proposed in Ref.~\cite{braun}),
and demonstrated that the former are clearly dominant in the 
(temperature and disorder) regime of experimental relevance.

%% file: var7.tex
%=============================================================================

%
% SECTION VII: ELECTRON-ELECTRON INTERACTIONS IN THE BALLISTIC REGIME
%
% file: var7.tex
%
% last version: 07/10
%

\section{Electron-electron interactions in the ballistic regime}
\label{sec:eeiitbr}

As discussed in the introduction, in Mesoscopic Physics the electron-electron
interactions are usually taken into account only by a renormalization of
single-particle quantities (like the effective mass or the effective potential 
felt by individual electrons) and by the effect of the quasiparticle life-time 
as a limiting factor of the phase-coherence length $L_{\Phi}$. This is the
approximation we have used so far in this work, which gives good account of
transport experiments, dealing with open systems. On the other hand, when we
go to dots weakly coupled to the electronic reservoirs, we enter in the 
regime of Coulomb blockade, where the effect of interactions becomes crucial 
for the conductance \cite{CBrev}. If we completely isolate the dots and 
measure thermodynamical properties, like the magnetic susceptibility, we expect 
important effects from the combined role of confinement and interactions.
In the context of persistent currents in disordered metals, the effect of
interactions was invoked \cite{ambegaokar,ESra} as a possible source for
the discrepancy found between the experimental results \cite{levy90,Webb1,Webb2}
and the non-interacting theories \cite{ensemble}.

In Secs.~\ref{sec:clean} and \ref{sec:disord} we concentrated ourselves on the
effect of confinement in ballistic cavities and, working within a single-particle
approach, we found a large orbital response which depends on the underlying
classical mechanics. In this section we demonstrate that interactions also give
rise to a large contribution, which has to be added to the non-interacting one.
We show that the semiclassical approach can be extended to the interacting 
problem and that the new contribution to the susceptibility is also depending
on the classical dynamics of non-interacting electrons \cite{ubroj98}.

\subsection{Screened Coulomb interaction in two dimensions}
\label{subsec:sci}

Interactions in a 2DEG formed at a semiconductor heterostructure
have been thoroughly studied since the early eighties \cite{Stern}, by 
adopting the techniques used for the 3D electron gas, like Landau's Fermi 
liquid theory and Random Phase Approximation (RPA) \cite{AGD,Fetter}. 
The latter approach yields an effective (momentum and frequency
dependent) interaction potential

\be
\label{eq:effpotrpa}
{\cal V}(\bq;\omega_m) = \frac{\hV(\bq)}{1-\hV(\bq)\Pi^0(\bq;\omega_m)} \ ,
\ee

\nin where $\hV(\bq)=(2 \pi e^2/\epsilon_\infty)(f_q/q)$ is the bare
potential. The 2D Fourier transform of the standard 3D Coulomb
interaction ($2 \pi e^2/q$) is reduced by two factors: the optical
(high-frequency) dielectric constant $\epsilon_\infty$ taking into
account the screening by the valence electrons and the form-factor
$f_q$ arising from the finite extent of the electron wave-function
in the direction perpendicular to the heterojunction \cite{Stern}.
The irreducible polarizability (particle-hole propagator) is given by

\be
\label{eq:Phi0}
\Pi^0(\bq;\omega_m) = - \frac{\gs}{\beta} \ \sum_{\epsilon_n} \int \frac{\dif \bp}{(2\pi)^2} 
\ {\cal G}(\bp;{\omega_m}) \ {\cal G}(\bp + \bq;{\omega_m}+\epsilon_n) \ ,
\ee

\nin in terms of the non-interacting finite-temperature Green functions
\cite{AGD,Fetter} (for which we develop a semiclassical approximation latter 
in this section). The Matsubara frequencies associated with one-particle
(two-particle) propagators are of fermionic (bosonic) kind: 
$\epsilon_n = (2n+1) \pi / \beta$ ($\omega_m=2m\pi/\beta$). 

In two dimensions, the zero-temperature (time-ordered) irreducible polarizability
admits the simple analytical form \cite{Stern}

\be
\label{eq:Phi0TO}
\Pi^0_{\rm T}(\bq;\omega) = - \ \frac{n}{\EF} \ \frac{\kf}{q} \left(\frac{q}{\kf} \ -
\ \sqrt{a_{+}^2-1} \ + \ \sqrt{a_{-}^2-1} \ \right) \ ,
\ee

\nin with $n$ the 2D electron density, $a_{\pm}=(\omega+i \eta)/(q \vf) \pm q/(2\kf)$,
and the complex square roots taken in the branch with positive imaginary part. 
RPA is a high-density expansion in the dimensionless parameter 
$r_s = r_0 / a_0$ ($\pi r_0^2$ is the average area per electron, and $a_0$ is the 
Bohr radius in the semiconductor). In $GaAs/AlGaAs$ heterostructures we typically
have $r_s \smeq 2$, but as in standard metals, RPA gives good results
beyond its regime of validity, and it has been extensively used in calculations
of effective masses, inelastic scattering times, etc \cite{Stern,JDS}.

The long wave-length limit of the static effective potential  yields the
Thomas-Fermi (or screened) 2D potential:

\be
\hV_{\rm TF}(\bq) = \frac {2 \pi e^2 / \epsilon_\infty}  {q+q_s} 
\qquad  ,  \qquad
V_{\rm TF}(\br)  =  \int \frac{\dif \bq}{(2\pi)^2} \ \hV_{\rm TF}(\bq) \ \exp[i\bq\!\cdot\!\br] \ ,
\label{eq:VqTF}
\ee

\nin where $q_s = (g_s m e^2) / (\hbar^2  \epsilon_\infty)$ is the screening
wave-vector (we take $f_q=1$). $V_{\rm TF}(\br)$ is the electrostatic potential created 
by a test charge at 
the origin, and given by the sum of the bare (long range) Coulomb potential and a term
associated with the fact that electrons in the gas are pushed  away
from the [negative] external charge, resulting in the screening of
the original potential. The screening properties of the 2DEG are poorer than in the 
three dimensional case. For instance, in the former the potential $V_{\rm TF}(\br)$
decays for large distances as the third power of $r$ while in the later the 
dependence is exponential. In RPA the difference in the decay of
the effective potential is the factor $r^{-d}$ of the Friedel oscillations. The poor 
screening in the 2DEG will then be a limitation to keep in mind when using local
effective interactions.

When we consider the confinement into a quantum box, we will still
keep the screened interaction of Eq.~(\ref{eq:VqTF}). This is a reasonable
approximation since the size of the quantum boxes we want to describe
($a=4.5 \mu m$) is much larger than the typical screening length
($2\pi/q_s=0.03 \mu m$). We therefore expect the boundary effects on
the screened interaction to be small.

Even if the semiclassical approach can be adapted to work with the potential
$V_{\rm TF}(\br)$ of Eq.~(\ref{eq:VqTF}) \cite{longinterac}, we will make in this 
section a further (and strong) approximation and neglect the momentum 
dependence of $\hV_{\rm TF}$, which leads to the local potential

\be
U(\br) = \lambda_0 \ N(0)^{-1} \ \delta(\br) \ ,
\label{eq:locpot}
\ee

\nin with $N(0)\!=\!\EF/n\!=\!\gs m/(2 \pi\hbar^2)$ the density of states and $\lambda_0 = 1$
introduced to identify the order of perturbation.

\subsection{Thermodynamics and semiclassics of small interacting systems}
\label{subsec:tasosis}

\begin{figure}
\setlength{\unitlength}{1mm}
\begin{picture}(100,110)
\put(20,80)
{\epsfxsize=9cm\epsfbox{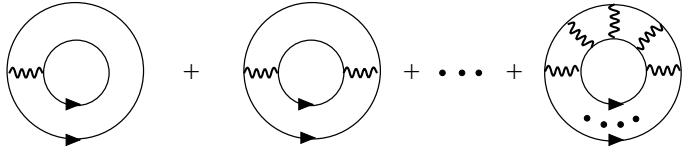}}
\end{picture}
\vspace{-7cm}
\label{fig:diagrams}
\caption{
Leading Cooper-channel diagrams for the interaction contribution to the
thermodynamic potential.
}
\end{figure}

The interaction induced thermodynamical properties are given by the contribution
$\Omega_i$ to the thermodynamical potential (Eq.~(\ref{eq:gcpott})). The
finite-temperature formalism for the grand-partition function yields a 
perturbative expansion for $\Omega_i$ \cite{AGD}. For local and spin
independent interactions, the leading order (Cooper channel) contribution
is represented diagrammatically in Fig.~\ref{fig:diagrams}, and given by \cite{doubleA} 

\begin{eqnarray}
\label{omega_ic}
\Omega_i^C & = & - {1 \over \beta} \sum_{k=1}^\infty
                \frac{(-\lambda_0)^k}{k} \sum_{\omega_m < \EF}
                \int d\br_1 \ldots d\br_k \Sigma(\br_2,\br_1;\omega_m)
                \ldots \Sigma(\br_1,\br_k;\omega_m)
                \nonumber \\
           & = & {1 \over \beta} \sum_{\omega_m < \EF} {\rm Tr} 
                \left\{ \ln [1+ \lambda_0 \Sigma(\omega_m) ] \right\} \ .
\end{eqnarray}

\nin The particle-particle propagator is given by \cite{AGD}

\be
\label{eq:sigma}
\Sigma(\brp,\br;\omega_m)={1\over \beta N(0) } \ 
\sum_{\epsilon_n < \EF} {\cal G}(\brp,\br;\epsilon_n) \
{\cal G}(\brp,\br;\omega_m-\epsilon_n) \; .
\ee

The trace over the space coordinates is a short way of expressing the
expansion in all orders in $\lambda_0 \Sigma$. The short-length (high-frequency) 
behavior is incorporated in the screened interaction, thus requiring a cutoff of 
the frequency sums at ${\EF}$ \cite{doubleA}. In Fig.~\ref{fig:diagrams} the
wavy lines represent the local interaction and the solid lines the non-interacting
Green function ${\cal G}$ in the presence of the confining potential. The concept 
of particle-particle propagator, as well as the Cooperon contribution, come from 
the Cooper pairs in the theory of superconductivity. The main difference with our 
case is that here the interaction is repulsive (thus the plus sign in the trace)
and that we have lost translational invariance (therefore we cannot
trade the operators for ordinary functions by going to momentum representation).
The factor $k$ in each term of the expansion (\ref{omega_ic}) is to be contrasted
with the $k!$ of the Feynman rules for the Green function, and it is responsible
for the poor convergence properties of perturbative expansions for $\Omega_i^C$ \cite{AGD}.
This is why higher-order diagrams are essential in the diagonal Cooper channel,
as known from the theory of superconductivity \cite{aslamazov} and persistent
currents \cite{ambegaokar,ESra}. 

The standard RPA contribution to $\Omega_i^C$ \cite{Fetter} is obtained by
reversing the direction of one of the loops in the diagrams of Fig.~\ref{fig:diagrams}
(or by using similar expressions to those of Eq.~(\ref{omega_ic}) with the
interchange of $\lambda_0 \Sigma$ by $U \Pi^0$). The first two diagrams of both series
coincide, but semiclassical arguments indicate that the RPA expansion gives a much
smaller contribution than that of the Cooper channel. The RPA susceptibility of a 3D
electron gas has been calculated by Vignale as an expansion in $r_s$ \cite{Vignale}. 
On the other hand, the Cooper channel gives for the 2DEG an interaction-induced 
susceptibility that overwhelms the Landau contribution and {\em increases} with 
$\kf$ \cite{aslamazov,bulk}. 

We have seen in the previous sections that the non-interacting grand-canonical 
susceptibility averaged over the microstructures vanishes in leading order in ${\bf N}$,
forcing us to make the distinction between the canonical and grand-canonical results.
Conversely, we will see that the interacting contribution from $\Omega_i$ does
not vanish upon average, and that is why in this section we work in the grand-canonical ensemble.

The non-interacting finite-temperature Green functions appearing in the diagrams 
of Fig.~\ref{fig:diagrams} (and in Eqs.~(\ref{eq:Phi0}) and (\ref{eq:sigma})) can be
written in terms of retarded and advanced  Green function as

\be
   \G(\brp,\br;\epsilon_n) = \theta(\epsilon_n) \ G^{(+)}(\brp,\br;\EF \!+\!i\epsilon_n) + 
                             \theta(-\epsilon_n) G^{(-)}(\brp,\br;\EF \!+\!i\epsilon_n) \ .
\label{eq:ftGreen}
\ee

\nin The complex energy-arguments force us to go to the analytic continuation.
However, if the Matsubara energies are much smaller than $\EF$, in a
semiclassical expansion of $G^{(+)}$ (Eq.~(\ref{eq:gfgutz})), we can 
expand the classical action around $\EF$. We can also include the
effect of a weak magnetic field by a perturbation in the action (as we
have consistently done in this work), and we arrive to the semiclassical
approximation for the finite-temperature Green function

\begin{eqnarray} 
\label{eq:ftsgf}
   \G(\brp,\br;\epsilon_n,B) & = & \frac{2\pi}{(2\pi i\hbar)^{3/2}} \left\{\theta (\epsilon_n)
\sum_{s(\br,\brp)} \sqrt {D_{s}} \exp{\left[\frac{i}{\hbar}S_{s}\!-\!i\frac{\pi}{2}\nu_{s}\right]} 
\exp\left[-\frac{\epsilon_n \tau_s}{\hbar}\right] 
\exp \left[ i \frac{B \Theta_s}{\Phi_0} \right] \right. \nonumber \\
\displaystyle
& + & \left. \theta(-\epsilon_n)
\sum_{s'(\brp,\br)} \sqrt {D_{s^{\prime}}} \exp{\left[-\frac{i}{\hbar}S_{s^{\prime}}\!+\!
i\frac{\pi}{2}\nu_{s^{\prime}}\right]} \exp\left[\frac{\epsilon_n \tau_{s^{\prime}}}{\hbar}\right] 
\exp{\left[- i \frac{B \Theta_{s^{\prime}}}{\Phi_0} \right]} \right\}
\end{eqnarray}

\nin where the unperturbed trajectories $s$ and $s'$ travel from $\br$ to $\brp$ in opposite directions,
with the energy $\EF$ and in the absence of magnetic field. $\tau_s = \partial S_s/ \partial E)$ is the
time associated with the trajectory $s$, and $\Theta_s$ is the effective area (Eq.~(\ref{eq:ea})).
Long trajectories are exponentially suppressed in the expansion.

The usefulness of Eq.~(\ref{eq:ftsgf}) goes far beyond the problem of
orbital magnetism that we discuss in this work, as it provides a calculational
approach to any perturbative problem where we have the knowledge of the
single-particle classical dynamics. For instance, it has been used to calculate the
interaction-induced renormalization of the density of states in open systems \cite{japan}.

\subsection{First order perturbation, diagonal and non-diagonal contributions}
\label{subsec:fopdandc}

The first-order (Hartree-Fock like) is obviously the simplest term to calculate in the diagrammatic
expansion (\ref{omega_ic}), and given by 

\be
   \label{first} 
   \Omega_i^{(1)} = { \lambda_0 \over \beta}
   \sum_{\omega_m}{\rm Tr}\, \{ \Sigma({\omega_m}) \} \ .
\ee

\begin{figure}
\setlength{\unitlength}{1mm}
\begin{picture}(100,140)
\put(30,-5)
{\epsfxsize=7cm\epsfbox{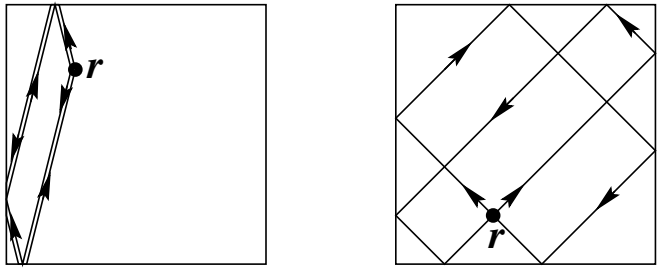}}
\end{picture}
\vspace{-8cm}
\label{fig:traj}
\caption{ 
Typical pairs of real-space trajectories that contribute to the 
average susceptibility to first order in the interaction in the 
diagonal channel (left) and the non-diagonal channel (right).}
\end{figure}

\nin Semiclassically, $\Sigma(\brp,\br;\omega_m)$ is a
sum over pairs of trajectories joining $\br$ to $\brp$.  However, most
pairs yield highly oscillating contributions 
which, after the spatial integrations, give higher order terms in $1/\kf a$.
To leading order, the only pairs that contribute to the susceptibility are
whose whose dynamical phases $\exp[i S_s(B\!=\!0)/\hbar]$ cancel while
retaining a magnetic-field dependence. One way this can be achieved is 
by pairing each orbit $s$ with its time reverse. The trace in
Eq.~(\ref{first}) yields a sum over closed but not necessarily periodic
trajectories (see Fig.~\ref{fig:traj} left, for a square). This {\em diagonal} contribution, 
is present independent of the nature of the classical dynamics.

In integrable systems, periodic orbits come in
families within which the action integral is constant.  If, as is generally
the case, two orbits of the same family cross at a given point, it is
possible to cancel the dynamical phases by pairing them (Fig.~\ref{fig:traj}, right).
This pair contributes to the trace in Eq.~(\ref{first}) because both orbits
are continuously deformable so that the phase is canceled throughout an
entire region of space. If we restrict ourselves to the family (1,1),
this {\em non-diagonal} first-order contribution is given by \cite{ubroj98}

\be
\label{nondiag}
   {\langle \chi^{(1)OD}_{i,11} \rangle \over \cl} =
   - \  {3 \kf a \over 4 \sqrt{2} \pi^3} \
   \ { d^2 {\cal C}_{11}^2(\varphi) \over d \varphi^2}
   \ R^2_T\left(\frac{L_{11}}{L_T}\right)
\ee

\nin The temperature dependence is governed by the function $R_T$ of Eq.~(\ref{eq:R_T}) and the 
field dependence by $ {\cal C}_{11}$ (Eq.~(\ref{Csimple})). The first-order contribution to $\chi$
in the diagonal channel $\langle \chi^{(1)D}_{i} \rangle$ has the same dependence on $\kf a$ as in
Eq.~(\ref{nondiag}) and a similar $T$ dependence; its magnitude is $\sim\! 1.4$ times larger.
Therefore, to first order in the interaction, the difference between chaotic systems
(for which there is only the diagonal term) and regular ones
(for which the non-diagonal term is also present) is numerical but not qualitative.
Disorder will not affect the diagonal contributions since we have paired time-reversed
trajectories. The suppression of $\langle \chi^{(1)OD}_{i,11} \rangle$ by a smooth
disorder potential will be, as in (\ref{subsec:applic2}), a power law.

\subsection{Higher order terms}
\label{subsec:hot}

As mentioned before, we should consider all diagrams of the Cooper channel
shown in Fig.~\ref{fig:diagrams}.  One
should sum all terms which (i)~do not vanish upon ensemble averaging,
(ii)~depend on $B$, and (iii)~are of leading order in $\hbar\sim 1/ \kf a$.
For instance, (iii) is checked by $\hbar$ power counting, since a pair of
Green functions scales as $ N(0)/ \hbar $, interactions as $ [N(0)]^{-1} $,
and Matsubara sums as $\hbar$.  Indeed, all terms in the series are of order
$\hbar$ despite the formal expansion in $\lambda_0$.  
Higher-order terms contain diagonal as well as non-diagonal contributions.
However, in the latter terms the location of the additional interaction points 
is severely limited: they must lie on both periodic orbits to cancel the 
dynamical phases and so must be near the intersections of the two orbits. 
Further analysis \cite{ubroj98,longinterac} shows that these contributions are
smaller by a factor of $1/\kf a$. Therefore, it is only the diagonal contribution 
that is strongly renormalized by higher-order terms.

The diagonal contribution $\Omega_i^{D}$ to $\Omega_i$ is given by
(\ref{omega_ic}) with the substitution of $\Sigma$ by its diagonal part

\be
\label{sigmaD}
\hspace{0.8cm} \Sigma^{D}(\br,\brp;\omega_m) = \frac{1}{(2 \pi \hbar)^2N(0)} 
   \sum_{s(\br,\br)}^{L_s > \Lambda_0} D_s {R(2 \tau_s/ \tau_T) \over \tau_s} 
   \ \exp{\left[i \frac{2 B}{\Phi_0} \Theta_s \right]} 
   \ \exp{\left[-\frac{\omega_m \tau_s}{\hbar}\right]} \ ,  
\ee

\nin expressed as a {\em single} sum over trajectories $s$ longer
than the cutoff $\Lambda_0 = \lf / \pi$ (associated with the upper
bound $\EF$ on the Matsubara sum in Eq.~(\ref{eq:sigma})]). 
While we cannot diagonalize $\Sigma^{D}$ analytically, it
has the nice property that (except for $\Lambda_0$) all variations occur
on classical scales: rapid quantum oscillations on the scale of $\lf$
have been washed out, greatly simplifying the original quantum problem.  
In this sense, $\Sigma^{D}$ is a ``classical'' operator. Hence,
we can discretize $\Sigma^{D}$ with mesh size larger than
$\lf$, sum over trajectories between cells, diagonalize, and so
compute $\Omega_i^{D}$ numerically.

\begin{figure}
\setlength{\unitlength}{1mm}
\begin{picture}(100,140)
\put(10,10)
{\epsfxsize=13cm\epsfbox{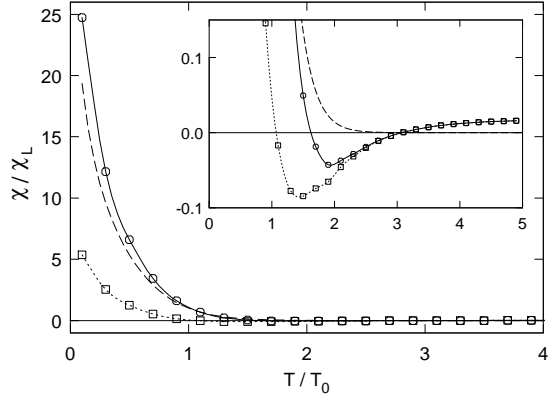}}
\end{picture}
\vspace{-8cm}
\label{fig:interac}
\caption{
Temperature dependence of the zero-field susceptibility (solid line) for
an ensemble of squares at $\kf a = 50$. The contribution of the
non-diagonal channel (dashed, family (11) and repetitions) exceeds that of
the diagonal Cooper channel (dotted) at low temperatures
(\protect{$k_{\rm B} T_0 = \hbar \vf/2\pi a$}). Inset: expanded scale shows
the change in sign as a function of $T$ (From Ref. \protect\cite{ubroj98}.)}
\end{figure}

In Fig.~\ref{fig:interac} we show the interaction contribution to the
susceptibility of a square billiard (solid) as the sum of the 
the non-diagonal contribution of the family (11) (and its repetitions)
(dashed) and the diagonal contribution $\chi_i^{D}$ (dotted) from the
numerical evaluation of $\Omega_i^{D}$. The non-diagonal contribution 
dominates $\chi_i^{D}$. Thus, {\em the existence of a family of periodic orbits-- a
characteristic of the non-interacting classical dynamics-- has a crucial effect
on the interaction contribution to the susceptibility.}
We can distinguish three regimes for $\chi^{D}$. At low-temperature it is 
{\it paramagnetic} and decays on a scale similar to the non-diagonal contribution,
but has a significantly smaller amplitude. In the intermediate range,
$\chi^{D}$ is small and {\it diamagnetic}. Finally, at high temperatures
it is again {\it paramagnetic}, but very small. This is naturally
understood by associating each regime with an order in the perturbation
series. The low-$T$ part corresponds to the first-order term
(orbits of the type in Fig.~\ref{fig:traj}, left) which is exponentially suppressed by
the temperature factor $R$ when $L_T$ becomes smaller than the shortest
closed orbit. At this point the second-order term, due to closed paths of
two trajectories connected by interactions, takes over. There is no minimum
length of these paths, and hence the second-order term is less rapidly
suppressed by $T$. For repulsive interactions, the sign is opposite
of the first-order term, thus the sign change in $\chi^{D}$.  At even
higher temperatures once $L_T\ll a$, this term is a surface contribution and
the third-order term takes over. The latter is a bulk contribution
\cite{aslamazov,bulk} since with three interactions flux can be enclosed without
bouncing off the boundary.

The previous interpretation of Fig.~\ref{fig:interac} should be reconsidered 
since, the final result for the diagonal channel at low $T$ is much smaller 
than the first-order diagonal contribution. Moreover, we observe 
numerically that the terms in the perturbation series increase in magnitude 
with order and therefore we do not have a good convergence of the series. As
often found with ill-convergent series, a renormalization mechanism contains
the basic physical ingredients. The essential idea is to reorder the 
perturbation expansion (\ref{omega_ic}) by gathering short paths as to 
produce effectively lower-order contributions \cite{ubroj98,longinterac}.
This analysis leads to a reduction of $1/\ln(\kf a)$ of 
$\langle \chi^{(1)D}_{i} \rangle$ and explains why the diagonal contribution
is dominated by $\langle \chi^{(1)OD}_{i,11} \rangle$.
The semiclassical ideas and the renormalization scheme can also be applied
to the calculation of the orbital response of the 2DEG \cite{bulk} and of
disordered mesoscopic conductors \cite{physicaE,longinterac}.

\begin{table}
\begin{tabular}{|c|cc|cc|}
\hline
& \hspace{1cm} $\chi^{(t)}/\protect\cl$  &  
& \hspace{1cm} $\langle \overline{\chi} \rangle/\protect\cl$ &  \\
\hline
& regular & chaotic & regular & chaotic \\
\hline
non-interacting & & & & \\
contribution & $(\kf a)^{3/2}$ & $(\kf a)$ & $(\kf a)$ & $(\kf a)^0$ \\
\hline
interacting & & & & \\
contribution &     $(\kf a)$ & $(\kf a)/\ln{(\kf a)}$ & $(\kf a)$ & $(\kf a)/\ln{(\kf a)}$ \\
\hline
smooth-disorder & & & & \\
damping  &     $(\xi/a)^{1/2}(l/L)^{1/4}$ & 1 & $(\xi/a)(l/L)^{1/2}$ & 1 \\
\hline
\end{tabular}
\vspace{0.25cm}
\caption{$\protect (\kf a)$ dependence of the non-interacting and interacting contributions
to the magnetic response and the parametric dependence of their dampings due to the
presence of smooth disorder for billiard-like microstructures in the presence
(regular case) and absence (chaotic case) of families of periodic orbits.
$\protect \chi^{(t)}$ and $\protect \langle \overline{\chi} \rangle$ are respectively the typical
and  energy-disorder averaged susceptibilities. We express them in units
of the 2D Landau susceptibility $\protect\cl=e^2/(12 \pi m c^2)$. $\protect \kf$ is the
Fermi wave-vector, $\protect a$ the typical size of the microstructure, $\protect l$ the elastic
mean-free-path, $\protect \xi$ the correlation length of the disorder potential, and
$\protect L$ is the length of the shortest
flux-enclosing periodic trajectories. }
\label{table:summary}
\end{table}

In table \ref{table:summary} we summarize the various contributions to the
susceptibility for the cases of chaotic and regular geometries. 
As discussed in (\ref{subsec:exp}) the non-interacting theory is roughly in
agreement with the experimental results (at the lowest temperatures). Incorporating 
the interaction-induced
contribution, which is of the same order as the non-interacting one, still keeps
us within the range of the measurements \cite{levy93} (which have an uncertainty of 
a factor of 4). The disagreement on the temperature 
dependence of the susceptibility is reduced respect to that of the non-interacting 
case, but the temperature scale $T_0$ in Fig.~\ref{fig:interac} is still
significantly smaller than that in the experiment.

%% file: var8.tex
%=============================================================================

%
% SECTION VIII: CONCLUSIONS
%
% file: var8.tex
%
% last version: 8/10
%

\section{Conclusions}
\label{sec:conc}

The central line of this review has been the use of semiclassical methods in
problems of Mesoscopic Physics. When applied to mesoscopic systems, the semiclassical
analysis needs to be adapted according to physical considerations, often 
introducing cut off lengths (elastic mean-free-path, coherence length, etc). The 
ability of semiclassical methods to deal with problems that are not translational
invariant gives them a very special place among the analytical tools used in the study 
of small systems.

In this article we have seen both the strength and weakness of the semiclassical tool.
Semiclassics is able to yield the signatures of the underlying classical dynamics on
quantum properties. It allows us to study important features of ballistic transport, like
conductance fluctuations and weak localization, predicting the line-shape of 
the correlation function and the weak localization peak. Technically, only
one part of these effects (the diagonal component) is easily obtained. For 
line-shapes in chaotic structures it is reasonable to expect that the diagonal part has
the same information of the total effect, and we have verified this conjecture with
numerical calculations. The magnitude of the conductance fluctuations and weak 
localization effects are not easily obtained from semiclassics. Random matrix theory,
whenever it is applicable, presents a better route to extract universal values. 

Each integrable system has its own specificity, but semiclassics can be applied to
all cases and therefore remains the only analytical tool in dealing with transport
through integrable cavities. The semiclassical analysis predicts important differences
between the quantum interference phenomena in transport through integrable and chaotic
cavities. Such a distinction is supported by many numerical calculations.

The experimental developments in quantum transport have been largely motivated by the
theoretical results. In the case of chaotic cavities, good agreement is obtained between
the measurements and the theoretical predictions. Signatures of integrable dynamics have
been obtained experimentally in the power spectrum of the conductance fluctuations and in
the (non-Lorentzian) weak localization peak. But the integrable behavior is much more 
fragile than in the chaotic case and it does not show up in some of the experiments on
cavities with regular geometry. The experimental attention has somehow shifted to the 
chaotic case, where the semiclassical and random matrix theory predictions have been
tested and used as a way of extracting information on the phase coherence length. 
However, we feel that regular microstructures deserve more attention if we want to achieve
a thorough understanding of their physics.

There has been a growing interest in the transport through cavities with mixed
dynamics lately. Further theoretical and experimental work seems necessary to make progress
in our understanding of this generic case. 

The semiclassical tool is also helpful to study thermodynamical properties of 
microstructures, like the magnetic susceptibility. The orbital response is a direct
consequence of the magnetic field dependence of the density of states. The latter
is known to be very sensitive to the underlying classical dynamics, exhibiting larger
fluctuations in the integrable case than in the chaotic one. This behavior
translates into an enhanced susceptibility for integrable cavities (which is mainly
given by the effect of short orbits).

We have presented a formalism capable of incorporating weak-disorder effects arising from
smooth potentials (slowly varying on the scale of the Fermi wave-length). Such a formalism is 
expressed in terms of one and two-particle Green functions and can be applied to 
transport and thermodynamical quantities. For the magnetic susceptibility it predicts
a weak (power-law) suppression of the clean values.

It might have not been evident to foresee, but the semiclassical tool can also be 
adapted to treat problems of electron-electron interaction, and the resulting
contribution to the magnetic susceptibility turns out to depend on the classical
dynamics of non-interacting electrons. The agreement between theory and experiment
concerning the magnetic response of ballistic microstructures is only qualitative
(and at the lowest temperatures), and much less satisfactory than in the case of
transport. The difficulties from the theoretical side are probably related to the
approximated way in which we are able to take interactions into account. We expect
that the few available experimental results of orbital magnetism in the 
ballistic regime will soon be complemented by detailed studies.  

In this work we have presented a few examples illustrating the link between Mesoscopic 
Physics and semiclassical theories, and we have left aside other problems of sustained 
interest at present, like the Coulomb blockade in quantum dots \cite{CBrev}, the tunneling 
through quantum wells with tilted magnetic fields \cite{not}, and the electron transport with 
inhomogeneous magnetic fields \cite{Ye}, where semiclassics is also a very helpful approach.

\vspace{3cm}

\nin {\bf Acknowledgements}

The work described in these lectures is mainly the result of long and
enjoyable collaborations with H.U.~Baranger, K. Richter, A.D. Stone
and D. Ullmo. I further acknowledge fruitful collaborations in other 
aspects of Mesocopic Physics (covered or not in this article) with
Y. Alhassid, C.W.J. Beenakker, D.P. DiVincenzo, G.-L.~Ingold, I.K. Marmorkos, E.R. Mucciolo,
F.~von~Oppen, J. Petitjean, J.-L.~Pichard, P.~Pichaureau, P.~Schmitteckert, M.~Schreier, 
and D.~Weinmann.

Support from the CNRS (Coop\'eration CNRS/DFG), the French-German program PROCOPE, and
the ``Institut Universitaire de France" are gratefully acknowledged.

%% file: varref.tex
% Bibliography

%% file: varenna.bbl
\begin{thebibliography}{0}

%Introduction
%%%%%%%%%%%%%%

\bibitem{LesHou89} Proceedings of the 1989 Les Houches Summer School 
on {\em Chaos and Quantum Physics}, edited  by M.-J.~Giannoni, A.~Voros, 
and J.~Zinn-Justin (North-Holland, Amsterdam, 1991).

\bibitem{Fer91} Proceedings of the 1991 Enrico Fermi International
School of Physics on {\em Quantum Chaos}, edited by G.~Casati, I.~Guarneri
and U.~Smilansky (North-Holland, Amsterdam, 1993).

\bibitem{ozor:book} A.M. Ozorio de Almeida, {\em Hamiltonian Systems:
Chaos and Quantization} (Cambridge University Press, 1988).

\bibitem{gutz_book} M.C.~Gutzwiller, {\it Chaos in Classical and Quantum
Mechanics} (Springer-Verlag, Berlin, 1990).

\bibitem{FriWin} H.~Friedrich and D.~Wintgen, Phys. Rep. {\bf 183},
37 (1989). 

\bibitem{Stockrev} H.-J.~St\"ockmann, J.~Stein, and M.~Kollmann, in
{\em Quantum Chaos}, edited by G.~Casati and B.~Chirikov
(Cambridge University Press, Cambridge, 1995).
  
\bibitem{Jal90} R.A.~Jalabert, H.U.~Baranger and A.D.~Stone,
Phys. Rev. Lett. {\bf 65}, 2442 (1990).

\bibitem{Mar92} C.M.~Marcus, A.J.~Rimberg, R.M.~Westervelt, P.F.~Hopkins
and A.C.~Gossard, Phys. Rev. Lett. {\bf 69}, 506 (1992).

\bibitem{Chaos} C.M.~Marcus, R.M~Westervelt, P.F.~Hopkings,
and A.C.~Gossard, Chaos, {\bf 3}, 643 (1993).

\bibitem{Chaost} H.U.~Baranger, R.A.~Jalabert, and A.D.~Stone,
Chaos, {\bf 3}, 665 (1993).

\bibitem{revha} H.U.~Baranger in {\it Nanotechnology}, edited by 
G.~Timp (AIP Press, 1995).

\bibitem{LesHouSt} A.D.~Stone, in Ref. \cite{LesHou94}, p.~325.

\bibitem{csf} {\it Chaos and Quantum Transport in Mesoscopic Cosmos},
special issue, Chaos, Solitons\&Fractals, edited by K.~Nakamura
{\bf 8}, 971-1412 (1997).

\bibitem{RUJ95}
K.~Richter, D.~Ullmo and R.A.~Jalabert, Phys.~Rep. {\bf 276}, 1 (1996).

\bibitem{Klaus} K.~Richter, {\it Semiclassical Theory of Mesoscopic
Quantum Systems}, Springer Tracts in Modern Physics (Springer-Verlag, 
in press).

\bibitem{ALWrev} {\em Mesoscopic Phenomena in Solids}, edited by 
B.L.~Altshuler, P.A.~Lee, and R.A.~Webb, (North-Holland,
Amsterdam, 1991).

\bibitem{LesHou94} Proceedings of the 1994 Les Houches Summer School
on {\em Mesoscopic Quantum Physics}, edited by E.~Akkermans, 
G.~Montambaux, J.-L.~Pichard, and J.~Zinn-Justin (North-Holland, 
Amsterdam, 1995).

\bibitem{Datta} S.~Datta, {\it Electronic Transport in Mesoscopic Systems}
(Cambridge University Press, Cambridge, 1995).

\bibitem{Imry} Y.~Imry, {\it Introduction to Mesoscopic Physics} 
(Oxford University Press, Oxford , 1997).

\bibitem{WashWebb} S.~Washburn and R.A.~Webb, Adv. Phys. {\bf 35}, 
375 (1986); Rep. Prog. Phys. {\bf 55}, 1311 (1992).

\bibitem{LeeRam} P.A.~Lee and T.V.~Ramakrishnan, Rev. Mod. Phys. 
{\bf 57}, 287 (1985).

\bibitem{BvHra} C.W.J.~Beenakker and H.~van~Houten in
{\it Solid State Physics}, Vol. 44, edited by H. Ehrenreich and D.
Turnbull (Academic Press, New York, 1991). 

\bibitem{DSS} S.~Das~Sarma and F.~Stern, Phys. Rev. B {\bf 32},
8442 (1988).

\bibitem{Rouk} M.L.~Roukes {\em et al.}, Phys. Rev. Lett. {\bf 59}, 
3011 (1987).

\bibitem{Ford} C.J.B.~Ford {\em et al.}, J. Phys. C {\bf 21},
L325 (1988); C.J.B.~Ford, S.~Washburn, M. B\"uttiker, C.M.~Knoedler, 
and J.M.~Hong, Phys. Rev. Lett. {\bf 62}, 2724 (1989).

\bibitem{Bohigas} 
O.~Bohigas, M.-J.~Giannoni, and C.~Schmit, Phys. Rev. Lett. {\bf 52}, 1 
(1984); O. Bohigas, in Ref. \cite{LesHou89}, p.~87.

\bibitem{levy93} L.P.~L\'evy, D.H.~Reich, L.~Pfeiffer, and K.~West,
Physica B {\bf 189}, 204 (1993).

\bibitem{BenMailly} D.~Mailly, C.~Chapelier, and A.~Benoit,
                    Phys. Rev. Lett. {\bf 70}, 2020 (1993).

\bibitem{marcusgroup1} C.M.~Marcus, R.M.~Westervelt, P.F.~Hopkins
and A.C.~Gossard, Phys. Rev. B {\bf 48}, 2460 (1993); R.M.~Clarke
{\em et al.}, Phys. Rev. B {\bf 52}, 2656 (1995).

\bibitem{Kel94}
M.W.~Keller, O. Millo, A. Mittal, D.E.~Prober, and R. N. Sacks,
Surf. Sci., {\bf 305}, 501 (1994); M.W.~Keller {\em et al.}, 
Phys.\ Rev.\ B {\bf 53}, R1693 (1996).

\bibitem{berry94} M.J.~Berry, J.H.~Baskey, R.M.~Westervelt, and
A.C.~Gossard, Phys. Rev. B {\bf 50}, 8857 (1994); M.J.~Berry, 
J.A.~Katine, R.M.~Westervelt, and A.C.~Gossard, Phys. Rev. B 
{\bf 50}, 17721 (1994).

\bibitem{chang94} A.M.~Chang, H.U.~Baranger, L.N.~Pfeiffer, and
K.W.~West, Phys. Rev. Lett. {\bf 73}, 2111 (1994); A.M.~Chang,
in Ref.~\cite{csf}, p.~1281.

\bibitem{marcusgroup2} I.H.~Chan, R.M.~Clarke,
C.M.~Marcus, K.~Campman, and A.C.~Gossard, Phys. Rev. Lett. {\bf 74},
3876 (1995); C.M.~Marcus {\em et al.}, in Ref.~\cite{csf}, p.~1261. 

\bibitem{bird} J.P.~Bird {\em et al.},  Phys.\ Rev.\ B {\bf 52}, 
R14336 (1995); and in Ref.~\cite{csf}, p.~1299..

\bibitem{persson} M.~Persson {\em et al.},  Phys.\ Rev.\ B {\bf 52},
8921 (1995).

\bibitem{lutj} G.~L\"utjering {\em et al.} Surf. Sci. {\bf 361/362}, 
709 (1996).

\bibitem{lee97} Y.\ Lee, G.\ Faini, and D.\ Mailly, Phys.\ Rev.\ B, 
{\bf 56}, 9805 (1997); and in Ref.~\cite{csf}, p.~1325.

\bibitem{zozou97} I.V.~Zozoulenko, R.~Schuster, K.-F.~Berggren, and 
K.~Ensslin, Phys. Rev. B {\bf 55}, R10209 (1997).

\bibitem{marcusgroup3} A.G.~Huibers {\em et al.} Phys. Rev. Lett. {\bf 81}, 1917 (1998);
cond-mat/9904274. 

\bibitem{NRC} R.P.~Taylor {\em at al.}, Phys. Rev. Lett. {\bf 78}, 
1952 (1997); A.S.~Sachrajda {\em at al.}, Phys. Rev. Lett. {\bf 80}, 
1948 (1998).

\bibitem{Mirlin} A.~Mirlin, in this issue.

\bibitem{Haake} F.~Haake, in this issue.

\bibitem{brack_book} M.~Brack and R.K.~Bhaduri, {\it Semiclassical Physics}
(Addison-Wesley, Reading, 1997).

\bibitem{KaLa} D.E.~Khmelnitskii and A.I.~Larkin, Ups. Fiz. Nauk.
{\bf 136}, 533 (1982) [Sov. Phys. Usp. {\bf 25}, 185 (1982)].

\bibitem{ChSm} S.~Chakravarty and A.~Schmid, Phys.~Rep. {\bf 140}, 195 (1986).

\bibitem{Ashcroft} N. W. Ashcroft and N. D. Mermin, {\it Solid State Physics}
(Holt, Rhinehart, and Winston, New York, 1976).

\bibitem{Blu88}
R. Bl\"umel and U. Smilansky, Phys. Rev. Lett.  {\bf 60}, 477 (1988);
Physica D {\bf 36}, 111 (1989).

\bibitem{Bar93} H.U.~Baranger, R.A.~Jalabert, and A.D.~Stone,
Phys. Rev. Lett. {\bf 70}, 3876 (1993).

\bibitem{paul} P.~Pichaureau and R.A.~Jalabert,
Eur. Phys. J. B {\bf 9}, 299 (1999).

\bibitem{ingold} M.~Schreier, K.~Richter, G.-L.~Ingold, and R.A.~Jalabert,
Eur. Phys. J. B {\bf 3}, 387 (1998).

\bibitem{vO94} F.~von~Oppen, Phys.~Rev.~B {\bf 50}, 17151 (1994).

\bibitem{URJ95}  D.~Ullmo, K.~Richter, and R.A.~Jalabert,
Phys. Rev.~Lett.~{\bf 74}, 383 (1995).

\bibitem{RapComm} K.~Richter, D.~Ullmo, and R.A.~Jalabert,
Phys.~Rev.~B, {\bf 56}, R5619 (1996).

\bibitem{ubroj98} D.~Ullmo, H.U.~Baranger, K.~Richter, F.~von~Oppen,
and R.A.~Jalabert, Phys.~Rev.~Lett.\ {\bf 80}, 895 (1998).


%Sec. II
%%%%%%%%%%%%%%

\bibitem{Landauer} R. Landauer, Phil. Mag. {\bf 21}, 863 (1970).

\bibitem{But86}
M. B\"uttiker, Phys. Rev. Lett. {\bf 57}, 1761 (1986).

\bibitem{arnold:book} V.I.~Arnold, {\em Mathematical Methods of
Classical Mechanics} (Springer-Verlag, New York, 1984).

\bibitem{jung} C.~Jung and H.-J.~Scholz, J. Phys. A {\bf 21}, 2301 (1988).

\bibitem{tel} T.~T\'el in {\it Direction in Chaos, Vol. 3,} edited by Hao 
Bai Lin (World Scientific, Singapore, 1990) p.~149.

\bibitem{LesHouSm}U.~Smilansky in Ref. \cite{LesHou89}, p.~371.

\bibitem{Gas89} P.~Gaspard and S.A.~Rice, J. Chem. Phys. {\bf 90}, 2225;
2242; 2255 (1989).

\bibitem{Jen91} R.V.~Jensen, Chaos {\bf 1}, 101 (1991).

\bibitem{Mac92}
R.B.S. Oakeshott and A. MacKinnon,
Superlat. and Microstruc. {\bf 11}, 145 (1992).

\bibitem{RamLech} P.~Lecheminant, J. Phys. I France {\bf 3}, 299 (1993).

\bibitem{Ber86}
M.V. Berry and M. Robnik, J. Phys. A {\bf 19}, 649 (1986).

\bibitem{Dor91} E.~Doron, U.~Smilansky, and A.~Frenkel, Phys. Rev. Lett. 
{\bf 65}, 3072 (1990); Physica D {\bf 50}, 367 (1991); in 
Ref.~\cite{Fer91}.

\bibitem{Bau91}
W. Bauer and G. F. Bertsch, Phys. Rev. Lett. {\bf 65}, 2213 (1990).

\bibitem{Lai92} Y.-C. Lai, R. Bl\"umel, E. Ott, and C. Grebogi,
Phys. Rev. Lett. {\bf 68}, 3491 (1992).

\bibitem{Lin93} W.A.~Lin, J.B.~Delos and R.V.~Jensen, Chaos {\bf 3}, 665 (1993).

\bibitem{Leg90} O. Legrand and D. Sornette, Physica D {\bf 44}, 229 (1990);
Phys. Rev. Lett. {\bf 66}, 2172 (1991).

\bibitem{BvH88}
C.W.J.~Beenakker and H.~van~Houten, Phys. Rev. B {\bf 37}, 6544 (1988).

\bibitem{Bar91}
H.U.~Baranger, D.P.~DiVincenzo, R.A.~Jalabert, and A.D.~Stone,
Phys. Rev. B {\bf 44}, 10637 (1991).

\bibitem{BaMe96} H.U.~Baranger and P.A.~Mello,
Phys. Rev. B {\bf 54}, R14297 (1996).

\bibitem{LanAbra} D.C.~Langreth and E.~Abrahams, Phys. Rev. B {\bf 24}, 2978 (1981).

\bibitem{BarSto} H.U.~Baranger and A.D.~Stone, Phys. Rev. B {\bf 40}, 8169 (1989).

\bibitem{FishLee} D.S.~Fisher and P.A.~Lee, Phys. Rev. B {\bf 23}, 6851 (1981).

\bibitem{Szafer}
A.D. Stone and A. Szafer, IBM J. Res. Dev. {\bf 32}, 384 (1988).

\bibitem{LeeFish} P.A.~Lee and D.S.~Fisher, Phys. Rev. Lett. {\bf 47}, 882 (1981). 

\bibitem{MacK} A.~MacKinnon, Z.~Phys.  B {\bf 59}, 385 (1985). 

\bibitem{Shepard} K.~Shepard, Phys. Rev. B {\bf 43}, 11623 (1991).

\bibitem{NSB} J.U.~N\"ockel, A.D.~Stone, and H.U.~Baranger, 
Phys. Rev. B {\bf 48}, 17569 (1993).

\bibitem{correction} Note a sign mistake in the original
work of Ref.~\cite{Jal90} and in Eq.~(9) of Ref.~\cite{Chaost}.
This point was noticed and corrected by W.A.~Lin, see Ref.~\cite{LinJen}.

\bibitem{Jen93} J.H.~Jensen, Phys. Rev. Lett. {\bf 73},
244 (1994).

\bibitem{Mil74} W.H.~Miller, Adv. Chem. Phys. {\bf 25}, 69 (1974).

\bibitem{ber76} M.V. Berry, M. Tabor, Proc. R. Soc. Lond.
                A.~{\bf 349}, 101 (1976).

\bibitem{LinJen} W.A.~Lin and R.V.~Jensen, Phys.\ Rev.\ B 
{\bf 53}, 3638 (1996); W.A.~Lin, in Ref.~\cite{csf}, p.~995.

\bibitem{ishio95} H.\ Ishio and J.\ Burgd\"orfer, Phys.\ Rev.\ B {\bf 51},
2013 (1995).

\bibitem{schwi96} C.\ D.\ Schwieters, J.\ A.\ Alford, and J.\ B.\ Delos,
Phys.\ Rev.\ B {\bf 54}, 10652 (1996).

\bibitem {vattay} G.~Vattay, J.~Cserti, G.~Palla, and G.~Sz\'alka,
in Ref.~\cite{csf}, p.~1031.

\bibitem{Gutz83}
M. C. Gutzwiller, Physica D {\bf 7}, 341 (1983).

\bibitem{Eric60}
T. Ericson, Phys. Rev. Lett. {\bf 5}, 430 (1960).

\bibitem{borguar} F.~Borgonovi and I.~Guarneri, J.~Phys. A {\bf 25}, 
3239 (1992); Phys. Rev. E {\bf 48}, R2347 (1993).

\bibitem{Iid90}
S.~Iida, H.A.~Weidenm\"uller, and J.A.~Zuk, Annals of Phys. {\bf 200}, 219 (1990).

\bibitem{Wei91}
C.H.~Lewenkopf and H.A.~Weidenm\"uller, Annals of Phys. {\bf 212}, 53 (1991).

%Sec. III
%%%%%%%%%%%%%%

\bibitem{Wirtz} L.~Wirtz, J.-Z.~Tang, and J.~Burgd\"orfer, 
Phys.\ Rev.\ B {\bf 56}, 7589 (1997).

\bibitem{khinchin} A.Ya.~Khinchin, {\em Continued Fractions} (University of
Chicago Press, Chicago, 1964).

\bibitem{correction2} Notice the missing $\varepsilon_n^{\prime}$ in
the exponent of Eq.~(19) of Ref.~\cite{paul}.

%Sec. IV
%%%%%%%%%%%%%%

\bibitem{Westervelt} R.M.~Westervelt, in Ref.~\cite{revha}.

\bibitem{But88}
M. B\"uttiker, Phys. Rev. B {\bf 33}, 3020 (1986); IBM J. Res. Dev. {\bf 32}, 63 (1988).

\bibitem{Bruus94}
H.~Bruus  and A.D.~Stone Phys.\ Rev.\ B {\bf 50}, 18275 (1994).

\bibitem{efetov} K.B.Efetov, Phys. Rev.~Lett.~{\bf 74}, 2299 (1995).

\bibitem{Moha} P.~Mohanty, E.M.Q.~Jariwala, and R.A.~Webb, Phys. Rev. Lett. 
{\bf 78}, 3366 (1997).

\bibitem{Ullmo} D.~Ullmo, unpublished.

\bibitem{BeenRMP} C.W.J.~Beenakker, Rev. Mod. Phys. {\bf 69}, 731 (1997).

\bibitem{WeiPR} T.~Guhr, A.M.~M\"uller-Groeling, and H.~A.~Weidenm\"uller,
Phys. Rep. {\bf 283}, 37 (1998).

\bibitem{Blu90} R. Bl\"umel and U. Smilansky, Phys. Rev. Lett.  {\bf 64},
241 (1990).

\bibitem{JP} R. A. Jalabert and J.-L. Pichard, J. Phys. (France) {\bf
5}, 287 (1995). 

\bibitem{BarMel} H.U.~Baranger and  P.A.~Mello,
Phys. Rev.~Lett.~{\bf 73}, 142 (1994).

\bibitem{JPB} R.A.~Jalabert, J.-L.~Pichard, and C.W.J.~Beenakker,
Europhys.~Lett.~{\bf 27}, 255 (1994).

\bibitem{BaMeFL} H.U.~Baranger and P.A.~Mello,
Phys. Rev. B {\bf 51}, 4703 (1995).

\bibitem{BrBeFL} P.W.~Brouwer and C.W.J.~Beenakker, 
Phys. Rev. B {\bf 51}, 7739 (1995); {\bf 55}, 4695 (1997).

\bibitem{Weid94} Z.~Pluha\u{r}, H.A.~Weidenm\"uller, J.A.~Zuk, and
C.H.~Lewenkopf, Phys. Rev.~Lett.~{\bf 73}, 2115 (1994).

\bibitem{DoroSm} E.~Doron and U. Smilansky, Nucl. Phys. A {\bf 545},
455c (1992).

\bibitem{MeBa} P.A.~Mello and  H.U.~Baranger, Europhys.~Lett.~{\bf 33}, 465 (1996).

\bibitem{Argetal} N. Argaman {\em et al.}, Phys. Rev.~Lett.~{\bf 71}, 4326 (1993).

\bibitem{Argaman} N. Argaman, Phys. Rev.~Lett.~{\bf 75}, 2750 (1995); Phys.\ Rev.\ B {\bf 53}, 7035 (1996).

\bibitem{AlLa} I.A.~Aleiner and A.I.~Larkin, Phys. Rev. B {\bf 54}, 14423 (1996).

\bibitem{roland} R. Ketzmerick, Phys.\ Rev.\ B {\bf 54}, 10841 (1996).

\bibitem{bodo} B. Huckestein, R. Ketzmerick, and C.H.~Lewenkopf
cond-mat/9908090.

\bibitem{casati} G.~Casati, I.~Guarneri, and G.~Maspero, unpublished 1999.

\bibitem{Wilk} M.~Wilkinson, J. Phys. A {\bf 20} 2415 (1987).

\bibitem{Fleisch} R.~Fleischmann, T.~Geisel and R. Ketzmerick, 
Phys. Rev.~Lett.~{\bf 68}, 1367 (1992).

\bibitem{KlausEPL} K.~Richter, Europhys.~Lett.~{\bf 29}, 7 (1995).

\bibitem{Greg} G.~Hackenbroich and F.~von~Oppen, Europhys.~Lett.~{\bf 29}, 151 (1995);
Z.~Phys. B {\bf 97} 157 (1995).

\bibitem{Weiss93} D. Weiss {\em et al.}, Phys. Rev. Lett., {\bf 70}, 4118 (1993).


%Sec. V
%%%%%%%%%%%%%%

\bibitem{Land} L.D.~Landau, Z.~Phys. {\bf 64}, 629 (1930).

\bibitem{Peierls} R.E.~Peierls, {\it Surprises in Theoretical Physics}
(Princeton University Press, Princeton NJ, 1979).

\bibitem{schmid} Yu.N~Ovchinnikov, W.~Lehle, and A.~Schmid,
Ann. Physik {\bf 6}, 487 (1997). 

\bibitem{AGD} A.A.~Abrikosov, L.P.~Gorkov, and I.E.~Dzyaloshinski,
{\em Methods of Quantum Field Theory in Statistical Physics}
(Prentice-Hall, Englewood Cliffs, 1963).

\bibitem{Fetter} A.L.~Fetter and J.D.~Waleka, {\em Quantum
Theory of Many-Particle Systems} (McGraw-Hill, New York, 1971).

\bibitem{BM} H.~Bouchiat and G.~Montambaux, J.~Phys. (Paris) {\bf 50},
2695 (1989).

\bibitem{ensemble}
A.~Schmid, Phys. Rev. Lett. {\bf 66}, 80 (1991);
F.~von Oppen and E.K.~Riedel, {\it ibid} 84;
B.L.~Altshuler, Y.~Gefen, and Y.~Imry, {\it ibid} 88.

\bibitem{Kubo64} R.~Kubo, J.~Phys.~Soc.~Japan {\bf 19}, 2127 (1964).

\bibitem{Prado} S.D.~Prado, M.A.~M.~de~Aguiar,
J.P.~Keating and R.~Egydio de Carvalho, J.~Phys.~A {\bf 27}, 6091 (1994).

\bibitem{sursci} R.A.~Jalabert, K.~Richter, and D.~Ullmo,
Surf. Sci., {\bf 361/362}, 700 (1996).

\bibitem{Sha} B.~Shapiro, Physica A, {\bf 200}, 498 (1993).

\bibitem{aga94} 0.~Agam, J.~Phys. I (France) {\bf 4} 694 (1994).

\bibitem{KlaBer} K.~Richter and B.~Mehlig, 
Europhys.~Lett.~{\bf 41}, 587 (1998).

%Sec. VI
%%%%%%%%%%%%%%

\bibitem{JMP} K.~Richter, D.~Ullmo, and R.A.~Jalabert,
J.~Math.~Phys. {\bf 37}, 5087 (1996).

\bibitem{Davies} J.A.~Nixon and J.H.~Davies, Phys. Rev. B {\bf 41},
7929 (1990).

\bibitem{Stopa} M.~Stopa, Phys. Rev. B {\bf 53}, 9595 (1996).

\bibitem{Mirlin96} A.~D.~Mirlin, E.~Altshuler, and P.~W\"olfle,
Ann. Physik {\bf 5}, 281 (1996).

\bibitem{braun} Y.~Gefen, D.~Braun, and G.~Montambaux,
Phys. Rev.~Lett.~{\bf 73}, 154 (1994).

\bibitem{McCann} E.~McCann and K.~Richter, Europhys.~Lett.~{\bf 43}, 241 (199),
Phys. Rev. B {\bf 59}, 13026 (1999).


%Sec. VII
%%%%%%%%%%%%%%

\bibitem{CBrev} M.~Kastner, Rev. Mod. Phys. {\bf 64}, 849 (1992).

\bibitem{ambegaokar} V.~Ambegaokar and U.~Eckern, Phys.~Rev.~Lett.
{\bf 65}, 381 (1990); U.~Eckern, Z.~Phys. {\bf B 42}, 389 (1991).

\bibitem{ESra} U.~Eckern and P.~Schwab, Adv.~Phys.~{\bf 44}, 387 (1995).

\bibitem{levy90} L.P.~L\'evy, G.~Dolan, J.~Dunsmuir, and H.~Bouchiat,
Phys. Rev. Lett. {\bf 64}, 2074 (1990).

\bibitem{Webb1} V.~Chandrasekhar, R.A.~Webb, M.J.~Brady, M.B.~Ketchen,
W.J.~Gallagher, and A.~Kleinsasser,
Phys. Rev. Lett. {\bf 67}, 3578 (1991).

\bibitem{Webb2} P.~Mohanty, E.M.Q.~Jariwala, M.B.~Ketchen,
            and R.A.~Webb, in {\em Quantum Coherence and Decoherence},
            edited by K.~Fujikawa and Y.A.~Ono (Elsevier, 1996).

\bibitem{Stern} T.~Ando, A.B~Fowler, and F.~Stern, Rev. Mod. Phys. 
{\bf 54}, 437 (1982).

\bibitem{JDS}R.~Jalabert and S.~Das~Sarma Phys. Rev. B {\bf 40},
9723 (1989).

\bibitem{aslamazov} L.G.~Aslamazov and A.I.~Larkin,
Sov.~Phys.-JETP {\bf 40}, 321 (1975).

\bibitem{doubleA} B.L.~Altshuler and A.G.~Aronov,
in {\it Electron-electron interactions in Disordered systems,}
edited by A.L.~Efros and M.~Pollak (North-Holland, Amsterdam, 1985).

\bibitem{longinterac} D.~Ullmo, H.U.~Baranger, K.~Richter, F.~von~Oppen,
and R.A.~Jalabert, unpublished.

\bibitem{Vignale} G.~Vignale, Phys. Rev. B {\bf 50}, 7668 (1994).

\bibitem{bulk}  K.~Richter, H.U.~Baranger, F.~von~Oppen, and D.~Ullmo,
unpublished.

\bibitem{japan} Y.~Takane, J.\ Phys.\ Soc.\ Japan  {\bf 67}, 3003 (1998).

\bibitem{physicaE}
         D.~Ullmo, K.~Richter, H.U.~Baranger, F.~von~Oppen,
         and R.A.~Jalabert, Physica E {\bf 1}, 268 (1997).

%Sec. VIII
%%%%%%%%%%%%%%

\bibitem{not} T.M.~Fromhold {\em et al.}, Phys. Rev.~Lett.~{\bf 72}, 2608 (1994).

\bibitem{Ye} P.D.~Ye {\em et al.}, Phys. Rev.~Lett.~{\bf 74}, 3013 (1995).

\end{thebibliography}
